\documentclass[aps,prd,onecolumn,11pt,tightenlines,floatfix,nofootinbib,superscriptaddress]{revtex4-2}
\usepackage[T1]{fontenc}
\usepackage{amsmath,amssymb,bm,graphicx}
\usepackage[colorlinks=true,citecolor=blue,linkcolor=blue,urlcolor=blue]{hyperref}
\newcommand{\vect}[1]{\bm{#1}}

\begin{document}
\title{Quantum noise and torque correlations in a torsion-bar Sagnac speed meter}
\author{Yuki Inoue}
\email[Corresponding author: ]{iyuki@ncu.edu.tw}
\affiliation{Department of Physics, National Central University, Taoyuan, Taiwan}
\affiliation{Center for High Energy and High Field (CHiP), National Central University, Taoyuan, Taiwan}
\affiliation{Institute of Physics, Academia Sinica, Taipei, Taiwan}
\affiliation{Institute of Particle and Nuclear Studies, High Energy Acceleration Research Organization (KEK), Tsukuba, Japan}

\author{Mario Onglao}
\affiliation{Center for High Energy and High Field (CHiP), National Central University, Taoyuan, Taiwan}
\affiliation{Department of Physics, National Central University, Taoyuan, Taiwan}
\affiliation{National Institute of Physics, University of the Philippines - Diliman, Quezon City 1101, Philippines}

\author{Daiki Tanabe}
\affiliation{Center for High Energy and High Field (CHiP), National Central University, Taoyuan, Taiwan}
\affiliation{Department of Physics, National Central University, Taoyuan, Taiwan}
\affiliation{Institute of Particle and Nuclear Studies, High Energy Acceleration Research Organization (KEK), Tsukuba, Japan}

\author{Masashi Hazumi}
\affiliation{Department of Physics,National Central University, Taoyuan, Taiwan}
\affiliation{Center for High Energy and High Field (CHiP), National Central University,  Taoyuan, Taiwan}
\affiliation{INFN Sezione di Ferrara, Via Saragat 1, 44122 Ferrara, Italy}

\author{Vivek Kumar}
\affiliation{Department of Physics, National Central University, Taoyuan, Taiwan}
\affiliation{Center for High Energy and High Field (CHiP), National Central University, Taoyuan, Taiwan}

\author{Apimook Watcharangkool}
\affiliation{National Astronomical Research Institute of Thailand (Public Organization), Chiang Mai 50180, Thailand}
\date[]{}
\begin{abstract}
Quantum fluctuations entering through weakly transmitting mirrors can
limit a torsion-bar Sagnac speed meter for low-frequency gravitational-wave
detection, even when its optical layout suppresses radiation-pressure back action. We show that, in the
{C}ryogenic sub-{H}z cROss torsion bar detector with
quantum NOn-demolition {S}peed meter (CHRONOS), vacuum fields
entering from behind the end and middle test masses can dominate the
low-frequency quantum noise despite cancellation of the torques on opposite
ends of each bar. For a reference configuration with mirror power
transmissions of only $10^{-7}$, these fields contribute 89.7\% of the total
quantum-noise power spectral density at 0.1 Hz. Omitting them would
underestimate the amplitude spectral density by a factor of 3.11.
Our input--output model connects the correlations produced by successive
optical interactions to the forces on each bar and the measured signal.
This connection shows why strong torque cancellation does not, by itself,
ensure a corresponding improvement in detector sensitivity. Replacing mirror
transmission with absorption at fixed reflectivity leaves the excess quantum
noise essentially unchanged for the two operating points studied.
The predicted signal response and quantum noise agree with independent
FINESSE 3 calculations as recycling detuning and homodyne angle are varied.
These results establish rear vacuum fields as a limiting noise source in
this torsion-bar speed meter and provide quantitative requirements for its
mirror and readout design.
\end{abstract}
\maketitle
\raggedbottom
\section{Introduction}
Laser-interferometric gravitational-wave detectors, including Advanced
LIGO~\cite{Aasi2015}, Advanced Virgo~\cite{Acernese2015}, and
KAGRA~\cite{Akutsu2019}, measure differential optical path lengths between test
masses separated by kilometers to detect gravitational-wave strain. The first
direct detection of gravitational waves from a binary black-hole merger by
Advanced LIGO~\cite{Abbott2016} opened a new observational channel to the
Universe. Extending observations below the principal band of these instruments
into 0.1--10 Hz could provide access to intermediate-mass black-hole binaries
and stochastic backgrounds~\cite{CHRONOSProgram2026,CHRONOSScience2026}.

A torsion-bar antenna measures low-frequency gravitational waves as angular motion~\cite{Ando2010}. Reaching its quantum-noise limit requires separate measures against classical noise: cryogenic operation and low mechanical loss reduce thermal noise~\cite{Saulson1990,CHRONOS2025}, while vibration isolation and control of translation-to-rotation coupling suppress seismic coupling. Gravity-gradient noise instead arises from direct gravitational coupling to environmental density fluctuations, requiring site assessment and environmental monitoring or subtraction~\cite{Harms2019}. As these classical contributions are reduced, radiation-pressure back action becomes an important sensitivity limit. TOBA has been used for stochastic-background searches
with one and two detectors~\cite{Ishidoshiro2011,Shoda2014}, as well as readout
of multiple rotational degrees of freedom~\cite{Shoda2017}. The Torsion
Pendulum Dual Oscillator (TorPeDO) has also been developed as a low-frequency
gravity sensor that optically measures the differential rotation of two torsion
pendulums~\cite{McManus2017}.

The speed-meter principle, in which resonators extract a signal proportional to velocity~\cite{Braginsky2000}, has been extended to gravitational-wave interferometers~\cite{PurdueChen2002}. In a Sagnac speed meter, the same optical field samples the two arms in opposite orders. Correlations between these successive interactions produce a velocity-sensitive response and reduce back action~\cite{Chen2003,Danilishin2019}. CHRONOS combines this topology with cryogenic torsion bars, ring cavities, and dual recycling cavities~\cite{CHRONOS2025}, so the model must also account for forces acting at opposite ends of each bar. The quantum model must therefore track two kinds of correlation: correlations between successive optical interactions in the Sagnac paths, and correlations between the left and right forces that determine the torque on a shared bar. Both arise from quantum fluctuations. The measurement--back action cross term in the detected noise is obtained after this torque propagates through the rotational response and interferometer readout.

The CHRONOS science program encompasses low-frequency gravitational-wave
observations~\cite{CHRONOSProgram2026,CHRONOSScience2026}, searches for
oscillating forces from dark matter~\cite{Graham2016,InoueDM2026}, measurements
of geophysical gravity-gradient variations~\cite{OnglaoGeophysics2026}, and
tests of Yukawa-type interactions using Newtonian torque cancellation
(Beyond Gravity)~\cite{InoueYukawa2026}. Converting these measurements into
constraints on physical quantities or coupling constants requires a
target-specific signal response, whereas the underlying quantum noise of
angular readout can be evaluated within a common framework.

Previous CHRONOS studies have evaluated the full noise
budget~\cite{CHRONOSNoise2026} and torque cancellation of classical intensity
noise~\cite{Tanabe2026}. Additional vacuum inputs, however, can limit
speed-meter sensitivity~\cite{Danilishin2004} and modify quantum correlations
in nonideal Sagnac interferometers~\cite{Danilishin2015}; small mirror
transmission alone therefore does not establish a negligible readout-noise
contribution. Here we distinguish vacuum fields entering from behind the end test
masses (ETMs) and middle test masses (MTMs) of the rings from the main input,
and quantify their individual quantum-noise contributions using input--output
relations that include cavity propagation, bar rotational dynamics, and
signal calibration.

We begin with creation and annihilation operators on each optical path and
their scattering and propagation relations. The linear input--output equations
are transformed to the two-photon representation~\cite{CavesSchumaker1985} and
organized into a common denominator $M$, input-specific coefficients $C_q$, and
a signal vector $\vect D$~\cite{BuonannoChen2001}. 

The paper is organized as follows. Section~\ref{sec:fields} defines the optical
configuration and rotational degrees of freedom, and Sec.~\ref{sec:readout}
derives the quantum input--output relations, signal transfer function, and
quantum noise. Section~\ref{sec:pi-operating-point} selects two operating points
using PI curves, and Sec.~\ref{sec:rear} evaluates rear vacuum fields and correlations
of the end torques. Section~\ref{sec:validation} compares the model with FINESSE 3,
followed by a comparison of transmission and absorption at fixed reflectivity
in Sec.~\ref{sec:absorption-comparison}. Section~\ref{sec:discussion} discusses
the input-resolved noise contributions, the effects of PRC detuning on power and relative readout phase, and full-noise sensitivity. Section~\ref{sec:conclusions} presents our conclusions.

\section{Optical configuration and rotational degrees of freedom}
\label{sec:fields}
\subsection{Configuration, inputs, and reference point}
\begin{figure}[t]
\centering\includegraphics[width=\textwidth]{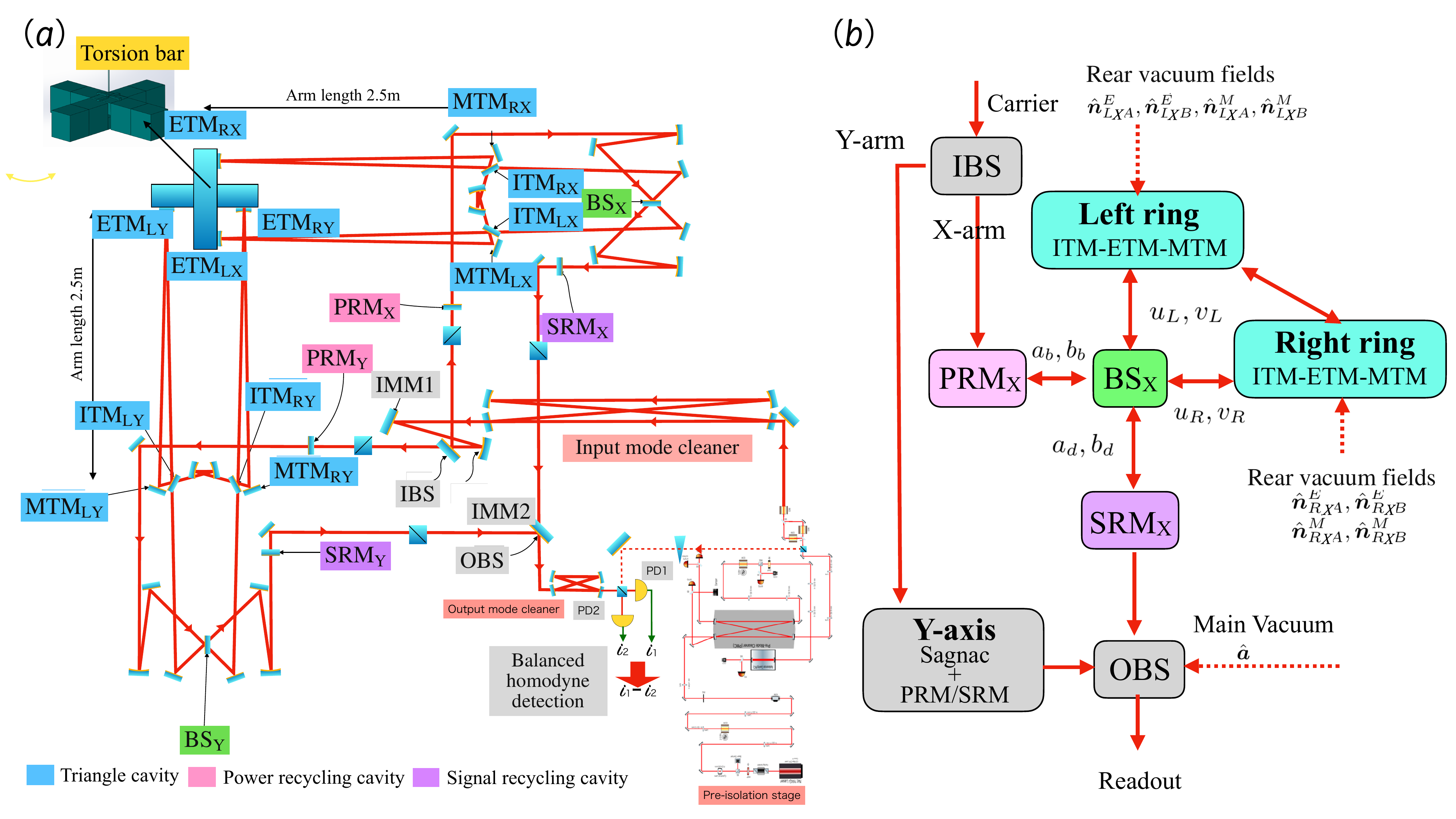}
\caption{Optical configuration and quantum inputs of CHRONOS. (a) Two-axis ring-Sagnac interferometer with PRMs, SRMs, and balanced homodyne readout, adapted from Inoue et al.~\cite{CHRONOS2025}. Each axis contains left and right ITM--ETM--MTM rings. (b) Expanded block diagram of the X axis. The IBS and OBS are shared; the Y-axis Sagnac and recycling system is represented by one block. The main vacuum $\hat{\vect a}$ enters at the OBS, while rear vacuum fields $\hat{\vect n}^{E,M}$ enter from behind the ETMs and MTMs of both rings. A and B denote the two circulation directions. Local mirror operators and ring propagation are detailed in Appendix~\ref{app:operator_locations}.}
\label{fig:topology}
\end{figure}

Figure~\ref{fig:topology} shows the optical configuration and a block diagram
for one axis. Two independent torsion bars are suspended on separate wires;
their optical systems are labeled by $j=X,Y$. Each bar has left and right
rings, each consisting of an input test mass (ITM), an ETM, and an MTM. The two
paths traverse the rings in opposite orders and recombine at BS$_j$. The left
and right test masses are connected by the same torsion bar. The two axes share
the input and output beam splitters (IBS and OBS). The power- and
signal-recycling mirrors (PRM and SRM) close the bright- and dark-port feedback
paths to form the power- and signal-recycling cavities (PRC and SRC).

We retain quantum fields entering from behind the ETMs and MTMs, while treating the
folding mirrors in the SRC, Sagnac, output, and local-oscillator (LO) paths as
ideal reflectors. The omitted vacuum contributions are assessed in this work for the symmetric differential readout (Appendix~\ref{app:validation}); they are below 1\% in the tested configurations. Their propagation lengths are nevertheless retained
because the associated delays affect optical phase. Suspension, damping, and
control are outside the quantum model. The main-laser spot lever is
$\ell=0.54365\,\mathrm m$; the
spot separation $2\ell=1.0873\,\mathrm m$ is also used in the thermal-noise
conversion. Table~\ref{tab:key-parameters} lists the principal parameters.
The decomposition into main-input, ETM-rear, and MTM-rear noise contributions, and the torque-cancellation calculation, use the operating point that minimizes the minimum of the all-noise power-law-integrated (PI) curve. Transfer-function and sensitivity comparisons also include the point that minimizes its logarithmic mean, as defined in Sec.~\ref{sec:pi-operating-point}.
The non-reflected ETM and MTM fields are transmitted in the reference model.
Replacement of transmission with absorption is examined separately in
Sec.~\ref{sec:absorption-comparison}.
Our main objective is an analytical quantum model of the interferometer.
Classical noise is included when selecting operating points and in the
full-noise comparison in the Discussion; details of the underlying detector and
noise assumptions are given in the CHRONOS
studies~\cite{CHRONOS2025,CHRONOSNoise2026}.
\begin{table}[!htbp]
\caption{Principal reference parameters. Phases are optical reflection phases, and calculations retain unrounded values.}
\label{tab:key-parameters}
\begin{ruledtabular}\begin{tabular}{ll}
Quantity & Reference value\\
\hline
Wavelength $\lambda$; input power $P_{\rm in}$ & $1064\,\mathrm{nm}$; $1\,\mathrm W$\\
Carrier frequency $\omega_0=2\pi c/\lambda$ & $1.77035\times10^{15}\,\mathrm{rad/s}$\\
Single-bar moment of inertia $I$ & $19.86\,\mathrm{kg\,m^2}$\\
Spot lever $\ell$ & $0.54365\,\mathrm m$\\
Differential moment $I_d=I/2$ & $9.930\,\mathrm{kg\,m^2}$\\
ITM power transmission $T_I$ & $10^{-4}$\\
ETM/MTM transmissions $T_E,T_M$ & $10^{-7}$ each\\
PRM / SRM power reflectivity & $0.9$ / $0.5$\\
Ring round-trip length $L_{\rm rt}$ & $5.083\,\mathrm m$\\
PR / SR path length & $5.069$ / $2.622\,\mathrm m$\\
$\Phi_{\rm PR}$; $\Phi_{\rm SR}$ & $-42.565^\circ$; $83.669^\circ$\\
LO phase $\zeta_{\rm LO}$ & $-39.042^\circ$\\
\end{tabular}\end{ruledtabular}
\end{table}

We denote the PR and SR mirror setting angles by $\delta_{\rm PR}$ and $\delta_{\rm SR}$, and the independent LO optical setting by $\zeta_{\rm LO}$. Their relation to the optical phases is $\Phi_{\rm PR}=\pi/2-2\delta_{\rm PR}$, $\Phi_{\rm
SR}=\pi/2-2\delta_{\rm SR}$, and $\Phi_h=\pi/2-\zeta_{\rm LO}$. The SRC
feedback is resonant at $\delta_{\rm SR}=0$; the corresponding PRC resonance is
at $\delta_{\rm PR}=90^\circ$. These conditions include the return phase of the
Sagnac, not only the mirror phase (Appendix~\ref{sec:sr-phase-reference}). The
optical setting $\zeta_{\rm LO}$ is independent of SRC detuning and distinct
from the quadrature projection angle $\zeta$.

\subsection{Creation and annihilation operators}
Let $e$ label an optical element and $k$ one of its connected paths. In the ring, $k=I,E,M$ identifies a path toward the ITM, ETM, or MTM, respectively; rearA and rearB label the two rear paths. Appendix~\ref{app:local} maps these physical paths to the row and column indices of each mirror matrix. For example,
$e=\mathrm{ETM}_{LX}$ is the ETM of the left ring on the X axis.
Tables~\ref{tab:element-labels} and~\ref{tab:main-symbols} summarize the
element labels and frequently used quantities.
\begin{table}[!htbp]
\caption{Optical elements labeled by $e$. The indices $i=L,R$ denote the left and right rings, and $j=X,Y$ labels the bar axis.}
\label{tab:element-labels}
\begin{ruledtabular}\begin{tabular}{lp{0.26\textwidth}}
Element & Function\\
\hline
$\mathrm{ITM}_{ij}$, $\mathrm{ETM}_{ij}$, $\mathrm{MTM}_{ij}$ & Ring input,
end, and middle mirrors\\
$\mathrm{BS}_j$ & Sagnac splitting and recombination\\
$\mathrm{PRM}_j$, $\mathrm{SRM}_j$ & Bright-/dark-port recycling mirrors\\
IBS, OBS & Two-axis input splitting/output combining\\
$\mathrm{BS}_{\rm LO}$, $\mathrm{BS}_{\rm HD}$ & Local-oscillator pickoff; homodyne mixing\\
\end{tabular}\end{ruledtabular}
\end{table}

We define $\hat c_{e,k}^{\rm in/out}(\omega)$ as the annihilation operator at
frequency $\omega$ on path $k$ of element $e$, and $\hat c_{e,k}^{\rm
in/out\dagger}(\omega)$ as the creation operator of the same mode. For example,
$\hat c_{\mathrm{ETM}_{LX},I}^{\rm in}$ describes light incident on the ETM
from the ITM in the left X ring. Independent external vacuum inputs $q$ satisfy $[\hat
c_q(\omega),\hat
c_{q'}^\dagger(\omega')]=2\pi\delta_{qq'}\delta(\omega-\omega')$. Internal
operators follow from propagation of these inputs and are not additional
independent reservoirs. Figure~\ref{fig:operator-summary} in
Appendix~\ref{app:operator_locations} identifies the corresponding front,
rear, and intracavity paths for each operator.

In a frame rotating at the carrier angular frequency $\omega_0$,
the field envelope is written in terms of the local optical power $P$ and
carrier phase $\varphi$ as

\begin{equation}
 \hat c(t)=\sqrt{\frac{P}{\hbar\omega_0}}e^{i\varphi}+\delta\hat c(t).
 \label{eq:carrier-power-phase}
\end{equation}

Thus $P/(\hbar\omega_0)$ is the stationary photon flux.
The often-used coherent amplitude $\bar\alpha=\langle\hat c\rangle$
equals $\sqrt{P/(\hbar\omega_0)}e^{i\varphi}$; it is an amplitude,
not a power. Below we expose $P$ and $\varphi$ in the signal, force,
and readout coefficients. A path label specifies where the power is
evaluated. We denote by $\Sigma_{e;kl}$ the complex amplitude reflection or transmission
coefficient from incident path $l$ to outgoing path $k$ of mirror $e$.
Its magnitude gives the field-amplitude ratio, and its argument gives the
phase change. In our convention, reflection entries are $r_e e^{\pm i\Phi_e}$
and transmission entries are $it_e$, where $r_e=\sqrt{R_e}$ and
$t_e=\sqrt{T_e}$; $R_e$ and $T_e$ are power reflectance and transmittance, respectively.
Let $\hat\theta_e$ be the local angle associated with the mirror spot.
Linearization about the carrier gives
\begin{equation}
 \begin{aligned}
 \delta\hat c^{\rm out}_{e,k}
   &=\sum_l\Sigma_{e;kl}\delta\hat c^{\rm in}_{e,l}
       +\gamma_{e,k}\hat\theta_e,\\
 \delta\hat c^{\rm out\dagger}_{e,k}
   &=\sum_l\Sigma_{e;kl}^*\delta\hat c^{\rm in\dagger}_{e,l}
       +\gamma_{e,k}^*\hat\theta_e,
 \end{aligned}
 \label{eq:annihilation-local}
\end{equation}
where the sum includes all incident modes, including auxiliary vacuum fields,
and the angular source follows directly from the incident power and phase:

\begin{equation}
 \gamma_{e,k}=\sum_l\frac{\partial\Sigma_{e;kl}}{\partial\theta_e}
 \sqrt{\frac{P^{\rm in}_{e,l}}{\hbar\omega_0}}e^{i\varphi^{\rm in}_{e,l}}.
 \label{eq:angle-source-power}
\end{equation}
 The network is constructed by connecting these coefficients and
propagation delays $\delta\hat c^{\rm in}(t)=\delta\hat c^{\rm out}(t-L/c)$
over distance $L$. Stationary carrier propagation phases are included in the
reference planes and $\Phi_e$. Appendix~\ref{app:local} derives the two- and four-port reflection and
transmission matrices, their angular couplings, and the vacuum normalization
used in this local input--output relation.

We then transform the upper and lower sidebands $\hat c_\pm(\Omega)=\hat
c(\omega_0\pm\Omega)$ to amplitude and phase
quadratures~\cite{CavesSchumaker1985}. With the Fourier convention $e^{+i\Omega
t}$ and $\Omega=2\pi f$, we write
\begin{equation}
 \begin{aligned}
 \hat{\vect d}&=\begin{pmatrix}\hat c_-\\\hat c_+^\dagger\end{pmatrix},\qquad
 \mathsf U_{\rm q}=\frac1{\sqrt2}\begin{pmatrix}1&1\\-i&i\end{pmatrix},\\
 \hat{\vect u}_{e,k}&=\sqrt{\hbar\omega_0}\,\mathsf U_{\rm q}\hat{\vect d}^{\rm in}_{e,k},\qquad
 \hat{\vect v}_{e,k}=\sqrt{\hbar\omega_0}\,\mathsf U_{\rm q}\hat{\vect d}^{\rm out}_{e,k}
 \end{aligned}
 \label{eq:annihilation-to-quadrature}
\end{equation}
Under this unitary transformation, each complex amplitude coefficient in Eq.~\eqref{eq:annihilation-local} becomes $\mathsf U_{\rm
q}\operatorname{diag}(\Sigma,\Sigma^*)\mathsf U_{\rm q}^\dagger$, and each
delay becomes $e^{-i\Omega L/c}\mathbb I_2$. The two-photon transformation above, Eq.~\eqref{eq:annihilation-to-quadrature}, therefore expresses the same creation--annihilation-operator calculation in another basis. Here $\operatorname{diag}(\Sigma,\Sigma^*)$ contains the annihilation-side coefficient $\Sigma$ and its creation-side conjugate $\Sigma^*$ as two separate diagonal entries; their order follows the definition of $\hat{\vect d}$.

The external inputs are independent quantum vacuum fields labeled by their injection boundaries. Measurement--back action correlations arise from the optical and mechanical responses to a shared input. We denote
the annihilation operator entering the open OBS boundary by $\hat a(\omega)$,
and those entering from behind each ETM and MTM by $\hat n^{E}_{ij\nu}(\omega)$ and
$\hat n^{M}_{ij\nu}(\omega)$, respectively; their Hermitian conjugates are the
creation operators. Here $i=L,R$ labels the ring, $j=X,Y$ the bar axis, and
$\nu=A,B$ the circulation direction after transmission. Direction A is
ITM$\to$ETM$\to$MTM$\to$ITM, and B is the reverse. Figure~\ref{fig:topology}(b) identifies the rear vacuum fields for each mirror and circulation direction.

Applying the same $\mathsf U_{\rm q}$ to these sidebands gives the
vacuum-normalized two-photon operators $\hat{\vect a}$, $\hat{\vect
n}^{E}_{ij\nu}$, and $\hat{\vect n}^{M}_{ij\nu}$. On each of the X and Y arms, the two ETMs and two MTMs give four mirrors in total, with two rear vacuum fields per mirror: four ETM inputs and four MTM inputs. Combining both arms therefore gives eight ETM inputs and eight MTM inputs, in addition to the single main input.  We retain these seventeen fields individually, indexed by $q$, and omit other vacuum inputs with negligible contributions to the differential readout. Each is an
independent zero-mean vacuum with unit one-sided symmetrized spectral density.
Omission affects only the sum of vacuum-noise contributions: the stationary carrier, optical boundary conditions, and bar dynamics used to calculate every retained response are kept fixed. The optical return coefficients are introduced in Sec.~\ref{sec:ring-connections}, and the coupled optomechanical denominator in Eq.~\eqref{eq:chen-common-denominator}. Appendix~\ref{app:validation} quantifies the omission error. Hats denote
operators and overbars denote carrier fields. We use $P$ for
optical power and $I$ for moment of inertia; $\Pi$ denotes propagation.
$P_{\rm in}$ is the laser input power, $P_{\rm ring}$ a one-direction
ring power, and $P_{\rm LO}$ the power incident on the homodyne mixer
from the LO path. These powers refer to different locations. The ideal-limit
correspondence with Chen's Sagnac notation~\cite{Chen2003} is collected in
Appendix~\ref{sec:chen-notation}; the common-denominator notation
$a,b,C,D,M$ follows Buonanno and Chen~\cite{BuonannoChen2001}.

\begin{table}[!htbp]
\caption{Principal notation. Matrices used only in intermediate derivations are defined in the corresponding appendixes.}
\label{tab:main-symbols}
\begin{ruledtabular}\begin{tabular}{lp{0.30\textwidth}}
Symbol & Meaning\\
\hline
$\hat c_{e,k}^{\rm in/out},\hat c_{e,k}^{\rm in/out\dagger}$ & Path
annihilation/creation operators\\
$\hat{\vect u}_{e,k},\hat{\vect v}_{e,k}$ & Incident/outgoing two-photon
fields\\
$P$, $\varphi$ & Local carrier power [W] and phase\\
$\hat{\vect a}$ & Main vacuum at the open OBS boundary\\
$\hat{\vect n}^{E}_{ij\nu},\hat{\vect n}^{M}_{ij\nu}$ & Independent ETM/MTM
rear vacuum fields\\
$\theta_j$, $\theta_{\rm sig}$ & Bar angle; external differential angle\\
$I_d=I/2$, $\theta_{\rm SQL,d}$ & Differential moment of inertia; angular SQL ASD\\
$\beta_{\rm ring},\beta_{\rm sag}$ & Ring-reflection and bare-Sagnac phases\\
$M,C_q,\vect D$ & Common denominator, noise matrix, signal vector\\
$\lambda_{\rm sig},\lambda_{\rm LO}$ & Carrier-dependent readout rows\\
$H_\theta$, $S_\theta$ & Angular TF [W/rad]; angular noise PSD\\
\end{tabular}\end{ruledtabular}
\end{table}

\subsection{Connecting the rings, Sagnac, and recycling cavities}
\label{sec:ring-connections}
Each ring supports the path ITM$\to$ETM$\to$MTM and its reverse. At every
mirror we include the transmitted rear vacuum field, as well as the phase modulation
due to the ETM angle, and solve the boundary condition for the field returning
to the ITM. The complex reflectivity of the resonant reference ring is
\begin{equation}
 r_{\rm cav}(\Omega)=
 \frac{r_I-r_Er_Me^{-i\Omega L_{\rm rt}/c}}
 {1-r_Ir_Er_Me^{-i\Omega L_{\rm rt}/c}}
 \label{eq:ring-scalar-reflection}
\end{equation}
Here $r_I,r_E,r_M$ are the amplitude reflectivities of the ITM, ETM, and MTM defined in Sec.~\ref{sec:fields}. Rear vacuum fields share this
circulation denominator, but their numerators and propagation distances depend
on their injection locations.

The Sagnac recombines at BS$_j$ the fields that traverse the two rings in
opposite orders. For the symmetric reference geometry with the mirrors held fixed, the reflection responses at the bright and dark ports are
\begin{equation}
 S_{bb}=S_{dd}=-e^{-i\Omega\tau_{\rm link}}r_{\rm cav}^2J,
 \qquad S_{bd}=S_{db}=0,
 \label{eq:sagnac-scalar-reflection}
\end{equation}
where $J=R(\pi/2)$ is a constant quadrature rotation and $\tau_{\rm
link}=(2L_b+L_c)/c$ is the propagation time outside the rings. $L_b$ is the length of each symmetric BS$_j$--ring branch and $L_c$ is the connection length between the two rings; these paths are connected explicitly in Appendix~\ref{app:optical}. Writing $r_{\rm
cav}=-|r_{\rm cav}|e^{-2i\beta_{\rm ring}}$ and continuously unwrapping the
phase from $\beta_{\rm ring}(0)=0$ gives
\begin{equation}
 \beta_{\rm sag}=2\beta_{\rm ring}
       +\frac{\Omega\tau_{\rm link}}2+\frac\pi2
 \label{eq:sagnac-phase-exact}
\end{equation}
Finally, the return fields from the PRM and SRM are connected to the bright and
dark ports. Recycling therefore closes a phase-sensitive optical feedback loop whose complex response determines both carrier buildup and sideband propagation. We
eliminate the circulating fields while retaining separate columns for each
input and the angular signal, then reconstruct the internal fields required for
radiation pressure by back substitution. Appendix~\ref{app:optical} derives Eqs.~\eqref{eq:ring-scalar-reflection}--\eqref{eq:sagnac-phase-exact} and the two-axis optical response used in the rotational dynamics of Appendix~\ref{app:feedback}.

\subsection{Bar motion and forces at opposite ends}
For a spot lever $\ell$ measured from the rotation axis, the normal
displacements of the left and right ETMs are $\ell\theta_j$ and
$-\ell\theta_j$. The reflection-phase change $2k_0\ell\theta_j$, with
$k_0=\omega_0/c$, produces the signal field. Interference of the same carrier
with the vacuum fluctuations produces mirror forces whose difference drives the
bar:
\begin{equation}
 \hat\tau_j=\ell(\hat F_{Lj}-\hat F_{Rj}),\qquad
 \hat\theta_j=-\frac{\hat\tau_j}{I\Omega^2}+s_j\theta_{\rm sig},
 \label{eq:bar-torque}
\end{equation}
Here $\hat F_{Lj}$ and $\hat F_{Rj}$ are the fluctuating radiation-pressure forces on the left and right ETMs of arm $j$, expressed along the common displacement convention used above, and $s_X=1/2,s_Y=-1/2$. Appendix~\ref{app:feedback} derives the ETM forces from optical momentum flux and solves Eq.~\eqref{eq:bar-torque} together with the optical relations to obtain rotational feedback and internal fields. The quantity $\theta_{\rm sig}$ is the externally
imposed differential angle before optical feedback, not the response to a
gravitational wave of arbitrary direction. The differential coordinate
$\theta_d=\theta_X-\theta_Y$ has moment of inertia $I_d=I/2$, so its SQL normalization is
\begin{equation}
 \theta_{\rm SQL,d}=\sqrt{\frac{2\hbar}{I_d\Omega^2}}
                  =\sqrt{\frac{4\hbar}{I\Omega^2}}.
 \label{eq:differential-sql}
\end{equation}
The standard quantum limit (SQL) is obtained by balancing measurement imprecision and radiation-pressure back action for an ideal continuous measurement of this free rotational coordinate when their cross correlation vanishes~\cite{Danilishin2012}. We use its ASD to normalize the signal in Eq.~\eqref{eq:chen-multiport-output}, allowing readout-noise levels to be compared on the scale set by $I_d$ and $\Omega$. A nonzero measurement--back action correlation can reduce the total noise below this uncorrelated reference.

\section{Quantum input--output relations with a common denominator}
\label{sec:readout}
\subsection{Input-specific noise and signal coefficients}
We solve the optical equations together with Eq.~\eqref{eq:bar-torque} and
eliminate the internal fields. The symmetric two-axis system admits a common
denominator, giving the signal field immediately before detection as
\begin{equation}
 \begin{aligned}
 \hat{\vect b}_{\rm sig}\simeq\frac1M\Biggl[&e^{-2i\beta_{\rm sag}}\Bigl(
 C_{a,\rm sig}\hat{\vect a}\\
 &+\sum_{ij\nu} C^{E}_{ij\nu,\rm sig}\hat{\vect n}^{E}_{ij\nu}
  +\sum_{ij\nu} C^{M}_{ij\nu,\rm sig}\hat{\vect n}^{M}_{ij\nu}\Bigr)\\
 &+e^{-i\beta_{\rm sag}}\vect D_{\rm sig}
       \frac{\theta_{\rm sig}}{\theta_{\rm SQL,d}}\Biggr].
 \end{aligned}
 \label{eq:chen-multiport-output}
\end{equation}
Each $C$ is a $2\times2$ matrix whose rows label output quadratures and columns
label input quadratures; $\vect D$ is a two-component column. The first term is
the main input $a$, and the two sums describe the ETM and MTM rear vacuum fields. The
approximate equality indicates omission of the remaining inputs. We use $C_q$
collectively for the coefficients of each of the seventeen retained inputs and
calculate propagation to the LO side for each of the same inputs. The signal
gain is included in $\vect D$, without assuming a single effective
optomechanical coupling.

The common denominator contains both optical return paths and rotational
feedback. With bright and dark ports denoted by $b,d$ and their optical
feedback matrices by $\mathsf A_b,\mathsf A_d$, respectively,
\begin{equation}
 M=(1-\chi\kappa_0)\Delta_b\Delta_d
       -\chi(\Delta_dh_b+\Delta_bh_d).
 \label{eq:chen-common-denominator}
\end{equation}
where $\Delta_r=\det\mathsf A_r$ is an optical circulation denominator and
$\chi=-1/(I\Omega^2)$ is the free-rotor susceptibility. The coefficient
$\kappa_0$ converts angle to torque with the return fields held fixed; $h_r$ is
the numerator of the angle--light--torque feedback through port $r$. Appendix~\ref{app:chen-derivation} derives $M$ and the input-specific
numerators from the coupled optical-field and bar-angle equations.
Appendix~\ref{app:chen-components} substitutes the injection and
circulation coefficients into scalar expressions for $b_1,b_2$, all four
components of $C_q$, and both components of $\vect D$. It separates the
direct optical response from the vacuum-driven torque and shows an ETM
rear vacuum-field example explicitly.
For each retained input, the coefficients follow by exact algebraic rearrangement of the boundary conditions; only the input selection is approximate.

Let $\hat b_1$ and $\hat b_2$ be the two quadrature components of $\hat{\vect b}_{\rm sig}$ in Eq.~\eqref{eq:chen-multiport-output}. Their homodyne projection is $\hat b_\zeta=\hat b_1\sin\zeta+\hat b_2\cos\zeta$. For a fixed input $q$, write $C_{\mu\nu}=(C_{q,\rm sig})_{\mu\nu}$, where $\mu=1,2$ labels the output quadrature and $\nu=1,2$ the input quadrature. The projected coefficients are then $C_{11}\sin\zeta+C_{21}\cos\zeta$ and $C_{12}\sin\zeta+C_{22}\cos\zeta$. For a finite-amplitude LO, the detector output is $\hat o=\lambda_{\rm sig}\hat{\vect b}_{\rm sig}+\lambda_{\rm LO}\hat{\vect b}_{\rm LO}$. Appendix~\ref{app:detector} derives the carrier-dependent rows $\lambda_{\rm sig}$ and $\lambda_{\rm LO}$ from the difference between the two photodetector outputs. A residual carrier at the signal port couples LO fluctuations to this output~\cite{Steinlechner2015}, so both optical fields are retained.

Defining the projected numerators $\mathcal C_q=\lambda_{\rm sig}C_{q,\rm
sig}+\lambda_{\rm LO}C_{q,\rm LO}$ and $\mathcal D=\lambda_{\rm sig}\vect
D_{\rm sig}+\lambda_{\rm LO}\vect D_{\rm LO}$, the sensing function and total
quantum-noise ASD are
\begin{equation}
 \begin{aligned}
 H_\theta&\equiv\frac{\partial\langle\hat o\rangle}{\partial\theta_{\rm sig}}
       =\frac{e^{-i\beta_{\rm sag}}\mathcal D}{M\theta_{\rm SQL,d}},\\
 \sqrt{S_\theta}&\simeq\theta_{\rm SQL,d}
          \frac{\sqrt{\sum_q\mathcal C_q\mathcal C_q^\dagger}}{|\mathcal D|}.
 \end{aligned}
 \label{eq:sensing-function-chen}
\end{equation}
The complex transfer function $H_\theta$ has units W/rad and retains its phase.
Signal- and LO-path amplitudes from each input are combined coherently before the independent-input PSDs are summed.

The sensitivity curves in Figs.~\ref{fig:rearcompare}, \ref{fig:transfer}--\ref{fig:all-noise-operating-points}, \ref{fig:angle-minimum}, and \ref{fig:angle-logmean} are referred to the dimensionless strain signal $h(t)$. We use the differential-angle response $\theta_{\rm sig}(t)=\eta h(t)$, with the geometric coupling $\eta=0.93656$ for the adopted test-mass geometry~\cite{CHRONOS2025}. This is the free-rotation, long-wavelength calibration for the antenna-projected strain; the signal response for a specified source direction and polarization must be included when interpreting a source amplitude. Accordingly, $\sqrt{S_h(f)}=\sqrt{S_\theta(f)}/|\eta|$ and $S_h^{\rm cross}=S_\theta^{\rm cross}/|\eta|^2$. The cross term is defined in Eq.~\eqref{eq:quantum-budget}. Every input group and noise component uses this calibration. Since $h$ is dimensionless and its PSD is defined per unit frequency, $S_h$ has units $\mathrm{Hz}^{-1}$ and its square root has units $\mathrm{Hz}^{-1/2}$. Transfer-function plots retain $H_\theta$ [W/rad]. The common calibration factor cancels in relative noise changes, input fractions, normalized correlations, and relative model residuals.

Input locations remain explicit after readout: $\mathcal C_a$, $\mathcal
C^{E}_{ij\nu}$, and $\mathcal C^{M}_{ij\nu}$ are defined with the same detector
rows. Thus
\begin{equation}
 \begin{aligned}
 S_\theta&\simeq S_{\theta,a}+S_{\theta,E}+S_{\theta,M},\\
 S_{\theta,a}&=\frac{\theta_{\rm SQL,d}^2}{|\mathcal D|^2}\|\mathcal C_a\|^2,\\
 S_{\theta,E(M)}&=\frac{\theta_{\rm SQL,d}^2}{|\mathcal D|^2}
                  \sum_{ij\nu}\|\mathcal C^{E(M)}_{ij\nu}\|^2.
 \end{aligned}
 \label{eq:input-group-budget}
\end{equation}
Equation~\eqref{eq:input-group-budget} applies the independent-input PSD sum within and across the three groups, retaining correlations among paths driven by each input.

An equivalent form that displays the optical phase and mechanical feedback is
\begin{equation}
 H_\theta(f)=\frac{G_1(f)\mathcal F_-(f)-G_2(f)\mathcal F_+(f)}
 {\mathcal F_+(f)\mathcal F_-(f)-\Xi(f)/f^2}.
 \label{eq:sensing-trigonometric}
\end{equation}
Here $\mathcal F_\pm$ are two optical phase factors, $G_{1,2}$ are gains
projected onto the detection direction, and $\Xi/f^2$ is the rotational
feedback term. Appendix~\ref{app:sensing} factorizes the optical denominator
of Eq.~\eqref{eq:sensing-function-chen} to obtain
Eq.~\eqref{eq:sensing-trigonometric}, defining $\mathcal F_\pm,G_{1,2},\Xi$
in terms of the original input--output coefficients. Their full frequency dependence is retained rather
than replaced by constant poles and zeros.

\subsection{Shot noise, back action, and their correlation}
Let the detected response to input $q$ be $\vect g_q=e^{-2i\beta_{\rm
sag}}\mathcal C_q/M$. We denote by $\vect g_{0q}$ the response with mechanical
motion suppressed in noise propagation, and define $\vect g_{{\rm ba},q}=\vect
g_q-\vect g_{0q}$. Then
\begin{equation}
 \begin{aligned}
 S_\theta^{\rm shot}&=\frac{\sum_q\|\vect g_{0q}\|^2}{|H_\theta|^2},\qquad
 S_\theta^{\rm ba}=\frac{\sum_q\|\vect g_{{\rm ba},q}\|^2}{|H_\theta|^2},\\
 S_\theta^{\rm cross}&=\frac{2\operatorname{Re}\sum_q\vect g_{0q}
                     \vect g_{{\rm ba},q}^{\dagger}}{|H_\theta|^2},
 \end{aligned}
 \label{eq:quantum-budget}
\end{equation}
and $S_\theta\simeq S_\theta^{\rm shot}+S_\theta^{\rm ba}+S_\theta^{\rm
cross}$~\cite{Danilishin2012}. For each optical configuration, all three terms are calibrated with
the same full signal TF. Detuning or LO changes can alter both the noise
readout coefficients and the signal TF, so uncalibrated noise alone does not
identify the cause of a sensitivity change. Section~\ref{sec:prdetuning}
uses this decomposition to separate power and readout-phase effects. The cross term is a signed PSD describing interference between $\vect g_{0q}$ and $\vect g_{{\rm ba},q}$ for the same input $q$. It is evaluated before summing the independent inputs, and can either reduce or increase their total PSD.
The same decomposition applies within each input group. For any input set
$\mathcal J$, its contribution is $S_{\theta,\mathcal J}=\sum_{q\in\mathcal
J}\|\vect g_q\|^2/|H_\theta|^2$, using the same response coefficients.

\section{Two operating points selected by PI curves}
\label{sec:pi-operating-point}
\begin{figure}[tbp]
\centering\includegraphics[width=0.70\textwidth]{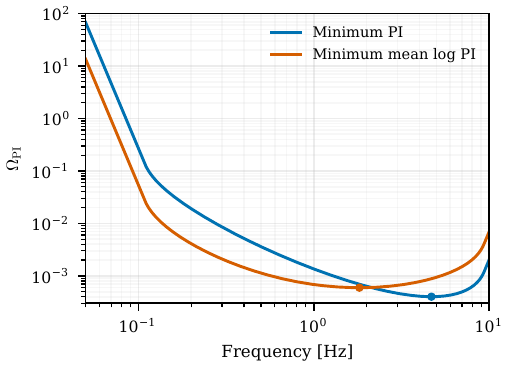}
\caption{PI curves selected by two objectives on the same axes. The blue solid curve minimizes the PI minimum, and the orange solid curve minimizes the logarithmic mean. Dots mark each curve's minimum. The displayed envelope starts at 0.05 Hz; integration and optimization remain restricted to 0.1--10 Hz. Both use the same full-noise model and observation assumptions: one detector, known noise, one year of observation, SNR 1, and an assumed response $\mathcal R=1$.}
\label{fig:pi-selection}
\end{figure}

Operating points are selected using power-law-integrated (PI)
curves~\cite{ThraneRomano2013} formed from the full noise budget. We add
coating and bar thermal noise, seismic noise, Newtonian noise, and intensity-noise PSDs from the detector noise budget~\cite{CHRONOSNoise2026,Tanabe2026} to the quantum noise, and vary PR, SR, and LO phases over
the 0.1--10 Hz analysis band. The optical configuration, 1 W input, $I$, and
$\ell$ are fixed; intensity noise and signal TF are recalculated at each
operating point.

We distinguish two objective functions. For $f_1=0.1$ Hz and $f_2=10$ Hz, we minimize
\begin{equation}
 \begin{aligned}
 J_{\min}&=\min_{f_1\leq f\leq f_2}\Omega_{\rm PI}(f),\\
 J_{\log}&=\frac{1}{\ln(f_2/f_1)}\int_{f_1}^{f_2}
             \ln\Omega_{\rm PI}(f)\,d\ln f.
 \end{aligned}
 \label{eq:pi-objectives}
\end{equation}
The second objective weights logarithmic frequency uniformly, and $\exp
J_{\log}$ is the geometric mean of the PI values.

\begin{table}[t]
\caption{Operating points selected by the two PI objective functions. The $J_{\min}$ column is the principal reference. $P_{\rm ring}$ is the circulating power in one direction.}
\label{tab:pi-operating-point}
\begin{ruledtabular}\begin{tabular}{lrr}
Quantity & Minimum $J_{\min}$ & Minimum $J_{\log}$\\
\hline
$\delta_{\rm PR}$ [deg] & $66.282$ & $25.301$\\
$\delta_{\rm SR}$ [deg] & $3.166$ & $0.705$\\
$\zeta_{\rm LO}$ [deg] & $-39.042$ & $-19.163$\\
$P_{\rm ring}$ [W] & $1609.75$ & $320.09$\\
$\min\Omega_{\rm PI}$ & $4.0142\times10^{-4}$ & $5.9611\times10^{-4}$\\
Frequency of minimum [Hz] & $4.687$ & $1.821$\\
$\exp J_{\log}$ & $2.4034\times10^{-3}$ & $1.5067\times10^{-3}$\\
\end{tabular}\end{ruledtabular}
\end{table}

Relative to the minimum-PI choice, the logarithmic-mean choice lowers the
geometric mean by 37.3\% but raises the minimum by 48.5\%
(Fig.~\ref{fig:pi-selection}). The criteria therefore select different sensitivity tradeoffs within the common 0.1--10 Hz band; its width is not optimized. Figure~\ref{fig:transfer} shows
the quantum noise and TF at both points, and their neighborhoods are tested
separately. Diagnostics of rear vacuum fields and torque cancellation use the $J_{\min}$
point.

\section{Rear vacuum fields and correlations of the end torques}
\label{sec:rear}
\subsection{Error from omitting rear vacuum fields}
We evaluate the minimum-PI reference point over 0.05 Hz--10 kHz. Input-specific
PSD fractions and their physical interpretation are discussed later; here we
quantify the sensitivity error caused by omitting the rear vacuum fields.

\begin{figure}[t]
\centering\includegraphics[width=0.95\textwidth]{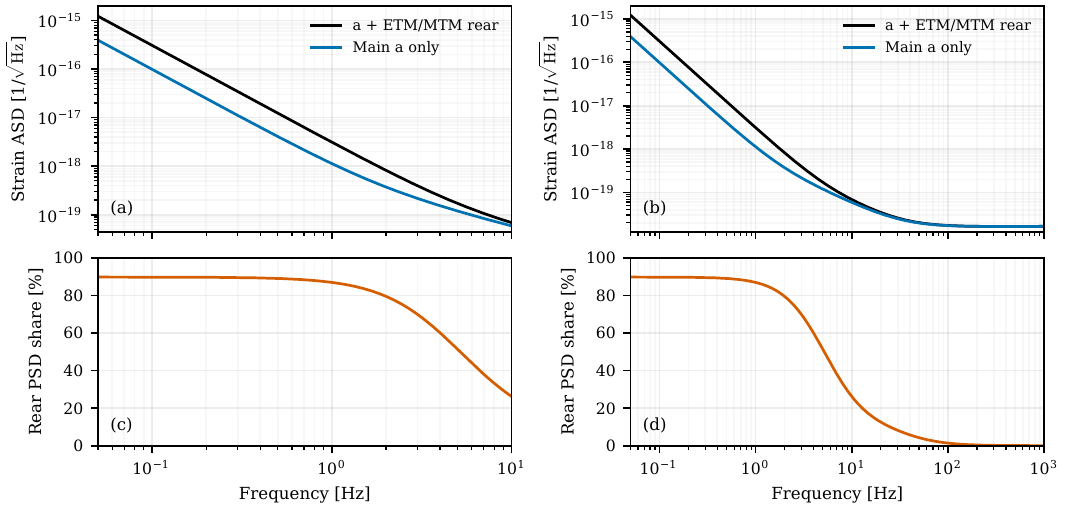}
\caption{Rear vacuum-field contribution. Black: strain ASD from the main input $a$ and sixteen ETM/MTM rear vacuum fields. Blue: main input only, with the sixteen rear contributions omitted. Finite mirror transmission is retained in both curves. Lower panels show the rear fraction of the three-group PSD. The left panels cover 0.05--10 Hz and the right panels extend to 1 kHz. All curves use the same signal TF and retain correlations within each input response.}
\label{fig:rearcompare}
\end{figure}

Figure~\ref{fig:rearcompare} shows the strain ASDs and fractional rear vacuum-field contribution to the total PSD. For the main-input curve, only the sixteen rear vacuum-field terms are omitted from the PSD sum; the optical matrices, carrier, signal TF, and finite mirror transmissions are kept fixed. At 0.1 Hz, the ASD of the three retained groups is
$3.063\times10^{-16}\,\mathrm{Hz}^{-1/2}$, whereas omission of the
rear vacuum fields gives $9.836\times10^{-17}\,\mathrm{Hz}^{-1/2}$, an
underestimate by a factor of 3.11. The factors are 2.76 at 1 Hz and 1.16 at 10
Hz. These inputs cannot be neglected even for $T_E=T_M=10^{-7}$, because their
contribution depends on cavity propagation, rotational susceptibility, and
angular calibration, not only on local transmission amplitudes.

Unlike this diagnostic omission, a physical change in mirror properties alters the optical model: perfect reflectors change the carrier and cavity response, while absorption at fixed reflectivity introduces an environmental vacuum field (Sec.~\ref{sec:absorption-comparison}). Figure~\ref{fig:portbudget} in Appendix~\ref{app:validation} shows the
frequency-dependent input fractions; that appendix also assesses the error
from omitting inputs outside the three retained groups.

\subsection{Cancellation of forces at opposite ends}
\label{sec:torque-cancellation}
Because the two mirror spots have opposite lever arms, an identical fluctuating force at both ends would give zero torque in Eq.~\eqref{eq:bar-torque}. Quantum-force fluctuations need not satisfy this condition: their relative phase and covariance must be calculated from the optical propagation.
To distinguish the Sagnac traversal-order correlation from the difference of
forces at the two bar ends, we first calculate force responses with the mirrors
fixed, denoted by the superscript $(0)$. For one independent vacuum input, let
$S_{LL},S_{RR},S_{LR}$ be the left and right auto-PSDs and their cross-PSD.
Then
\begin{equation}
 S_\tau^{(0)}=\ell^2(S_{LL}+S_{RR}-2\operatorname{Re}S_{LR}).
 \label{eq:torque-cancellation}
\end{equation}
Equal auto-PSDs do not by themselves imply perfect cancellation; the force
responses must match in both quadratures and phase.

\begin{figure}[tbp]
\centering\includegraphics[width=0.70\textwidth]{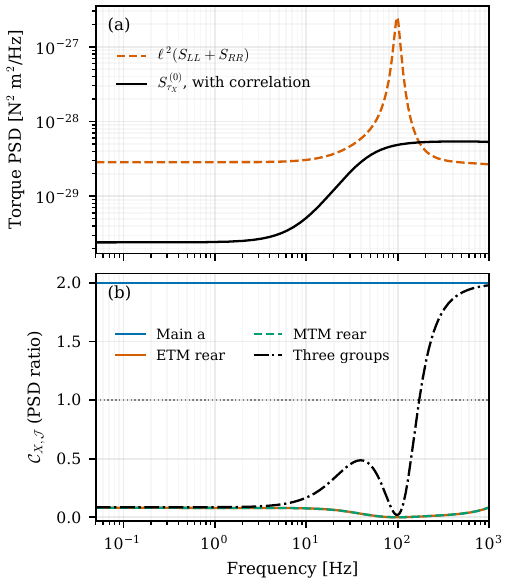}
\caption{Driving-torque correlations with the mirrors held fixed, for the main input $a$ and sixteen ETM/MTM rear vacuum fields. (a) Sum of the left and right torque auto-PSDs and the driving-torque PSD including their cross term. (b) Ratio $\mathcal C_{X,\mathcal J}$ for the X bar and input set $\mathcal J$, obtained after summing the PSDs within that set. Values below unity indicate cancellation; values above unity indicate enhancement. The Y bar gives the same result. It quantifies the change in driving-torque PSD; readout sensitivity additionally depends on the mechanical response and signal calibration.}
\label{fig:torquecancellation}
\end{figure}

We quantify the force correlation by the PSD ratio $\mathcal C=S_\tau^{(0)}/[\ell^2(S_{LL}+S_{RR})]$, shown in Fig.~\ref{fig:torquecancellation}. The plotted $\mathcal C_{X,\mathcal J}$ is this ratio for the X bar, after summing the numerator and denominator separately over the input set $\mathcal J$ (main input, ETM rear, MTM rear, or all three groups). It satisfies $0\leq\mathcal C\leq2$,
with values below unity indicating correlation-induced torque suppression. We
restrict this diagnostic to the seventeen inputs retained in the main model. At
0.1 Hz on the X bar, the auto-PSD sum $2.845\times10^{-29}\,\mathrm{N^2m^2/Hz}$
becomes $2.418\times10^{-30}\,\mathrm{N^2m^2/Hz}$ after including the cross
term, giving a ratio of 0.08498. The separately summed ETM- and MTM-rear groups each give 0.08152, whereas the main input gives 2; a single cancellation factor therefore cannot describe all three groups.

The force correlation determines the driving torque, while subsequent
mechanical feedback and readout correlations determine the detected noise. The
correlations in Eqs.~\eqref{eq:torque-cancellation}
and~\eqref{eq:quantum-budget} must therefore be distinguished.
Appendix~\ref{app:feedback} gives the mechanical response connecting them.

\section{Comparison with FINESSE 3}
\label{sec:validation}
The analytical coefficients constructed from local boundary conditions are
compared with an independent FINESSE 3 (version 3.0.2) model~\cite{FinesseSoftware}. It
uses the same optical parameters and connects the left and right ETMs through
the shared-bar force-to-displacement response
\begin{equation}
 \frac{\ell^2}{I s^2}\begin{pmatrix}1&-1\\-1&1\end{pmatrix},
 \qquad s=i\Omega.
\end{equation} Comparisons use the two points selected in
Sec.~\ref{sec:pi-operating-point} and the optical phases in
Table~\ref{tab:pi-operating-point}.

\begin{figure}[tp]
\includegraphics[width=\textwidth]{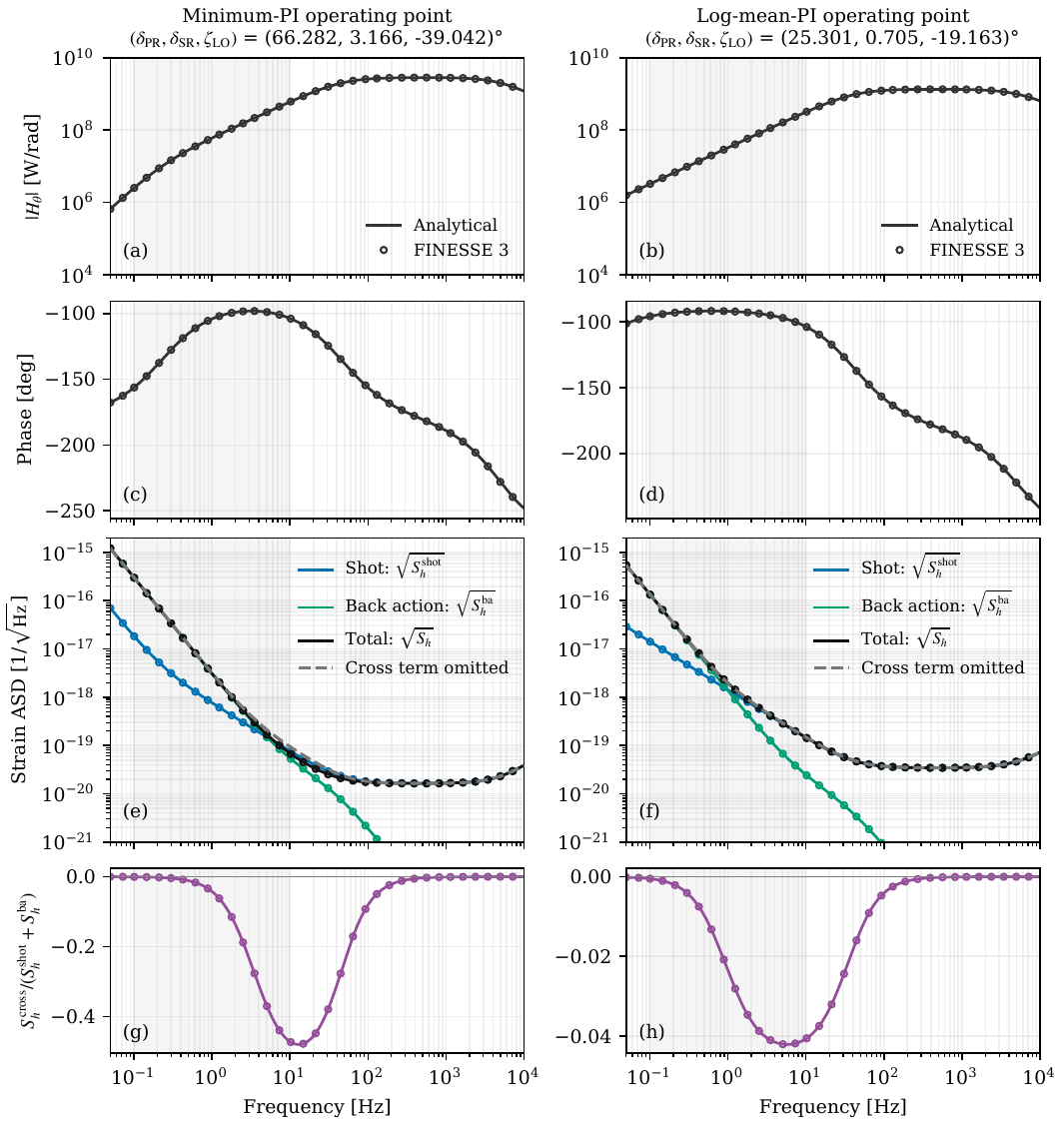}
\caption{Analytical model and FINESSE 3 at the two PI reference points. Left: minimum-PI point. Right: minimum-logarithmic-mean point. Lines are analytical results and circles are selected FINESSE 3 samples. Gray shading marks 0.1--10 Hz. (a),(b) Differential-angle TF magnitude; (c),(d) phase in the common Fourier convention; (e),(f) Shot $\sqrt{S_h^{\rm shot}}$ (blue), back action $\sqrt{S_h^{\rm ba}}$ (green), and total quantum noise including correlations $\sqrt{S_h}$ (black). The gray dashed curve shows $\sqrt{S_h^{\rm shot}+S_h^{\rm ba}}$, with only the cross term omitted. For both plotted points, the negative cross term reduces the black total ASD relative to this curve. (g),(h) The purple curves show the signed correlation term $S_h^{\rm cross}/(S_h^{\rm shot}+S_h^{\rm ba})$. The panels display the three terms of Eq.~\eqref{eq:quantum-budget}, calibrated by $S_h=S_\theta/|\eta|^2$; the square root of the summed PSD terms gives the black total ASD. The complementary input-group decomposition in Eq.~\eqref{eq:input-group-budget} is shown in Fig.~\ref{fig:portbudget}. The ASD axis extends down to $10^{-21}\,\mathrm{Hz}^{-1/2}$.}
\label{fig:transfer}
\end{figure}

Figure~\ref{fig:transfer} compares differential-angle TFs and quantum noise at
both reference points. Although classical noise enters the PI selection, the
sensitivities compared here are quantum noise calibrated with the same full TF.
We compare the complex TF over 0.05 Hz--10 kHz. Markers show selected computed frequency samples for readability; error estimates use every sample on the calculation grid.
The cross term in Eq.~\eqref{eq:quantum-budget} is a signed PSD and can be
negative, so its square root cannot be plotted as an independent positive ASD
on a logarithmic axis. We therefore add the gray dashed comparison curve
$\sqrt{S_h^{\rm shot}+S_h^{\rm ba}}$ to Fig.~\ref{fig:transfer}(e),(f), omitting
only the cross term while retaining the same optical response and signal
calibration. At both operating points, the cross term is negative throughout the displayed band, placing the black total-noise curve below the gray dashed curve; the two become visually indistinguishable where this contribution is small.
Panels (g),(h) display the sign and relative size of the cross term as
$S_h^{\rm cross}/(S_h^{\rm shot}+S_h^{\rm ba})$.

To test agreement around each operating point, we vary $\delta_{\rm PR}$ by
$\pm15^\circ$, $\delta_{\rm SR}$ by $\pm2^\circ$, and $\zeta_{\rm LO}$ by
$\pm15^\circ$ around each point. One angle changes at a time; input power,
moment of inertia, spot lever, and transmissions remain fixed. In particular, the LO
optical phase $\Phi_h$ is held fixed during SRC-detuning scans, keeping these
two controls independent. The carrier and signal TF are recalculated for every
case.

We compare the two reference cases and twelve variations using the maxima
of $\epsilon_H(f)$ and $\epsilon_N(f)$ defined in
Eq.~\eqref{eq:validation-residuals}. Shot-noise and back action ASD errors are
defined in the same way as the total-ASD error.

Across all fourteen cases, maximum relative differences are $1.04\times10^{-8}$
for the complex TF, $2.95\times10^{-8}$ for the shot and back action ASDs, and
$4.23\times10^{-10}$ for the total quantum ASD. The cross-PSD difference,
normalized by the sum of shot and back action PSDs, remains below
$1.29\times10^{-10}$.

Figures~\ref{fig:angle-minimum} and~\ref{fig:angle-logmean} in Appendix~\ref{app:validation} show the curves and frequency-resolved residuals over 0.05--100 Hz. Restricting the noise sum to the three groups in Eq.~\eqref{eq:input-group-budget} preserves the agreement reported above.

We also compare the direct solution of the coupled optical and mechanical boundary equations with their common-denominator representation in Eqs.~\eqref{eq:chen-multiport-output}--\eqref{eq:sensing-function-chen}, using the same frequency grid: the relative differences are below
$2.9\times10^{-10}$ for the complex TF and $2.3\times10^{-13}$ for the total
quantum ASD. Appendix~\ref{app:validation} also assesses input omission and the component and trigonometric representations. These comparisons establish numerical consistency at the specified frequencies and angles; dynamical stability at other operating points requires separate analysis.

\section{Transmission and absorption at fixed reflectivity}
\label{sec:absorption-comparison}
\begin{figure}[tbp]
\centering\includegraphics[width=0.70\textwidth]{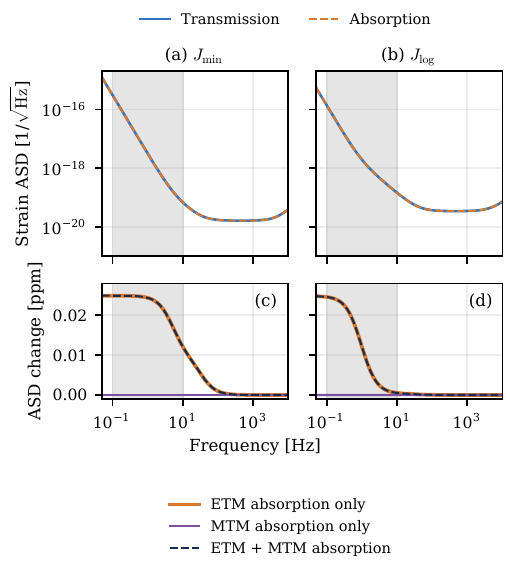}
\caption{Comparison of ETM and MTM transmission with absorption at fixed reflectivity. The left and right columns correspond to the $J_{\min}$ and $J_{\log}$ operating points, respectively. (a),(b) Total quantum strain ASD, including correlations, for transmission (solid blue) and absorption at both mirrors (dashed orange); the curves overlap at this scale. (c),(d) Relative ASD changes from the transmission model. Absorption at the ETMs alone gives almost the same change as absorption at both mirror types; replacing only MTM transmission gives a difference within numerical precision. These are changes in total quantum ASD calibrated with each condition's signal TF, rather than changes in mean force. One ppm denotes a relative change of $10^{-6}$. The shaded band is 0.1--10 Hz. Classical noise and thermal effects of absorption are excluded.}
\label{fig:transmission-absorption}
\end{figure}

In the reference model, the non-reflected fraction at each ETM and MTM is
transmitted to a rear port, whose reciprocal path admits an independent vacuum
field. Here we compare this reference with absorption of the same non-reflected
fraction. Defining the power absorptivity as $A_e$, we impose $R_e+T_e+A_e=1$.
For all ETMs and MTMs, we fix $R_e=1-10^{-7}$ and compare
$(T_e,A_e)=(10^{-7},0)$ with $(0,10^{-7})$. The two operating points in
Table~\ref{tab:pi-operating-point} retain the same optical phases, input power,
moment of inertia, and beam position. The PI curves are not reoptimized for absorption.

Absorption is represented by the coupling of additional environmental vacuum fields to the optical fields~\cite{BuonannoChen2001}. Let $\Sigma_e^{\rm opt}$ denote the
scattering matrix between the four physical paths at element $e$, with
$r_e=\sqrt{R_e}$ and $t_e=\sqrt{T_e}$ in the coefficients of
Eq.~\eqref{eq:annihilation-scattering} of Appendix~\ref{app:local}. At fixed $R_e$, replacing transmission with absorption sets $t_e=0$ in this optical matrix and $A_e=1-R_e$ in the bath coupling. Thus the external rear-path coupling vanishes while the absorption-vacuum term remains. The local linearized relation is
\begin{equation}
 \begin{split}
 \delta\hat c_{e,k}^{\rm out}
  ={}&\sum_{l=1}^{4}(\Sigma_e^{\rm opt})_{kl}
           \delta\hat c_{e,l}^{\rm in}
       +\gamma_{e,k}\hat\theta_e\\
    &+\sqrt{A_e}\,\hat v_{e,k}^{\rm abs}
 \end{split}
 \label{eq:absorption-local}
\end{equation}
and the creation-operator relation is its Hermitian conjugate. Here $\hat
v_{e,k}^{\rm abs}$ annihilates an environmental mode coupled through
absorption, distinct from the external rear vacuum fields $\hat n^{E,M}_{ij\nu}$. We
assume independent, zero-mean vacuum modes for different mirrors and paths,
with the same commutator normalization as the external inputs. 

We define the absorption-bath phase so that its local coupling is the
positive real coefficient $\sqrt{A_e}$. For a general coefficient
$\sqrt{A_e}e^{i\Phi_{\rm abs}}$, the redefinition
$\hat v_{e,k}^{\rm abs}=e^{i\Phi_{\rm abs}}\hat v_{e,k}$ absorbs the phase
into the bath operator and preserves its commutator. The quadrature covariance
of independent vacuum is also invariant under rotation:

\begin{equation}
 R(\Phi_{\rm abs})\mathbb I_2R^{\mathsf T}(\Phi_{\rm abs})=\mathbb I_2.
 \label{eq:absorption-vacuum-phase}
\end{equation}

The local coupling in the two-photon basis is therefore
$\sqrt{A_e}\mathbb I_2$ in this convention. This choice fixes the bath's phase
origin; it does not constrain a physical absorption phase to vanish. The
reflection and transmission rotations of Appendix~\ref{app:local}, and the
subsequent propagation and readout phases, are retained. Including the bath
input gives

\begin{equation}
 \Sigma_e^{\rm opt}\Sigma_e^{{\rm opt}\dagger}+A_e\mathbb I_4=\mathbb I_4.
 \label{eq:absorption-commutator}
\end{equation}

This condition preserves the output-field commutators and is consistent
with a unitary description including the environment. It does not imply that
the lossy optical part $\Sigma_e^{\rm opt}$ is itself unitary.

The bath fields are then propagated through the two-photon, optical, rotational, and homodyne relations defined above.

Absorption describes loss of light to an environment; vacuum specifies the
quantum state entering through that environmental coupling. In the present
model, each absorption input has the one-sided symmetrized quadrature PSD
$\mathbb I_2$, as does each rear vacuum field. Their injection sites and
propagation coefficients differ. Denoting the readout numerator for an
absorption input by $\mathcal C_{e,k}^{\rm abs}$ in the convention of
Eq.~\eqref{eq:sensing-function-chen}, its angular-noise PSD is
\begin{equation}
 S_{\theta,\rm abs}
 =\frac{\theta_{\rm SQL,d}^2}{|\mathcal D|^2}
   \sum_{e,k}\mathcal C_{e,k}^{\rm abs}
                 (\mathcal C_{e,k}^{\rm abs})^\dagger
 \label{eq:absorption-angle-psd}
\end{equation}
where the sum runs over the ETMs, MTMs, and their absorption couplings. Each
$\mathcal C_{e,k}^{\rm abs}$ includes the local factor $\sqrt{A_e}$, optical
propagation, mechanical feedback, and readout through both the signal and LO
paths. These coefficients and $\mathcal D$ are evaluated for the same
absorption condition. The PSD includes shot noise, back action, and their
correlation for the absorption inputs.

Because the complex reflection coefficients are fixed, the fixed-mirror
reflection and circulation response inside the rings is unchanged. Transmitted
vacuum and absorption-bath vacuum both have unit covariance, with optical
coupling magnitudes $\sqrt{T_e}$ and $\sqrt{A_e}$, respectively. Exchanging these channels at fixed non-reflected fraction therefore preserves the vacuum covariance added to the front paths.

Radiation pressure, however, depends on the difference between incoming
and outgoing momentum flux and need not be identical in the two cases. For
normal incidence from one side with power $P$, the mean force magnitude is
$F=(1+R_e-T_e)P/c=(2R_e+A_e)P/c$. The two coefficients are identical by
$R_e+T_e+A_e=1$; they are not distinct force models. At the same
$R_e=1-10^{-7}$ and incident power, transmission and absorption give

\begin{equation}
 \begin{aligned}
 F_{\rm tr}&=2R_eP/c,\qquad F_{\rm abs}=(1+R_e)P/c,\\
 \frac{F_{\rm abs}-F_{\rm tr}}{F_{\rm tr}}
 &=\frac{1-R_e}{2R_e}\simeq5.0\times10^{-8}.
 \end{aligned}
 \label{eq:absorption-force-scaling}
\end{equation}

The relative mean-force change is therefore about 0.050 ppm. Momentum
transfer from reflected photons is common to both cases; additional momentum
is deposited only by the fraction $10^{-7}$ of photons whose transmission is
replaced by absorption. This small mean-force difference provides a scale for
the noise change, but does not directly determine its ASD coefficient.

For noise, force and optical output are linearized with respect to the
same vacuum inputs to retain their correlations. Converting the terms in
Eq.~\eqref{eq:quantum-budget} to strain gives
$S_h=S_h^{\rm shot}+S_h^{\rm ba}+S_h^{\rm cross}$. A small change from
transmission to absorption then satisfies

\begin{equation}
 \frac{\Delta\sqrt{S_h}}{\sqrt{S_h}}
 \simeq\frac{\Delta S_h^{\rm shot}+\Delta S_h^{\rm ba}
                  +\Delta S_h^{\rm cross}}{2S_h}.
 \label{eq:absorption-asd-change}
\end{equation}

Here $\Delta$ denotes absorption minus transmission, and the denominator
on the right is the transmission model's total quantum PSD. Each PSD is
calibrated with the complete signal TF of its own optical condition, so TF
changes are included. The ASD change cannot be obtained simply by multiplying
the relative mean-force change by $1/2$.

Figure~\ref{fig:transmission-absorption} compares the models from 0.05 Hz to 10
kHz. Replacing transmission by absorption at both mirror types changes the
total quantum ASD by at most $2.49\times10^{-8}$ at the $J_{\min}$ point and
$2.48\times10^{-8}$ at the $J_{\log}$ point, approximately 0.025 ppm in either
case. The maximum relative changes in the complex TF are $4.89\times10^{-8}$
and $9.81\times10^{-9}$, respectively. Independent FINESSE 3 calculations,
including replacement at the ETMs or MTMs alone, agree within
$3.4\times10^{-10}$ for the total quantum ASD and $1.1\times10^{-8}$ for the
complex TF across all tested conditions.

The nearly constant low-frequency differences in
Figs.~\ref{fig:transmission-absorption}(c) and (d) occur where back action
dominates the total quantum noise: the back action ASD changes by only about
0.02484--0.02485 ppm over 0.1--1 Hz. The total ASD change follows
Eq.~\eqref{eq:absorption-asd-change} and decreases as shot noise gains relative
weight at higher frequency. Table~\ref{tab:absorption-low-frequency} shows
that back action still dominates at 1 Hz for $J_{\min}$, whereas
$S_h^{\rm ba}/S_h=49.43\%$ there for $J_{\log}$, reducing the total ASD
change to about 0.01216 ppm. A constant back action fraction therefore cannot
be assigned to the entire 0.1--1 Hz band at both operating points.

\begin{table}[tbp]

\caption{Ratio of back action PSD to total quantum PSD in the transmission model, and relative change in the total quantum ASD when both ETM and MTM transmissions are replaced by absorption. The total quantum PSD includes the cross term.}
\label{tab:absorption-low-frequency}
\begin{ruledtabular}
\begin{tabular}{lrrr}
Point & $f$ [Hz] & $S_h^{\rm ba}/S_h$ [\%] & ASD change [ppm]\\
\hline
$J_{\min}$ & 0.1 & 99.67 & 0.02492\\
$J_{\min}$ & 1.0 & 98.28 & 0.02437\\
$J_{\log}$ & 0.1 & 98.95 & 0.02459\\
$J_{\log}$ & 1.0 & 49.43 & 0.01216
\end{tabular}
\end{ruledtabular}
\end{table}

This back action ratio differs from the input-location fractions of
Eq.~\eqref{eq:input-group-budget}. The 89.7\% quoted in Sec.~\ref{sec:rear}
is the contribution of ETM and MTM rear vacuum fields to the total quantum PSD at
0.1 Hz for $J_{\min}$; it includes shot noise, back action, and their cross
term within those inputs.

The input-resolved calculation confirms that absorption-bath inputs nearly replace the contributions of rear vacuum fields, leaving the low-frequency excess noise essentially unchanged. This result
applies to the two tested operating points with small loss and does not assert
a general equivalence of transmission and absorption. Absorption is modeled as
an optical loss bath with instantaneous momentum transfer; heating, thermal
deformation, and photothermal torques are not included. In all sections of this paper other than the present absorption comparison, the sensitivity curves, vacuum-field labels, and reference parameters use the transmission model.

\section{Discussion}
\label{sec:discussion}
\begin{figure}[tbp]
\centering\includegraphics[width=0.70\textwidth]{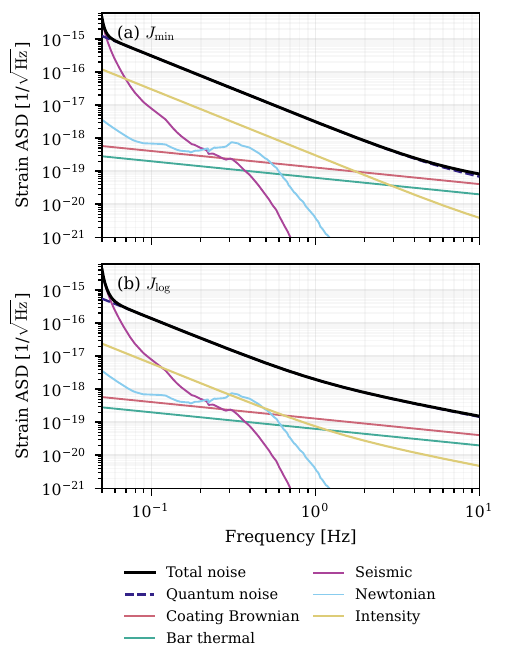}
\caption{Full-noise sensitivity at the two operating points in Table~\ref{tab:pi-operating-point}: (a) minimum PI value and (b) minimum logarithmic mean PI value. Solid black curves show total noise; dashed purple curves show quantum noise including correlations. The other components are coating Brownian, bar thermal, seismic, Newtonian, and intensity noise. All noise components are expressed as strain ASDs. The plotted band is 0.05--10 Hz; the PI selection uses 0.1--10 Hz. Components below the plotting floor are included in the total PSD. Bends in the seismic and Newtonian spectra reflect the tabulated environmental inputs and their propagation, as discussed in Sec.~\ref{sec:all-noise-discussion}.}
\label{fig:all-noise-operating-points}
\end{figure}

\subsection{Main-input and rear vacuum-field noise fractions}
Equation~\eqref{eq:input-group-budget} separates the detected noise into
contributions from the main input $a$, ETM rear vacuum fields, and MTM rear vacuum fields.
The following fractions refer to PSDs at the $J_{\min}$ point, rather than
ASDs. At 0.1 Hz, the ETM and MTM groups of rear vacuum fields each contribute 44.84\% of the
total quantum PSD. Each contributes 45.93\% of the back action PSD, while the
main input contributes 8.15\% (Appendix~\ref{app:validation}). An estimate
based only on the main input therefore misses the low-frequency excess at this
operating point.

The similar ETM and MTM fractions arise from distinct injection sites and optical paths, evaluated with the independent-input sum of Eq.~\eqref{eq:input-group-budget}. Their values depend on frequency and operating point, so other detunings or readout angles require recalculation.

\subsection{PRC power enhancement and phase rotation}
\label{sec:prdetuning}
We identify the origin of the sensitivity to PRC detuning by separating
the internal power from the readout phase relative to the carrier. At fixed
input power and SR and LO settings, changing the PR detuning changes the
low-frequency back action more strongly than the shot noise. We compare
$\delta_{\rm PR}=0^\circ,45^\circ,90^\circ$, retaining the SR and LO settings
of the $J_{\min}$ reference in Table~\ref{tab:pi-operating-point}, together
with the mirror properties, path lengths, and bar dynamics. The angle
$45^\circ$ is a reference for separating these effects, rather than a
PI-optimal operating point. All noise quantities below are quantum strain
noise, obtained by calibrating the three terms in Eq.~\eqref{eq:quantum-budget}
with the signal TF of each configuration.

PR detuning first changes the complex amplitude of the internal carrier.
Let $P_E$ denote the incident power summed over both directions at all four
ETMs. The feedback phase in Appendix~\ref{sec:sr-phase-reference} gives the
carrier denominator and positive feedback amplitude as

\begin{equation}
 \begin{aligned}
 d_{\rm PR}(\delta_{\rm PR})&=1+q_{\rm PR}e^{2i\delta_{\rm PR}},\\
 q_{\rm PR}&=|r_{\rm cav}(0)|^2r_{\rm PR}F_{\rm PR}(0)^2
             \simeq0.9411
 \end{aligned}
 \label{eq:pr-carrier-denominator}
\end{equation}

At fixed input power, the ratio of the carriers incident on an ETM is
\begin{equation}
 \frac{\bar E_E(\delta_{\rm PR})}{\bar E_E(45^\circ)}
 =\frac{d_{\rm PR}(45^\circ)}{d_{\rm PR}(\delta_{\rm PR})}
 =A_{\rm PR}e^{i\alpha_{\rm PR}}
 \label{eq:pr-carrier-ratio}
\end{equation}
Thus $P_E(\delta_{\rm PR})/P_E(45^\circ)=A_{\rm PR}^2$. Since
$|d_{\rm PR}|^2=(1-q_{\rm PR})^2+4q_{\rm PR}\cos^2\delta_{\rm PR}$, the
small feedback denominator near resonance at $\delta_{\rm PR}=90^\circ$
produces a strong power dependence on detuning. The same complex ratio also
rotates the carrier phase by $\alpha_{\rm PR}$, so both power and phase must
be tracked.

Power changes have a pronounced low-frequency effect because the
radiation-pressure torque generated by interference of vacuum fluctuations
with the carrier is read out through the free rotational susceptibility
$\chi=-1/(I\Omega^2)$. In this configuration, the ETM and MTM rear vacuum fields
make large contributions to the low-frequency torque noise that remains after
the cancellation due to speed-meter correlations. With a 1 W input, the
back action ASD ratios for the three detunings are 0.70743, 1, and 23.3239
relative to $45^\circ$, agreeing with the ratios of $\sqrt{P_E}$. The
overlap between the detuning-sensitive band and the band dominated by rear vacuum fields can therefore be explained by the effect of internal power on residual
back action. Shot noise also depends on the signal TF and readout projection;
its smaller variation alone does not exclude a power effect.

At a fixed LO optical setting, carrier rotation changes the detected quadrature and hence the shot term, cross term, and signal TF in Eq.~\eqref{eq:quantum-budget}~\cite{BuonannoChen2001,Kimble2001}. This phase effect applies separately to each rear vacuum field through its optical and torque-mediated paths.

To test this interpretation, we first adjust the input power to match
$P_E$ to its $45^\circ$ value while retaining the SR and LO settings. The
$0^\circ$ and $90^\circ$ results coincide, and the back action PSD is unchanged,
including that of each rear vacuum field. The total quantum ASD nevertheless exceeds
the reference by 0.107\%, 6.71\%, and 21.2\% at 0.1, 1, and 10 Hz,
respectively. At 1 Hz, normalizing to the total quantum PSD at $45^\circ$,
the cross term changes from $-0.0462$ to $+0.0633$, while the shot term
increases by 0.0292. Both terms therefore contribute to the difference at
matched power. We then keep $P_E$ fixed and apply
$\Phi_h\to\Phi_h-\alpha_{\rm PR}$ to restore the readout phase relative to
the carrier. The total quantum ASD returns to the reference curve within a
relative difference of $3.1\times10^{-13}$. The LO gain change accompanying
the input-power adjustment cancels in the signal-referred noise.

The scan thus attributes the large low-frequency change primarily to power and the matched-power residual to the carrier--LO phase. Figure~\ref{fig:prports} in Appendix~\ref{app:validation} distinguishes each input's noise fraction from its response to this phase change. Because the bright and dark ports decouple in the symmetric fixed-mirror network, the effect does not require an additional dark-port optical pole. This interpretation is limited to the symmetric configuration tested here.

\subsection{Sensitivity including the full noise budget}
\label{sec:all-noise-discussion}
Figure~\ref{fig:all-noise-operating-points} shows the sensitivity and noise
budget at the two operating points in Table~\ref{tab:pi-operating-point}. The
test-mass parameters are mass 171.32 kg,
moment of inertia 19.86 kg m$^2$, temperature 8 K, and main-beam spot separation 1.0873
m. Coating Brownian, bar thermal, seismic, Newtonian, and laser intensity noise
follow the noise-budget study used for PI selection~\cite{CHRONOSNoise2026}. These physical and environmental parameters are common to both operating points; only the PR and SR detuning angles and the LO setting angle are selected separately by the two PI objective functions.
The total strain PSD combines the quantum contribution of Eq.~\eqref{eq:input-group-budget}, using the calibration in Sec.~\ref{sec:readout}, with the classical spectra. Intensity noise is evaluated with the circulating power and signal TF at each operating point; no additional tuning or reoptimization is performed.

The small bends in the seismic and Newtonian curves reflect the
frequency structure of the tabulated ground-motion spectra used in the
noise-budget model~\cite{CHRONOSNoise2026}. Seismic noise is obtained by
interpolating the ground-motion input and applying the suspension response;
the ground Newtonian term uses a separately tabulated displacement spectrum,
interpolated in logarithmic frequency, with the gravity-gradient and bar
responses. Piecewise interpolation is continuous but can leave changes of
slope at the tabulated frequencies. Suspension resonances also shape the
seismic curve. These environmental spectra and their interpolation are common
to both operating points and are retained without smoothing; their small-scale
structure should not be interpreted as resonances of the quantum optical
response.

The logarithmic-mean operating point gives lower total noise at low
frequencies, whereas the minimum-PI operating point gives lower noise near the
upper band edge. This tradeoff corresponds to the two criteria in
Fig.~\ref{fig:pi-selection}. Quantum noise is the leading component over
0.1--10 Hz, but coating and bar thermal noise also contribute toward the upper
end at the minimum-PI point. Improving a quantum-noise curve and improving a PI
curve derived from the full noise budget are therefore distinct objectives.

\section{Conclusions}
\label{sec:conclusions}
Vacuum fields coupled through the small ETM and MTM transmissions can dominate
the residual low-frequency quantum noise of a torsion-bar Sagnac speed meter.
At the operating point selected by the minimum PI value, rear vacuum fields account
for 89.7\% of the total quantum-noise PSD at 0.1 Hz even with power
transmissions of $10^{-7}$ each. Omitting these inputs underestimates the noise
ASD by a factor of 3.11. Finite mirror transmission is therefore a leading
design constraint on the low-frequency quantum sensitivity of this
configuration. Assessing back action reduction for the main input must be
accompanied by an evaluation of vacuum propagation and torque driving from
the ETM and MTM inputs.

Strong cancellation of the left and right torques can coexist with a dominant residual contribution from rear vacuum fields. This distinction follows by connecting the temporal Sagnac and spatial force correlations to rotational dynamics and the detected measurement--back action correlation. The resulting operator-based input--output model evaluates residual torque, signal response, and readout noise together; a torque-cancellation factor alone cannot specify the sensitivity gain.

This result also constrains mirror design. Replacing ETM and MTM transmission
by absorption at the same reflectivity nearly replaces the contribution of rear vacuum fields with that of the environmental vacuum coupled to absorption. At the
two operating points studied, the maximum relative change in total quantum
ASD is only about $2.5\times10^{-8}$. Redistributing transmission into
absorption therefore does not reduce the low-frequency excess noise in this
small-loss model. Mirror specifications must account for the vacuum fields coupled
through transmission or absorption and for their propagation to bar rotation
and the measured output. This comparison concerns optical quantum noise and
excludes thermal effects of absorption.

For recycling and readout design, we separated the two effects of PRC
detuning on the internal carrier amplitude and phase. The large low-frequency
back action change in this scan primarily follows the ETM incident power,
whereas the quantum-sensitivity difference remaining at matched power vanishes
when the carrier--LO relative phase is restored. Dominance of the noise budget
by rear vacuum fields alone therefore does not attribute a detuning-induced change
to the cross term. For this symmetric configuration, distinguishing internal
power from the detected quadrature relative to the carrier provides a design
criterion for optimizing the operating point. The complex TF and quantum-noise predictions agree with
independent FINESSE 3 calculations at both full-noise PI operating points and
under changes of detuning and homodyne angle. The validated input-resolved budget translates these results into quantum constraints on mirror specifications and readout settings for low-frequency gravitational-wave, dark-matter, and geophysical observations.

\begin{acknowledgments}
We thank Ippei Obata, Ryuji Shibuya,Yang Hsin-Kuan, Avani Patel, and Jeff Kissel. Y.I. acknowledges support from NSTC, and CHiP in Taiwan under Grant No.115-2112-M-008 -022 -.
MH is partially supported by the Yushan Fellow Program by the Ministry of Education (MOE), Taiwan.
\end{acknowledgments}

\appendix

\section{Local fields and reflection--transmission relations}
\label{app:local}
We first identify the paths entering and leaving a mirror and write their
reflection and transmission in terms of creation and annihilation operators.
After specifying vacuum normalization, we transform to amplitude and phase
quadratures and connect the mirrors by propagation to obtain the carrier.
The same local boundary conditions apply to the 17 retained inputs and to the
auxiliary inputs used to assess the omission error.

The derivation proceeds from fields to force to readout.
Appendix~\ref{app:operator_locations} follows a field through the optical
layout; Appendix~\ref{app:optical} closes the ring and recycling paths.
The resulting local powers give the torque in Appendix~\ref{app:feedback}.
Appendix~\ref{app:chen-derivation} closes the mechanical response once,
and Appendix~\ref{app:chen-components} writes its output coefficients
explicitly, with the notation correspondence collected at its beginning.
Appendixes~\ref{app:sensing} and~\ref{app:detector} express the same
response in calibration and detector units. Numerical checks are in
Appendix~\ref{app:validation}.

\subsection{Optical paths and local operator relations}
A port is one optical path connected to a mirror. At the same port $k$, the
field approaching the mirror is $\hat c^{\rm in}_{e,k}$ and the field leaving
it is $\hat c^{\rm out}_{e,k}$. Incoming and outgoing directions are distinct
from the front and rear sides of the mirror; see Fig.~\ref{fig:operator-summary}(a).

A mirror at normal incidence has two ports, one path on each side. A mirror
at oblique incidence or a beam splitter has four ports, two paths on each
side. For four ports, label the front paths 1 and 2 and the rear paths 3 and
4. Reflection connects $1\leftrightarrow2$ and $3\leftrightarrow4$, whereas
transmission connects $1\leftrightarrow3$ and $2\leftrightarrow4$. This count
refers to paths, not to the two amplitude and phase quadratures introduced below.

Row $k$ labels an outgoing path and column $l$ an incoming path:
$\Sigma_{e;kl}$ is the amplitude coefficient from incoming path $l$ to
outgoing path $k$. With $r_e=\sqrt{R_e}$ and $t_e=\sqrt{T_e}$, define the
reflection coefficients $\rho_\pm=r_e e^{\pm i\Phi_e}$ and transmission
coefficient $\tau_e=it_e$. For two ports, 1 and 2 label opposite sides:
reflection returns to the same port and transmission reaches the other side.
The matrices in Eq.~\eqref{eq:annihilation-local} are then
\begin{equation}
 \begin{gathered}
 \Sigma_e^{(2)}=\begin{pmatrix}\rho_+&\tau_e\\\tau_e&\rho_-\end{pmatrix},\\
 \Sigma_e^{(4)}=\begin{pmatrix}
 0&\rho_+&\tau_e&0\\\rho_+&0&0&\tau_e\\
 \tau_e&0&0&\rho_-\\0&\tau_e&\rho_-&0
 \end{pmatrix}.
 \end{gathered}
 \label{eq:annihilation-scattering}
\end{equation}
A zero entry means that the corresponding paths have no direct coupling.
For example, the first two rows of the four-port ETM matrix give
\begin{equation}
 \begin{aligned}
 \delta\hat c^{\rm out}_{E,1}
 &=r_Ee^{i\Phi_E}\delta\hat c^{\rm in}_{E,2}
       +it_E\delta\hat c^{\rm in}_{E,3}+\gamma_{E,1}\hat\theta_E,\\
 \delta\hat c^{\rm out}_{E,2}
 &=r_Ee^{i\Phi_E}\delta\hat c^{\rm in}_{E,1}
       +it_E\delta\hat c^{\rm in}_{E,4}+\gamma_{E,2}\hat\theta_E.
 \end{aligned}
\end{equation}
The creation-operator equation contains the complex conjugates of all
coefficients, including a transmission coefficient $-it_E$. The incoming rear
modes are independent vacuum inputs; their reflected outgoing fields do not constitute
additional vacuum inputs. In the main-text notation, $\delta\hat c^{\rm
in}_{E,3}=\hat n^{E}_{ijB}$ and $\delta\hat c^{\rm in}_{E,4}=\hat n^{E}_{ijA}$.
At an MTM, paths 3 and 4 likewise correspond to directions B and A. The
two-photon transformation preserves this identification.

If an angular displacement $\theta_E$ changes the reflection phase by $2\sigma
k_0\ell_E\theta_E$, each reflected matrix element satisfies
$\partial\Sigma_{E;kl}/\partial\theta_E=2i\sigma k_0\ell_E\Sigma_{E;kl}$. Here
$\sigma=+1,-1$ denotes the front and rear surfaces, and the first derivative of
a transmission element is zero. Substitution gives $\gamma_{E,k}$ in the main
text. The angular-coupling column in the annihilation/creation basis is
$(\gamma_{E,k},\gamma_{E,k}^*)^{\mathsf T}\hat\theta_E$.

\subsection{Sidebands and vacuum normalization}
For the creation and annihilation operators defined in the main text, the
conventional two-photon quadratures are
\begin{equation}
 \hat q_1=\frac{\hat c_++\hat c_-^\dagger}{\sqrt2},\qquad
 \hat q_2=\frac{\hat c_+-\hat c_-^\dagger}{i\sqrt2}
\end{equation}
with the common labels $e,k,\mathrm{in/out}$
suppressed~\cite{CavesSchumaker1985}. Extending each component by $\hat
q_\alpha(-\Omega)=\hat q_\alpha^\dagger(\Omega)$, our $e^{+i\Omega t}$
convention uses $\hat{\vect z}(\Omega)=\hat{\vect q}(-\Omega)=\mathsf U_{\rm
q}\hat{\vect d}(\Omega)$. This relation connects the incident and outgoing
fields $\hat{\vect u},\hat{\vect v}$ to the conventional sideband definition.

An external vacuum vector $\hat{\vect n}_q=(\hat n_{q1},\hat n_{q2})^{\mathsf
T}$ denotes the corresponding mode of $\hat{\vect z}^{\rm in}$. Its one-sided
symmetrized PSD is defined by
\begin{equation}
 \begin{aligned}
 \frac12\langle&\hat n_{q\alpha}(\Omega)\hat n_{q'\beta}^{\dagger}(\Omega')
 +\hat n_{q'\beta}^{\dagger}(\Omega')\hat n_{q\alpha}(\Omega)\rangle\\
 &=\pi\delta(\Omega-\Omega')\delta_{qq'}\delta_{\alpha\beta}.
 \end{aligned}
\end{equation}
so each input has unit covariance. Stacking all inputs gives $\hat{\vect n}$.
Each retained vacuum contributes two quadratures to this vector.
The factor $\sqrt{\hbar\omega_0}$ converts its unit vacuum normalization to
the power-amplitude normalization of the local fields $\hat{\vect u}$ and
$\hat{\vect v}$.

\subsection{Mirror scattering and angular coupling}
The stationary carrier on each incident path is specified by
its power and phase. Equation~\eqref{eq:carrier-power-phase} gives

\begin{equation}
 \begin{aligned}
 \bar E^{\rm in}_{e,k}&=\sqrt{P^{\rm in}_{e,k}}e^{i\varphi^{\rm in}_{e,k}},\\
 \bar u_{e,k,1}&=\sqrt{2P^{\rm in}_{e,k}}\cos\varphi^{\rm in}_{e,k},\\
 \bar u_{e,k,2}&=\sqrt{2P^{\rm in}_{e,k}}\sin\varphi^{\rm in}_{e,k}.
 \end{aligned}
 \label{eq:carrier-power-quadratures}
\end{equation}

The outgoing carrier $\bar{\vect v}$ has the same form with
$P^{\rm out},\varphi^{\rm out}$. Bars therefore denote stationary
fields, while hats denote quantum fluctuations. The quadrature vectors
are a compact representation of these powers and phases. Define

\begin{equation}
 \begin{gathered}
 R(\psi)=\begin{pmatrix}\cos\psi&-\sin\psi\\\sin\psi&\cos\psi\end{pmatrix},\\
 J=R(\pi/2),\qquad R_\pm=R(\pm\Phi).
 \end{gathered}
 \label{eq:quadrature-rotation}
\end{equation} The basis transformation in Eq.~\eqref{eq:annihilation-to-quadrature} gives
\begin{equation}
 \begin{aligned}
 \mathsf U_{\rm q}\operatorname{diag}(r_e e^{i\Phi_e},r_e e^{-i\Phi_e})\mathsf U_{\rm q}^\dagger
 &=r_eR(\Phi_e),\\
 \mathsf U_{\rm q}\operatorname{diag}(it_e,-it_e)\mathsf U_{\rm q}^\dagger&=t_eJ.
 \end{aligned}
\end{equation}
Thus a phase factor becomes a rotation of the two quadratures, while the
incoming-to-outgoing path connections remain unchanged. For the absorption coupling in
Sec.~\ref{sec:absorption-comparison}, the local phase is absorbed into the bath
operator. Equation~\eqref{eq:absorption-vacuum-phase} then allows the coupling
to be written as $\sqrt{A_e}\mathbb I_2$ without changing the independent-vacuum
covariance.

 The two-port and
four-port matrices are
\begin{equation}
 \begin{gathered}
 S_e^{(2)}=\begin{pmatrix}rR_+&tJ\\tJ&rR_-\end{pmatrix},\\
 S_e^{(4)}=\begin{pmatrix}
 0&rR_+&tJ&0\\rR_+&0&0&tJ\\tJ&0&0&rR_-\\0&tJ&rR_-&0
 \end{pmatrix}.
 \end{gathered}
\end{equation}
Each $2\times2$ block acts on the amplitude and phase quadratures of one path. Thus $S_e^{(2)}$ is $4\times4$ and $S_e^{(4)}$ is $8\times8$, with $r=\sqrt R,t=\sqrt T$. For an ETM beam distance
$\ell_E$ and signed local angle $\theta_E=s_i\theta_j$, where $s_L=1,s_R=-1$,
\begin{equation}
 \begin{aligned}
 \hat{\vect v}_{E,1}&=r_ER_+\hat{\vect u}_{E,2}
   +t_EJ\sqrt{\hbar\omega_0}\hat{\vect n}_{E,3}
   +\ell_E\vect G_{E,1}\hat\theta_E,\\
 \hat{\vect v}_{E,2}&=r_ER_+\hat{\vect u}_{E,1}
   +t_EJ\sqrt{\hbar\omega_0}\hat{\vect n}_{E,4}
   +\ell_E\vect G_{E,2}\hat\theta_E.
 \end{aligned}
\end{equation}
The local angle $\theta_E$ is not an independent mirror degree of freedom. The
reflection-phase change $2k_0\ell_E\theta_E$ and $dR/d\psi=JR$ give the
coupling from incident path $l$ as $\ell_E\vect G_{E,k}=2\sigma
k_0\ell_EJ(S_E)_{kl}\bar{\vect u}_{E,l}$, where $\sigma=\pm1$ specifies the
reflecting surface. Transmission has no first-order displacement dependence,
and products of two fluctuations are omitted.

\subsection{Network connections and steady-state fields}
Two kinds of relation are now connected: a mirror determines its
outgoing fields from its incident fields, whereas a propagation path supplies
the incident field of the next mirror. Vacuum amplitudes and the prescribed
angle are sources; the circulating fields are the unknowns. The carrier is
found first at zero sideband frequency because it sets both phase-modulation
strength and radiation-pressure coupling.

Let $S$ be the direct sum of local scattering matrices and $\Pi$ the
propagation matrix. Each connection block is $\Pi_{\rm
link}(L,\Omega)=e^{-i\Omega L/c}\mathbb I_2$. For example, the ITM-to-ETM
connection is $\hat{\vect u}_{E,I}=\Pi_{\rm link}(L_1,\Omega)\hat{\vect
v}_{I,E}$. With external-input placement matrix $N$ and outgoing
auxiliary-input coefficients $L$,
\begin{equation}
 \begin{aligned}
 \hat{\vect u}&=\Pi\hat{\vect v}+\sqrt{\hbar\omega_0}N\hat{\vect n},\\
 \hat{\vect v}&=S\hat{\vect u}+\sqrt{\hbar\omega_0}L\hat{\vect n}+G\hat{\vect x}.
 \end{aligned}
\end{equation}
Here $\hat{\vect x}$ is only a column of dependent displacements
$\ell_E\hat\theta_E$; the matrix $L$ is distinct from a path length. Placing
the input carrier $\sqrt{2P}(\cos\varphi,\sin\varphi)^{\mathsf T}$ in a drive
column $\vect d_c$ gives $\bar{\vect v}=[\mathbb I-S\Pi(0)]^{-1}S\vect d_c$ and
$\bar{\vect u}=\Pi(0)\bar{\vect v}+\vect d_c$.

\section{Operators and optical ports}
\label{app:operator_locations}
\begin{figure}[!htbp]
 \centering\includegraphics[width=0.62\textwidth]{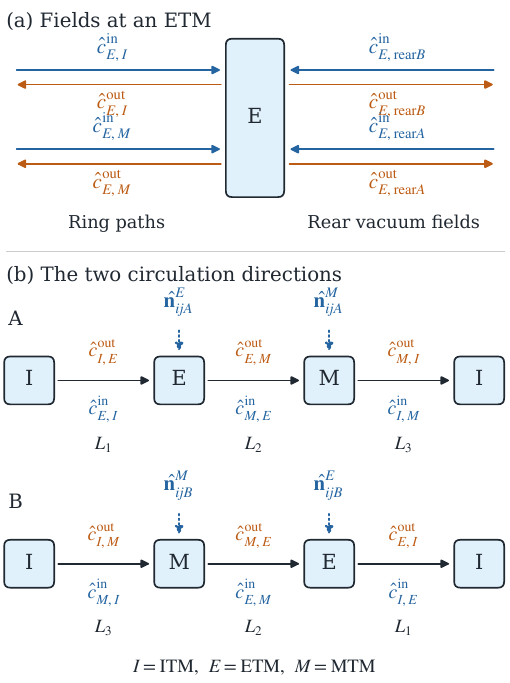}
 \caption{Local operators and ring propagation. (a) Incoming and outgoing fields on the four ETM paths. Blue and orange arrows and operator labels denote incoming and outgoing fields, respectively. Reflection and transmission are determined by the local scattering matrix. (b) Direction A follows ITM--ETM--MTM--ITM, and direction B the reverse order. Outgoing and incoming operators label the ends of each link, while separate arrows mark ETM and MTM rear vacuum fields. Links $L_1,L_2,L_3$ have delays $L_1/c,L_2/c,L_3/c$. The diagram specifies connectivity, not physical incidence angles.}
 \label{fig:operator-summary}
\end{figure}

Figure~\ref{fig:operator-summary} is read in the order in which light travels:
we connect annihilation operators along the path before collecting them into
a cavity response. The labels I, E, and M abbreviate ITM, ETM, and MTM;
$i=L,R$ specifies the ring and $j=X,Y$ the bar axis. Incoming and outgoing
fields travel toward and away from a mirror. The adjoint creates a photon
in the same propagating mode and does not denote reverse propagation.
The construction applies to all four rings.

\subsection{Following one field through the diagram}
Consider one circulation in direction A. The ITM at the right of
Fig.~\ref{fig:operator-summary}(b) is the same mirror as the ITM at the left.
The unfolded path shows light leaving the ITM, visiting the ETM and MTM,
and returning to the ITM. We denote these mirrors by I, E, and M and temporarily suppress the ring and bar indices $i,j$.

The two subscripts of a field identify the mirror at which it is defined
and the neighboring mirror to which its path leads. Thus
$\delta\hat c^{\rm out}_{I,E}$ and $\delta\hat c^{\rm in}_{E,I}$ describe the
same traveling field at the two ends of a link, not two vacuum sources.
Using the manuscript's Fourier convention for annihilation-operator
fluctuations about the carrier and $z_m=e^{-i\Omega L_m/c}$, the first flight is
\begin{equation}
 \delta\hat c^{\rm in}_{E,I}=z_1\delta\hat c^{\rm out}_{I,E}.
 \label{eq:map-first-flight}
\end{equation}
In the time domain, the ETM receives the ITM output after a delay $L_1/c$.
The stationary carrier phase is included in the reference planes as in the
main text.

\subsection{Vacuum fields and angular signals added at a mirror}
The field reflected by the ETM contains both vacuum transmitted from its
rear and the phase modulation of the carrier produced by rotation. With
$\rho_E=r_Ee^{i\Phi_E}$, one row of the local relation in
Appendix~\ref{app:local} reads
\begin{equation}
 \delta\hat c^{\rm out}_{E,M}
 =\rho_E\delta\hat c^{\rm in}_{E,I}
   +it_E\hat n^E_{ijA}+\gamma_{E,M}\hat\theta_E.
 \label{eq:map-etm-reflection}
\end{equation}
The terms describe reflection of the field arriving from the ITM, the ETM rear vacuum field, and the angular signal, respectively. The coefficient
$\gamma_{E,M}$ is obtained from the stationary field and the angular derivative
of the reflection phase; it is not an independent noise source. It includes
the distance from the rotation axis to the main-laser spot. The rearA label
in Fig.~\ref{fig:operator-summary}(a) identifies the rear path whose transmitted
field joins direction A.

The field next propagates through $L_2$, reflects at the MTM, and returns
through $L_3$. With $\rho_M=r_Me^{i\Phi_M}$,
\begin{equation}
 \begin{aligned}
 \delta\hat c^{\rm in}_{M,E}&=z_2\delta\hat c^{\rm out}_{E,M},\\
 \delta\hat c^{\rm out}_{M,I}&=\rho_M\delta\hat c^{\rm in}_{M,E}
                         +it_M\hat n^M_{ijA},\\
 \delta\hat c^{\rm in}_{I,M}&=z_3\delta\hat c^{\rm out}_{M,I}.
 \end{aligned}
 \label{eq:map-return-flight}
\end{equation}
The ETM and MTM rear vacuum fields are independent quantum inputs. However, a field
injected at the ETM remains the same operator after propagation to the MTM;
it must not be replaced by a new vacuum at that point.

\subsection{From one circulation to the cavity response}
Successive substitution of Eqs.~\eqref{eq:map-first-flight}--\eqref{eq:map-return-flight}
gives the field returning after one circulation:
\begin{equation}
 \begin{aligned}
 \delta\hat c^{\rm in}_{I,M}
 &=\rho_M\rho_E z_1z_2z_3\delta\hat c^{\rm out}_{I,E}\\
 &\quad+it_E\rho_Mz_2z_3\hat n^E_{ijA}
       +it_Mz_3\hat n^M_{ijA}\\
 &\quad+\gamma_{E,M}\rho_Mz_2z_3\hat\theta_E.
 \end{aligned}
 \label{eq:map-one-circulation}
\end{equation}
Each coefficient is the product of the propagations and reflections between
the point of injection and the ITM. ETM rear vacuum fields and the angular signal
undergo reflection at the MTM, whereas MTM rear vacuum fields return directly to
the ITM. Their different paths produce different frequency responses and phases.

At the ITM, part of the returning field is reflected back into the ring.
Substitution into the ITM boundary condition therefore sums successive
circulations and produces the cavity denominator.
Equation~\eqref{eq:ring-return-A} in Appendix~\ref{app:optical} is the
two-photon version of Eq.~\eqref{eq:map-one-circulation}; $D_A^{-1}$ sums
these repeated returns.

Direction B follows ITM--MTM--ETM--ITM by the same steps. Although the total
length is unchanged, the paths from each rear vacuum field to the ITM differ.
The fields $\hat n^{E,M}_{ijB}$ are independent of their direction-A
counterparts, not creation operators for those fields. Creation-operator
relations follow from Eq.~\eqref{eq:annihilation-local}; combining the two
sidebands then gives the two-photon representation. Rear outputs also follow
from the local relation and enter the momentum flux and torque in
Appendix~\ref{app:feedback}. This ordering first determines the optical
fields and then determines how they drive the bar.

\section{Optical response from the rings to dual recycling}
\label{app:optical}

Starting from the one-circulation field in Appendix~\ref{app:operator_locations},
we build the optical response of one ring, connect two rings into a Sagnac
interferometer, and then close the PRM and SRM feedback paths. At each stage,
we retain separately the coefficients of incident fields and those of rear
vacuum and angular signals added along the path. The successive results are
the ring responses $H_{A,B}$, the Sagnac coefficients
$S_{\rm sag},N_{\rm sag},G_{\rm sag}$, and the recycled coefficients
$S_{\rm DR},N_{\rm DR},G_{\rm DR}$. We finally connect the two axes and LO
path. Bar angles are prescribed inputs at this stage; the mechanical feedback
that determines them from optical torque is added in
Appendix~\ref{app:feedback}.

\subsection{One ring: paths and added fields}
\label{sec:ring}
In one ring, write ITM, ETM, and MTM as I, E, and M. The path lengths
are $L_1$ (I--E), $L_2$ (E--M), and $L_3$ (M--I), with
$L_{\rm rt}=L_1+L_2+L_3$ and $z_j=e^{-i\Omega L_j/c}$. Direction A in
Fig.~\ref{fig:operator-summary}(b) follows I$\to$E$\to$M$\to$I; direction B
reverses this order. We use the two-photon transformation of
Appendix~\ref{app:local} to express each field as a power-normalized
two-component operator, preserving the input identities defined with creation
and annihilation operators.

Label the internal ITM path toward E by A and that toward M by B.
The fields $\hat{\vect v}_{A,B}$ leave the ITM along those paths, whereas
$\hat{\vect u}_{A,B}$ approach it. External input $\hat{\vect a}_A$
transmits through the ITM to drive $\hat{\vect v}_A$. After one circulation
in direction A, the field returns from the M side as $\hat{\vect u}_B$;
its transmitted part contributes to external output $\hat{\vect c}_B$.
Interchanging A and B gives the other direction. Thus
$\hat{\vect v}_A\to\hat{\vect u}_B$ describes one circulation in a single
direction, not conversion into the opposite propagation direction.

At the ETM, paths 1 and 2 face the ITM and MTM, respectively;
at the MTM they face the ETM and ITM. Rear paths 3 and 4 transmit into front
paths 1 and 2. Reflection connects the two front paths. A direction-A field
therefore enters E and M on path 1 and leaves on path 2, whereas a direction-B
field enters on path 2 and leaves on path 1. This input--output distinction
sets the subscripts on the added fields below.

The rotations $R(\Phi)$ and $J=R(\pi/2)$ are defined in
Appendix~\ref{app:local}. Write reflection as $\mathsf R_E=r_ER(\Phi_E)$
and $\mathsf R_M=r_MR(\Phi_M)$, and transmission as $\mathsf T_e=t_eJ$
for $e=I,E,M$. The internal and external ITM reflections are $\mathsf
R_I=r_IR(\sigma_I\Phi_I)$ and $\mathsf R_I^{\rm ext}=r_IR(-\sigma_I\Phi_I)$,
with $\sigma_I=+1$ on the left and $-1$ on the right. These quadrature matrices
are distinct from the scalar power coefficients $R,T$. For $k=1,2$, define the
fields added at E and M in each direction as
\begin{equation}
 \begin{aligned}
 \hat{\vect e}_k&=\mathsf T_E\sqrt{\hbar\omega_0}\hat{\vect n}_{E,k+2}
                    +\ell_E G_{E,k}\hat\theta_E,\\
 \hat{\vect m}_k&=\mathsf T_M\sqrt{\hbar\omega_0}\hat{\vect n}_{M,k+2}.
 \end{aligned}
\end{equation}
Here $k$ labels the outgoing path. The symbols $\hat{\vect e}_k$ and
$\hat{\vect m}_k$ collect the rear vacuum-field and angular contributions added at the ETM and the rear vacuum-field contribution added at the MTM, respectively;
they are not new independent inputs. Direction A receives
$\hat{\vect e}_2,\hat{\vect m}_2$, and direction B receives
$\hat{\vect e}_1,\hat{\vect m}_1$. In the reference ring,
$R_E+T_E=R_M+T_M=1$, so rear vacuum fields enter through transmission.

\subsection{Solving the ring feedback and recovering internal fields}
\label{sec:ring-response}
We first calculate the field returning after one circulation, then
solve the repeated reflection at the ITM, and finally recover the fields at
the mirrors needed for radiation pressure. Matrix products act from the right,
in the order followed by the light.

\begin{subequations}\label{eq:ring-round-trip}
In direction A, the field follows I$\to$E$\to$M$\to$I, giving $\mathcal
F_A=z_3\mathsf R_Mz_2\mathsf R_Ez_1$. Propagating the fields added at E and M
back to the ITM gives
\begin{equation}
 \hat{\vect\eta}_A=z_3\mathsf R_Mz_2\hat{\vect e}_2+z_3\hat{\vect m}_2.
 \label{eq:ring-noise-A}
\end{equation}
and the returning field is
\begin{equation}
 \hat{\vect u}_B=\mathcal F_A\hat{\vect v}_A+\hat{\vect\eta}_A.
 \label{eq:ring-return-A}
\end{equation}

Direction B follows I$\to$M$\to$E$\to$I, so $\mathcal F_B=z_1\mathsf
R_Ez_2\mathsf R_Mz_3$. The field added at M is now reflected at E before
returning to the ITM, giving
\begin{equation}
 \hat{\vect\eta}_B=z_1\hat{\vect e}_1+z_1\mathsf R_Ez_2\hat{\vect m}_1.
 \label{eq:ring-noise-B}
\end{equation}
and the field returning to ITM port A is
\begin{equation}
 \hat{\vect u}_A=\mathcal F_B\hat{\vect v}_B+\hat{\vect\eta}_B.
 \label{eq:ring-return-B}
\end{equation}
\end{subequations}
Part of each returning field reflects at the ITM to circulate again
in the same direction. Adding the transmitted external input gives the ITM
boundary conditions
\begin{equation}
 \hat{\vect v}_A=\mathsf R_I\hat{\vect u}_B+\mathsf T_I\hat{\vect a}_A,
 \quad
 \hat{\vect v}_B=\mathsf R_I\hat{\vect u}_A+\mathsf T_I\hat{\vect a}_B.
\end{equation}
Substitute Eqs.~\eqref{eq:ring-return-A} and \eqref{eq:ring-return-B}
and collect the circulating field on the left. For the direction index
$\nu=A,B$, define $D_\nu=\mathbb I_2-\mathsf R_I\mathcal F_\nu$ to obtain

\[
 \hat{\vect v}_\nu=D_\nu^{-1}(\mathsf T_I\hat{\vect a}_\nu+\mathsf R_I\hat{\vect\eta}_\nu).
\]

The inverse $D_\nu^{-1}$ sums repeated circulation after reflection
at the ITM. The external output combines prompt reflection at the ITM with
transmission of the returning field: for direction A,
$\hat{\vect c}_B=\mathsf R_I^{\rm ext}\hat{\vect a}_A+\mathsf T_I\hat{\vect u}_B$.
Thus $\hat{\vect c}_B=H_A\hat{\vect a}_A+\hat{\vect\xi}_A$, and similarly
$\hat{\vect c}_A=H_B\hat{\vect a}_B+\hat{\vect\xi}_B$, with coefficients

\begin{equation}\label{eq:ring-external-response}
 \begin{aligned}
 H_\nu&=\mathsf R_I^{\rm ext}+\mathsf T_I\mathcal F_\nu D_\nu^{-1}\mathsf T_I,\\
 \hat{\vect\xi}_\nu&=\mathsf T_I(\mathbb I_2+\mathcal F_\nu D_\nu^{-1}\mathsf R_I)
                       \hat{\vect\eta}_\nu.
 \end{aligned}
\end{equation}

Here $H_\nu$ is the ring response to an external input, while
$\hat{\vect\xi}_\nu$ is the externally emerging contribution of vacuum and
angular signals added inside the ring. These are the ring relations needed
for the Sagnac connection. Equal circulation denominators in the two
directions do not make $\hat{\vect\xi}_A$ and $\hat{\vect\xi}_B$ the same
noise: their insertion sites and propagation order differ.

For the scalar reflectivities and phase rotations used here, $\mathsf
R_I\mathcal F_{A,B}=\rho zR(\psi)$, where $\rho=r_Ir_Er_M$, $z=e^{-i\Omega
L_{\rm rt}/c}$, and $\psi=\sigma_I\Phi_I+\Phi_E+\Phi_M$. Thus
\[
 \det D_{A,B}=1-2\rho z\cos\psi+\rho^2z^2.
\]
Using the $\operatorname{Cof}$ convention of
Appendix~\ref{app:chen-derivation}, the inverse is
$D_\nu^{-1}=\operatorname{Cof}(D_\nu)/\det D_\nu$. This rearrangement retains
the round-trip delay and detuning without a single-pole bandwidth approximation.

Radiation pressure requires the fields at the ETM, not only
the external ring output. Substituting along each path gives all four
incident front fields directly from the solved ITM fields:

\begin{equation}
 \begin{aligned}
 \hat{\vect u}_{E,1}&=z_1\hat{\vect v}_A,\qquad
 \hat{\vect u}_{M,2}=z_3\hat{\vect v}_B,\\
 \hat{\vect u}_{M,1}&=z_2\mathsf R_Ez_1\hat{\vect v}_A+z_2\hat{\vect e}_2,\\
 \hat{\vect u}_{E,2}&=z_2\mathsf R_Mz_3\hat{\vect v}_B+z_2\hat{\vect m}_1.
 \end{aligned}
 \label{eq:ring-internal-fields}
\end{equation}

The added ETM field reaches the MTM in direction A; the
added MTM field reaches the ETM in direction B. Outgoing fields then
follow from the local mirror relation without another network solve.
Rear outputs follow from the same local relations. For example, the field leaving ETM rear path 3 is

\begin{equation}
 \begin{aligned}
 \hat{\vect v}_{E,3}&=\mathsf T_E\hat{\vect u}_{E,1}
 +r_ER(-\Phi_E)\sqrt{\hbar\omega_0}\hat{\vect n}_{E,4}\\
 &\quad+\ell_E G_{E,3}\hat\theta_E.
 \end{aligned}
 \label{eq:ring-rear-output}
\end{equation}

This field also contributes to the momentum flux. Elimination from the external input--output relation does not discard
the field's momentum. The recovered front- and rear-path fields are used in
the torque calculation of Appendix~\ref{app:feedback}.

\subsection{Building the Sagnac response from two rings}
\label{sec:sagnac}
We now connect two rings using only their responses $H_{A,B}$ and
added fields $\hat{\vect\xi}_{A,B}$. The objective is a relation between the
bright and dark inputs and outputs of $\mathrm{BS}_j$. Each route is followed
through splitting at the beam splitter, the first ring, the central link, the
second ring, and recombination.

At the 50:50 beam splitter $\mathrm{BS}_j$ on axis $j$, define the field
incident from the PRC as the bright input $\hat{\vect a}_b$ and that incident
from the SRC as the dark input $\hat{\vect a}_d$. The beam splitter sends light
into path A of each ring, while the two path-B ports are interconnected. We
suppress the axis label in the following single-axis equations. Ideal folding
mirrors add no forward-reflection rotation in the adopted reference-plane
convention.

Let $\hat{\vect v}_{L,R}$ propagate from $\mathrm{BS}_j$ toward the rings and
$\hat{\vect u}_{L,R}$ return to it. Branch and central-link delays are
$z_b=e^{-i\Omega L_b/c}$ and $z_c=e^{-i\Omega L_c/c}$.
\begin{subequations}\label{eq:sagnac-paths}
\begin{samepage}
First follow $\hat{\vect v}_R$ toward the right ring. Its path-A output
traverses the central link and enters path B of the left ring: $\hat{\vect
a}_{B,L}=z_c\hat{\vect c}_{B,R}$. The coefficient for return to $\mathrm{BS}_j$
from the left is $\Lambda_{LR}=z_bH_{B,L}z_cH_{A,R}z_b$. The first subscript of $\Lambda$ labels the return side and the second
the departure side. Thus $LR$ means departure to the right and return from the
left; its index order differs from the chronological traversal $R\to L$. The added field is
\begin{equation}
 \hat{\vect\eta}_L=z_b(H_{B,L}z_c\hat{\vect\xi}_{A,R}
                              +\hat{\vect\xi}_{B,L}).
 \label{eq:sagnac-noise-RL}
\end{equation}
and the returning left field is
\begin{equation}
 \hat{\vect u}_L=\Lambda_{LR}\hat{\vect v}_R+\hat{\vect\eta}_L.
 \label{eq:sagnac-return-RL}
\end{equation}
\end{samepage}

\begin{samepage}
Conversely, $\hat{\vect v}_L$ traverses left-ring path A, the central link, and
right-ring path B. Substituting $\hat{\vect a}_{B,R}=z_c\hat{\vect c}_{B,L}$
gives $\Lambda_{RL}=z_bH_{B,R}z_cH_{A,L}z_b$ and
\begin{equation}
 \hat{\vect\eta}_R=z_b(H_{B,R}z_c\hat{\vect\xi}_{A,L}
                              +\hat{\vect\xi}_{B,R}).
 \label{eq:sagnac-noise-LR}
\end{equation}
so the returning right field is
\begin{equation}
 \hat{\vect u}_R=\Lambda_{RL}\hat{\vect v}_L+\hat{\vect\eta}_R.
 \label{eq:sagnac-return-LR}
\end{equation}
\end{samepage}
\end{subequations}
Finally, the two returning fields recombine at $\mathrm{BS}_j$. With
$r_b=t_b=1/\sqrt2$ and $T_b=t_bJ$, the relations
\begin{equation}
 \begin{aligned}
 \hat{\vect v}_{L}&=r_b\hat{\vect a}_b+T_b\hat{\vect a}_d,\\
 \hat{\vect v}_{R}&=T_b\hat{\vect a}_b+r_b\hat{\vect a}_d,\\
 \hat{\vect b}_b&=r_b\hat{\vect u}_{L}+T_b\hat{\vect u}_{R},\\
 \hat{\vect b}_d&=T_b\hat{\vect u}_{L}+r_b\hat{\vect u}_{R}.
 \end{aligned}
\end{equation}
are combined with Eqs.~\eqref{eq:sagnac-return-RL} and
\eqref{eq:sagnac-return-LR}. Collecting the coefficients of each incident
field gives four matrices $S_{rs}$, with the first index labeling the output
and the second the input:
\begin{equation}
 \begin{aligned}
 S_{bb}&=r_b\Lambda_{LR}T_b+T_b\Lambda_{RL}r_b,\\
 S_{bd}&=r_b\Lambda_{LR}r_b+T_b\Lambda_{RL}T_b,\\
 S_{db}&=T_b\Lambda_{LR}T_b+r_b\Lambda_{RL}r_b,\\
 S_{dd}&=T_b\Lambda_{LR}r_b+r_b\Lambda_{RL}T_b.
 \end{aligned}
\end{equation}
The added bright-output field is $r_b\hat{\vect\eta}_L+T_b\hat{\vect\eta}_R$;
the added dark-output field is $T_b\hat{\vect\eta}_L+r_b\hat{\vect\eta}_R$.
Separate their coefficients of independent vacuum inputs and mirror
displacements, and order the components as bright then dark:

\begin{equation}
 \begin{aligned}
 \hat{\vect a}_{\rm sag}&=\begin{pmatrix}\hat{\vect a}_b\\\hat{\vect a}_d\end{pmatrix},\quad
 \hat{\vect b}_{\rm sag}=\begin{pmatrix}\hat{\vect b}_b\\\hat{\vect b}_d\end{pmatrix},\\
 S_{\rm sag}&=\begin{pmatrix}S_{bb}&S_{bd}\\S_{db}&S_{dd}\end{pmatrix},\\
 \hat{\vect b}_{\rm sag}&=S_{\rm sag}\hat{\vect a}_{\rm sag}
        +N_{\rm sag}\hat{\vect n}+G_{\rm sag}\hat{\vect x}.
 \end{aligned}
 \label{eq:sagnac-open-response}
\end{equation}

Each input or output vector has four components, and $S_{\rm sag}$
is a $4\times4$ matrix. The vector $\hat{\vect n}$ lists each independent
vacuum operator once; $N_{\rm sag}$ contains its propagation coefficients
and $\sqrt{\hbar\omega_0}$. The vector $\hat{\vect x}$ collects ETM angular
displacements converted to linear displacements with their spot lever arms.
The shared-bar relation $\hat{\vect x}=B\hat{\vect\theta}$ is applied in
Appendix~\ref{app:feedback}. Here $G_{\rm sag}$ is the optical response to
prescribed angles. Adding the two traversal routes at the field level preserves
interference between paths driven by the same input. Back action is determined
only after this optical response is connected to torque and rotational dynamics.

\subsubsection{Angular signal coefficients in terms of the carrier}

The angle coefficients need not remain an unspecified column of
$G_{\rm sag}$. For the resonant reference rings they can be written
directly in terms of the incident ETM carriers. Suppress the axis label and
write $P_{ik}$ and $\varphi_{ik}$ for the carrier power and phase
incident on front path $k=1,2$ of the ETM in ring $i=L,R$.
These are local powers at the ETM, with propagation and finite
transmission retained. For compact quadrature expressions below,
$\bar u_{ik,1}=\sqrt{2P_{ik}}\cos\varphi_{ik}$ and
$\bar u_{ik,2}=\sqrt{2P_{ik}}\sin\varphi_{ik}$. The left/right ETM reflection signs
are $+/-$, while their spot displacements are $+\ell\theta_j$ and
$-\ell\theta_j$. These signs cancel in the local angle source.
Solving the ring return once, the output response per unit angle is
\begin{equation}
 \begin{aligned}
 g^{\rm ring}_{iA,\alpha}
 &=-\frac{2k_0\ell t_Ir_Er_Mz_2z_3}{d_{\rm ring}}\bar u_{i1,\alpha},\\
 g^{\rm ring}_{iB,\alpha}
 &=-\frac{2k_0\ell t_Ir_Ez_1}{d_{\rm ring}}\bar u_{i2,\alpha},\\
 d_{\rm ring}&=1-r_Ir_Er_Mz_1z_2z_3.
 \end{aligned}
 \label{eq:ring-angle-carrier}
\end{equation}
For direction A the angle-generated field still passes the MTM; for B it
returns directly to the ITM. This explains the different propagation factors.
Substituting these expressions into the two Sagnac traversal paths gives
all four returning-field coefficients. With the common
carrier-to-angle prefactor $g_0=-2k_0\ell t_Ir_Ez_b/d_{\rm ring}$,
\begin{equation}
 \begin{aligned}
 g_{b1}/g_0={}&r_br_{\rm cav}z_cr_Mz_2z_3\sqrt{2P_{R1}}\cos\varphi_{R1}\\
 &+r_bz_1\sqrt{2P_{L2}}\cos\varphi_{L2}\\
 &+t_br_{\rm cav}z_cr_Mz_2z_3\sqrt{2P_{L1}}\sin\varphi_{L1}\\
 &-t_bz_1\sqrt{2P_{R2}}\sin\varphi_{R2},\\
 g_{b2}/g_0={}&r_br_{\rm cav}z_cr_Mz_2z_3\sqrt{2P_{R1}}\sin\varphi_{R1}\\
 &+r_bz_1\sqrt{2P_{L2}}\sin\varphi_{L2}\\
 &-t_br_{\rm cav}z_cr_Mz_2z_3\sqrt{2P_{L1}}\cos\varphi_{L1}\\
 &+t_bz_1\sqrt{2P_{R2}}\cos\varphi_{R2},\\
 g_{d1}/g_0={}&-t_br_{\rm cav}z_cr_Mz_2z_3\sqrt{2P_{R1}}\sin\varphi_{R1}\\
 &-t_bz_1\sqrt{2P_{L2}}\sin\varphi_{L2}\\
 &-r_br_{\rm cav}z_cr_Mz_2z_3\sqrt{2P_{L1}}\cos\varphi_{L1}\\
 &+r_bz_1\sqrt{2P_{R2}}\cos\varphi_{R2},\\
 g_{d2}/g_0={}&t_br_{\rm cav}z_cr_Mz_2z_3\sqrt{2P_{R1}}\cos\varphi_{R1}\\
 &+t_bz_1\sqrt{2P_{L2}}\cos\varphi_{L2}\\
 &-r_br_{\rm cav}z_cr_Mz_2z_3\sqrt{2P_{L1}}\sin\varphi_{L1}\\
 &+r_bz_1\sqrt{2P_{R2}}\sin\varphi_{R2}.
 \end{aligned}
 \label{eq:sagnac-angle-carrier}
\end{equation}
The terms proportional to $r_{\rm cav}$ carry the first-ring modulation through the
second ring; the terms without $r_{\rm cav}$ are generated during the second
interaction. Their interference produces the angular signal response.
The stationary boundary conditions in Appendix~\ref{app:local}
determine both $P_{ik}$ and $\varphi_{ik}$. PRC buildup changes the
$\sqrt{P_{ik}}$ factors; carrier rotation changes their sine and cosine
weights. Replacing a complex amplitude by power alone would lose the
latter effect.

\subsubsection{Relation to the ideal Sagnac phase}
\label{sec:sagnac-phase}
We check how the constructed Sagnac response gives the reflection phase
used in the main text, first retaining finite transmission and then taking the
ideal limit. The limit checks correspondence with the ideal speed-meter model;
the recycling calculation below uses the finite-transmission response.

Equation~\eqref{eq:ring-scalar-reflection} follows from substituting
$r_I^2+t_I^2=1$ and $J^2=-\mathbb I_2$ into
Eq.~\eqref{eq:ring-external-response}. The reference-plane convention gives
$H_{A,L}=H_{B,L}=r_{\rm cav}\mathbb I_2$ and $H_{A,R}=H_{B,R}=-r_{\rm
cav}\mathbb I_2$; recombination at the beam splitter yields
Eq.~\eqref{eq:sagnac-scalar-reflection}. Removing the constant output rotation
$J$ requires applying $J^{-1}$ to the added field and $J$ to the detection row
simultaneously.

In the ideal limit of zero rear transmission, $|r_{\rm cav}|=1$ and
\begin{equation}
 \beta_{\rm ring}=\arctan\!\left[\frac{1+r_I}{1-r_I}
 \tan\!\frac{\Omega L_{\rm rt}}{2c}\right]
 \simeq\arctan(\Omega/\gamma),
 \label{eq:ring-phase-ideal}
\end{equation}
Here $\gamma=2c(1-r_I)/[(1+r_I)L_{\rm rt}]$, and the approximation requires
$|\Omega L_{\rm rt}/c|\ll1$. Neglecting the external transit time further gives
Chen's $\beta_{\rm sag}=2\beta_{\rm ring}+\pi/2$~\cite{Chen2003}. The ideal
difference between two measurements, $1-e^{-2i\beta_{\rm ring}}$, gives
$\mathcal K_{\rm sag}=4\mathcal K_{\rm arm}\sin^2\beta_{\rm ring}$: the sine
contains $\beta_{\rm ring}$, not $\beta_{\rm sag}$. Our calculation retains
finite rear transmission and propagation delays rather than substituting this
ideal formula for the full noise model.

\subsection{Closing the PRM and SRM return paths}
\label{sec:dr}
In Eq.~\eqref{eq:sagnac-open-response}, the bright and dark fields
incident on $\mathrm{BS}_j$ are still prescribed externally. Attaching PRM
and SRM returns part of each outgoing field to those inputs. We distinguish
these returning fields from the external inputs, then substitute them into the
Sagnac relation to close the recycling feedback.

The bright and dark sides of $\mathrm{BS}_j$ in Fig.~\ref{fig:topology}
connect to the inner surfaces of $\mathrm{PRM}_j$ and $\mathrm{SRM}_j$,
respectively. The laser enters the external PRM port and the external SRM
port connects to the output optics. We describe each path by its total length
$L_{\rm PR},L_{\rm SR}$ and one-way amplitude coefficient
$F_{\rm PR},F_{\rm SR}$. With static phases included in the reflection
phases at the reference planes,
\begin{equation}
 \begin{aligned}
 F_{\rm PR}&=F_{\rm PR}(0)e^{-i\Omega L_{\rm PR}/c},\\
 F_{\rm SR}&=e^{-i\Omega L_{\rm SR}/c}.
 \end{aligned}
\end{equation}
Here $F_{\rm PR}(0)=0.99999985$ and $F_{\rm SR}(0)=1$, with unrounded
values used in the calculation. These scalars multiply both quadratures.
Mirrors within each path are represented by its net amplitude coefficient
and propagation time, without introducing individual labels absent from the
optical layout. The external fields $\hat{\vect a}_{\rm ext}$ approach the recycling
mirrors from outside, and $\hat{\vect b}_{\rm ext}$ leave outward. Both are
ordered as the two PR-side components followed by the two SR-side components.
The matrix $R_o$ describes prompt external reflection, $R_i$ the return of a
Sagnac output after reflection at a recycling mirror, $T_i$ transmission from
outside to the Sagnac, and $T_o$ transmission in the reverse direction. With $R_{{\rm PR}\pm}=r_{\rm PR}R(\pm\Phi_{\rm
PR})$, $R_{{\rm SR}\pm}=r_{\rm SR}R(\pm\Phi_{\rm SR})$, $T_{\rm PR}=t_{\rm
PR}J$, and $T_{\rm SR}=t_{\rm SR}J$, define
\begin{equation}
 \begin{aligned}
 R_o&=\operatorname{diag}(R_{{\rm PR}+},R_{{\rm SR}-}),\\
 R_i&=\operatorname{diag}(F_{\rm PR}R_{{\rm PR}-}F_{\rm PR},F_{\rm SR}R_{{\rm
SR}+}F_{\rm SR}),\\
 T_i&=\operatorname{diag}(F_{\rm PR}T_{\rm PR},F_{\rm SR}T_{\rm SR}),\\
 T_o&=\operatorname{diag}(T_{\rm PR}F_{\rm PR},T_{\rm SR}F_{\rm SR}).
 \end{aligned}
\end{equation}
The two factors of $F$ in $R_i$ describe outward and return propagation
between $\mathrm{BS}_j$ and a recycling mirror. The boundary conditions for
the re-entering field and the external output are
\begin{equation}
 \begin{aligned}
 \hat{\vect a}_{\rm sag}&=T_i\hat{\vect a}_{\rm ext}
                  +R_i\hat{\vect b}_{\rm sag}+N_i\hat{\vect n},\\
 \hat{\vect b}_{\rm ext}&=R_o\hat{\vect a}_{\rm ext}
                  +T_o\hat{\vect b}_{\rm sag}+N_o\hat{\vect n}.
 \end{aligned}
\end{equation}
Denote added fields along the PR recycling-mirror-to-Sagnac path by $W_{f,{\rm
PR}}\hat{\vect n}$ and those on the return path by $W_{b,{\rm PR}}\hat{\vect
n}$, with analogous definitions on the SR side. The $W$ matrices are coefficients of vacuum injected along a path,
not additional vacuum operators. As for the ring fields
$\hat{\vect\eta}_{A,B}$, each column is obtained by propagating from its
injection point through the subsequent reflections. Splitting these added
fields into reflection back to the Sagnac and transmission to the exterior gives
\begin{equation}
 \begin{aligned}
 N_i&=\begin{pmatrix}
 F_{\rm PR}R_{{\rm PR}-}W_{b,{\rm PR}}+W_{f,{\rm PR}}\\
 F_{\rm SR}R_{{\rm SR}+}W_{b,{\rm SR}}+W_{f,{\rm SR}}
 \end{pmatrix},\\
 N_o&=\begin{pmatrix}T_{\rm PR}W_{b,{\rm PR}}\\T_{\rm SR}W_{b,{\rm
SR}}\end{pmatrix}.
 \end{aligned}
\end{equation}
ETM and MTM rear vacuum fields are already contained in $N_{\rm sag}$ and are
not introduced again through $N_i,N_o$. Path vacua omitted in the main model
are omitted by removing the corresponding input columns here as well. When
one retained vacuum reaches several terms, its coefficients are added for
the same operator.

Substituting the boundary condition for $\hat{\vect a}_{\rm sag}$
into Eq.~\eqref{eq:sagnac-open-response} places the same unknown
$\hat{\vect b}_{\rm sag}$ on both sides. Moving its return term to the left gives

\begin{equation}
 \begin{aligned}
 D_{\rm DR}\hat{\vect b}_{\rm sag}
 &=S_{\rm sag}T_i\hat{\vect a}_{\rm ext}\\
 &\quad+(N_{\rm sag}+S_{\rm sag}N_i)\hat{\vect n}
            +G_{\rm sag}\hat{\vect x},\\
 D_{\rm DR}&=\mathbb I_4-S_{\rm sag}R_i.
 \end{aligned}
 \label{eq:dr-feedback-source}
\end{equation}

The right-hand side is specified by external fields, retained vacuum,
and prescribed mirror angles. Solve for the Sagnac output with
$D_{\rm DR}^{-1}$, then substitute into the external-output boundary condition:
\begin{equation}
 \begin{aligned}
 S_{\rm DR}&=R_o+T_oD_{\rm DR}^{-1}S_{\rm sag}T_i,\\
 N_{\rm DR}&=N_o+T_oD_{\rm DR}^{-1}(N_{\rm sag}+S_{\rm sag}N_i),\\
 G_{\rm DR}&=T_oD_{\rm DR}^{-1}G_{\rm sag},\\
 \hat{\vect b}_{\rm ext}&=S_{\rm DR}\hat{\vect a}_{\rm ext}
                +N_{\rm DR}\hat{\vect n}+G_{\rm DR}\hat{\vect x}.
 \end{aligned}
\end{equation}
Here $S_{\rm DR}$ maps external inputs to external outputs,
$N_{\rm DR}$ propagates internally injected vacuum to those outputs, and
$G_{\rm DR}$ maps displacements due to mirror angles to the same outputs.
As in the ring calculation, the eliminated variables are internal optical
fields, not input operators or angular signals. This single-axis optical
response retains propagation phase, PRC/SRC detuning, and the retained vacua;
the bar equation of motion has not yet been closed.

\subsubsection{SR resonance defined by the feedback phase}
\label{sec:sr-phase-reference}
We now read the resonance phase reference from the feedback denominator
just constructed. In the symmetric reference configuration, fixed-mirror
bright and dark fields decouple, so carrier feedback on each side can be
expressed as a round-trip coefficient $\mathcal L$ and denominator
$1-\mathcal L$. This circulation includes both reflection at the recycling
mirror and return from the Sagnac. At the
carrier frequency in the reference geometry, the Sagnac bright- and dark-port
reflections are $S_{bb}(0)=S_{dd}(0)=-\rho_sJ$, where
Eq.~\eqref{eq:sagnac-scalar-reflection} gives $\rho_s=|r_{\rm
cav}(0)|^2=0.9920323$. The complex amplitude reflectivity is therefore
$-i\rho_s$. The inner SRM reflection is $r_{\rm SR} e^{i\Phi_{\rm SR}}$, so the
fixed-mirror carrier round-trip feedback and denominator are
\begin{equation}
 \begin{aligned}
 \mathcal L_{\rm SR}(0)&=(-i\rho_s)r_{\rm SR} e^{i\Phi_{\rm SR}}
                 =q_{\rm SR} e^{-2i\delta_{\rm SR}},\\
 d_{\rm SR}(0)&=1-q_{\rm SR} e^{-2i\delta_{\rm SR}},\qquad
 q_{\rm SR}=\rho_s r_{\rm SR}.
 \end{aligned}
 \label{eq:sr-carrier-reference}
\end{equation}
At $\delta_{\rm SR}=0$, the feedback phase vanishes, defining the SR-resonant
configuration. The reference $\delta_{\rm SR}=3.166^\circ$ is measured from
this resonance and gives a round-trip feedback phase of $-6.331^\circ$. We use
this resonant origin throughout. It follows from the complex Sagnac reflection;
transferring a mirror setting from another interferometer does not by itself
identify resonance. Finite-frequency and quantum-noise calculations retain the
full matrices, delays, vacuum inputs, and mechanical coupling derived above.

The inner PRM surface has the opposite reflection coefficient $r_{\rm PR}
e^{-i\Phi_{\rm PR}}$. At the same reference plane, $\mathcal L_{\rm
PR}(0)=-q_{\rm PR} e^{2i\delta_{\rm PR}}$, with the positive feedback amplitude
$q_{\rm PR}$ defined in Eq.~\eqref{eq:pr-carrier-denominator}. The PRC resonance is
therefore at $\delta_{\rm PR}=90^\circ$: zero PR and SR phase parameters do not
denote the same resonance condition. This distinction is retained in
Sec.~\ref{sec:prdetuning}.

\subsection{Connecting the axes to mechanics and readout}
We now have each axis's optical response as viewed from outside its
PRM and SRM. On the input side, IBS divides the laser between the two axes;
on the output side, OBS combines the fields leaving their SRMs. The main
vacuum $\hat{\vect a}$ enters the open OBS port, and the signal output
continues to the homodyne beam splitter. The LO picked off from the input
laser reaches that same beam splitter along its own path. These connections
use the same local reflection--transmission relations and propagation delays.

Because both axes share the laser and main vacuum, a shared input
must not be duplicated as independent vacuum fields. We retain its propagation
coefficients to both axes and to the signal and LO paths. The external fields
of the full optical system and its local mirror fields are then functions of
the same inputs and bar angles. Appendix~\ref{app:feedback} computes torque
from those internal fields and closes the rotational dynamics;
Appendix~\ref{app:detector} constructs differential detection of the signal
and LO. Appendix~\ref{app:chen-derivation} rearranges the coupled optical and
mechanical equations into the common denominator and input-specific
coefficients used in the main text.

\section{Rotational dynamics and internal-field recovery}
\label{app:feedback}
The optical calculation so far treats angle as an imposed
source. We now use the resulting fields to compute torque, and allow that
torque to change the angle. It is useful to keep these operations separate:
the optical response can then be checked with the bar fixed before any
mechanical feedback is added.

\subsection{From power fluctuations to torque}
Ordering the ETMs as LX, LY, RX, RY gives the angle-to-displacement and force-to-torque relations
\begin{equation}
 \begin{gathered}
 B=\ell\begin{pmatrix}1&0\\0&1\\-1&0\\0&-1\end{pmatrix},\\
 \hat{\vect x}=B\hat{\vect\theta},\qquad
 \hat{\vect\tau}=B^{\mathsf T}\hat{\vect F}.
 \end{gathered}
\end{equation} Expanding the photon-flux power $\hat P=\hbar\omega_0\hat
c^\dagger\hat c$ about the carrier gives $\delta\hat
P=\hbar\omega_0(\bar\alpha^*\delta\hat c+\bar\alpha\delta\hat c^\dagger)$.
Substituting the power and phase gives

\begin{equation}
 \begin{aligned}
 \delta\hat P&=\sqrt{\hbar\omega_0P}
   (e^{-i\varphi}\delta\hat c+e^{i\varphi}\delta\hat c^\dagger),\\
 \delta\hat P^{\rm in}_{m,k}&=\sqrt{2P^{\rm in}_{m,k}}
   (\cos\varphi^{\rm in}_{m,k}\hat u_{m,k,1}
    +\sin\varphi^{\rm in}_{m,k}\hat u_{m,k,2}).
 \end{aligned}
 \label{eq:power-fluctuation-explicit}
\end{equation}

The second line is the component form of
$\bar{\vect u}_{m,k}^{\mathsf T}\hat{\vect u}_{m,k}$; the outgoing
power has the same form. The momentum flux gives
\begin{equation}
 \hat F_m=-\frac1c\sum_k\sigma_{m,k}
           (\delta\hat P^{\rm in}_{m,k}+\delta\hat P^{\rm out}_{m,k})
 \label{eq:local-force}
\end{equation}
At the reference ETMs, $R_E+T_E=1$. The signs $\sigma_{m,k}$ are $(+1,-1)$ for
two-port and $(+1,+1,-1,-1)$ for four-port elements.

The local force can itself be simplified before any network matrices are
introduced. Consider an ETM with no incident carrier on its rear paths and
$r_E^2+t_E^2=1$, as in the transmission model. For front path 1 the two
outgoing fields are $r_ER(\Phi_E)\hat{\vect u}_{E,1}
+t_EJ\sqrt{\hbar\omega_0}\hat{\vect n}_{E,4}$ on front path 2 and
$t_EJ\hat{\vect u}_{E,1}+r_ER(-\Phi_E)
\sqrt{\hbar\omega_0}\hat{\vect n}_{E,4}$ on rear path 3, in addition to
the angle-generated phase modulation. Inserting these fields and their
carriers into Eq.~\eqref{eq:local-force} combines the front-field terms
into $1+r_E^2-t_E^2=2R_E$. The two interference terms driven by the
rear vacuum fields give $2r_Et_E$. Including both front paths yields the scalar
expression
\begin{equation}
 \begin{aligned}
 \hat F_E={}&-\frac{2R_E}{c}\sum_{k=1,2}\sqrt{2P_{E,k}}\Bigl[
    \cos\varphi_{E,k}\hat u_{E,k,1}\\
 &\hspace{37mm}+\sin\varphi_{E,k}\hat u_{E,k,2}\Bigr]\\
 &-\frac{2r_Et_E}{c}\sum_{(k,l)=(1,4),(2,3)}\sqrt{2\hbar\omega_0P_{E,k}}\Bigl[\\
 &\quad\sin(\Phi_E+\varphi_{E,k})\hat n_{E,l,1}\\
 &\quad-\cos(\Phi_E+\varphi_{E,k})\hat n_{E,l,2}\Bigr].
 \end{aligned}
 \label{eq:force-transmission-scalar}
\end{equation}
Here $P_{E,k},\varphi_{E,k}$ refer to the incident front carrier.
The term proportional to $R_E$ is the fluctuating intracavity radiation pressure. The terms proportional to $r_Et_E$ describe interference of rear vacuum fields with light
transmitted out of the cavity. The overall sign follows the displacement
convention of Eq.~\eqref{eq:local-force}. A local phase change alone has no
linear power change because $\bar{\vect u}^{\mathsf T}J\bar{\vect u}=0$;
its force effect appears when the modulated field propagates and returns in
$\hat u_{E,k,\alpha}$. Rear transmission therefore enters both the local
force and the cavity-circulated field. The bar torque is obtained by taking
$\ell(\hat F_{Lj}-\hat F_{Rj})$ after these amplitudes have been combined.
This reduction is for transmission; the absorption boundary conditions in
Sec.~\ref{sec:absorption-comparison} require their own momentum balance.

\subsection{From torque to the closed optical response}

For each axis, the local force gives
$\hat\tau_j=\ell(\hat F_{Lj}-\hat F_{Rj})$.
Insert this torque into the free-bar equation~\eqref{eq:bar-torque}.
The force depends on the returned optical fields, which in turn depend
on this same angle. Appendix~\ref{app:chen-derivation} solves those
coupled equations for the two returning fields and one angle; its scalar
solution is Eq.~\eqref{eq:scalar-driven-angle}. No independent
mechanical noise source is introduced. Each retained vacuum drives both
the direct optical output and the torque. Their shared origin must be
preserved when forming the interference term in
Eq.~\eqref{eq:quantum-budget}. The two axes are combined at the field
level in Eq.~\eqref{eq:scalar-obs-output}.

\section{Derivation of the common denominator and input numerators}
\label{app:chen-derivation}

\subsection{Five-component system for one axis}
\label{sec:chen-common}
We rearrange the input--output relations into a common denominator and
input-specific two-photon coefficients, as in Eq.~(5.6) of Buonanno and
Chen~\cite{BuonannoChen2001}. Only the order of internal-field elimination
changes; the physical conditions and input set remain the same. We first treat
one PRC, Sagnac, SRC, and bar, then connect the axes and readout.

Symmetry gives $S_{bd}=S_{db}=0$ for the fixed-mirror Sagnac. Define $\mathsf
R_b=F_{\rm PR}R_{{\rm PR}-}F_{\rm PR}$, $\mathsf R_d=F_{\rm SR}R_{{\rm
SR}+}F_{\rm SR}$, $\mathsf A_b=\mathbb I_2-S_{bb}\mathsf R_b$, and $\mathsf
A_d=\mathbb I_2-S_{dd}\mathsf R_d$. These describe optical feedback through the
bright and dark ports. The two-component responses $\vect g_b,\vect g_d$ to bar
angle are obtained by inserting the spot displacements $B_{:,j}\theta_j$ into
$G_{\rm sag}$.

Substituting the internal fields into Eq.~\eqref{eq:local-force} gives the
driving torque $\hat\tau=\vect f_b\hat{\vect b}_b+\vect f_d\hat{\vect
b}_d+\kappa_0\hat\theta+\hat\tau_{\rm in}$. The rows $\vect f_b,\vect f_d$ map
the returning bright and dark fields to torque; $\kappa_0$ maps angle to torque
with both returning fields held fixed. They follow from the internal fields of
Sec.~\ref{sec:ring-response} and $B^{\mathsf T}\vect F$, without an independently
postulated effective coupling. Define $\chi=-1/(I\Omega^2)$ and
$\mu=1-\chi\kappa_0$. Suppressing the axis label, only five unknown components
remain:
\begin{equation}
 \begin{pmatrix}
 \mathsf A_b&0&-\vect g_b\\
 0&\mathsf A_d&-\vect g_d\\
 -\chi\vect f_b&-\chi\vect f_d&\mu
 \end{pmatrix}
 \begin{pmatrix}\hat{\vect b}_b\\\hat{\vect b}_d\\\hat\theta\end{pmatrix}
=\begin{pmatrix}\hat{\vect\eta}_b\\\hat{\vect\eta}_d\\\hat\xi_\theta\end{pmatrix}.
 \label{eq:five-variable-system}
\end{equation}
Each returning field has two quadratures, and the bar contributes
one angle. The source fields $\hat{\vect\eta}_{b,d}$ contain the incident
SRM field and rear vacuum fields propagated to the bright and dark outputs with the
returning fields held fixed. They are sums of the original vacuum operators,
not additional inputs. The torque source is
$\hat\tau_{\rm in}=\sqrt{\hbar\omega_0}\sum_{q,\alpha}
t_{q\alpha}\hat n_{q\alpha}$ and
$\hat\xi_\theta=\theta_{{\rm sig},j}+\chi\hat\tau_{\rm in}$.
The coefficients $t_{q\alpha}$ follow by substituting those injected fields
into Eq.~\eqref{eq:force-transmission-scalar} at fixed returning fields.
The external differential angle is applied as
$\theta_{{\rm sig},j}=s_j\theta_{\rm sig}$, with $s_X=1/2$ and $s_Y=-1/2$.

For a $2\times2$ matrix, define the adjugate
\begin{equation}
 \begin{gathered}
 \operatorname{Cof}(A)=\begin{pmatrix}A_{22}&-A_{12}\\-A_{21}&A_{11}\end{pmatrix},\\
 A\operatorname{Cof}(A)=(\det A)\mathbb I_2.
 \end{gathered}
\end{equation}
With $\Delta_r=\det\mathsf A_r$ and $h_r=\vect f_r\operatorname{Cof}(\mathsf A_r)\vect g_r$, the
five-component determinant is Eq.~\eqref{eq:chen-common-denominator}. Its first
term combines optical feedback with the angular response at fixed returning
fields; the second describes paths through the bright and dark ports back to
the bar. Using $S_{bb}=S_{dd}=sJ$, where $s=-e^{-i\Omega\tau_{\rm link}}r_{\rm
cav}^2$, also gives a scalar expression for the optical determinants. The return matrices are
$\mathsf R_b=r_{\rm PR}F_{\rm PR}^2R(-\Phi_{\rm PR})$ and
$\mathsf R_d=r_{\rm SR}F_{\rm SR}^2R(\Phi_{\rm SR})$.
Their determinants are written directly in these mirror and propagation
parameters in Eq.~\eqref{eq:scalar-optical-factors}.

The elimination can be followed without expanding a
five-dimensional determinant. First solve each optical row for its returning
field at a prescribed angle. Then insert both fields into the torque row:

\begin{equation}
 \begin{gathered}
 \hat{\vect b}_r
 =\frac{\operatorname{Cof}(\mathsf A_r)}{\Delta_r}
                  (\hat{\vect\eta}_r+\vect g_r\hat\theta),\\
 \begin{aligned}
 &\left(\mu-\chi\frac{h_b}{\Delta_b}-\chi\frac{h_d}{\Delta_d}\right)
     \hat\theta
 \\
 &\quad=\hat\xi_\theta+
   \chi\sum_{r=b,d}\frac{\vect f_r\operatorname{Cof}(\mathsf A_r)
                       \hat{\vect\eta}_r}{\Delta_r}.
 \end{aligned}
 \end{gathered}
 \label{eq:five-variable-elimination-steps}
\end{equation}

Multiplication of the second line by $\Delta_b\Delta_d$
gives the common denominator $M$. Substituting the resulting angle into
the dark optical row completes the solution. Appendix~\ref{app:chen-components} carries out this back substitution
in scalar form and identifies each optical and torque contribution. At an isolated zero of an intermediate optical denominator,
the coupled result is evaluated by its limit or by the original system.

For identical axes, $M_X=M_Y=M$, so the OBS combination
preserves the same denominator. Unequal axes can instead be written over
$M_XM_Y$.

The original and rearranged expressions were compared from 0.05 Hz to 10 kHz.
Maximum relative differences were below $2.9\times10^{-10}$ for the complex TF,
$9.6\times10^{-11}$ for shot ASD, $5.0\times10^{-11}$ for radiation-pressure
ASD, and $2.8\times10^{-13}$ for total quantum ASD. The correlation-PSD
difference, normalized by the sum of shot and radiation-pressure PSDs, was also
below $2.8\times10^{-13}$. Changing the feedback-elimination order therefore
reproduces the reference results without changing the reference data.

\section{Scalar output coefficients after substitution}
\label{app:chen-components}

The matrix result in Appendix~\ref{app:chen-derivation} can be reduced
further. We first give the two outgoing quadratures as a direct optical
response plus the field generated by bar motion. We then substitute the
injection coefficients for the main and rear vacuum fields. No unevaluated matrix
products for injection or output propagation remain in these coefficient formulas.
The reduction uses the same resonant rings and symmetric single-axis
geometry as Eq.~\eqref{eq:sagnac-scalar-reflection}; PR and SR detuning,
all propagation delays, and the carrier phase remain arbitrary within this
geometry. Products are complex amplitudes, without conjugation. Complex
conjugation is applied when forming a noise PSD, not in these substitutions.

\subsection{Reference form and notation}
\label{sec:chen-notation}
Chen's ideal unrecycled Sagnac~\cite{Chen2003} is a useful reference
for the entries of Eq.~\eqref{eq:chen-multiport-output}. After matching
Fourier and reference-plane conventions, $\Psi_{\rm arm}$ and
$\Psi_{\rm sagnac}$ correspond to $\beta_{\rm ring}$ and
$\beta_{\rm sag}$; his dark-port $p_{1,2},q_{1,2}$ become
$a_{1,2},b_{1,2}$. His $I_c$ is total circulating power, equal to
$2P_{\rm ring}$ for an ideal equal-power ring, whereas $I$ here denotes
inertia. His homodyne angle obeys $\Phi=\pi/2-\zeta$ in the same basis.
Using a superscript Ch to distinguish this reference model, its coefficients
are simply
\begin{equation}
 \begin{gathered}
 M^{\rm Ch}=1,\qquad C^{\rm Ch}_{11}=C^{\rm Ch}_{22}=1,\\
 C^{\rm Ch}_{12}=0,\qquad C^{\rm Ch}_{21}=-\mathcal K_{\rm sag},\\
 D^{\rm Ch}_1=0,\qquad D^{\rm Ch}_2=\sqrt{2\mathcal K_{\rm sag}}.
 \end{gathered}
 \label{eq:chen-reference-coefficients}
\end{equation}
Here the signal is normalized by Chen's $h/h_{\rm SQL}$.
The diagonal entries transmit the incident vacuum, $C^{\rm Ch}_{21}$
is its motion-induced phase modulation, and $D^{\rm Ch}_2$ carries the
signal. In the present model, these same roles are separated explicitly
in Eqs.~\eqref{eq:C-elements} and~\eqref{eq:D-elements}, using the
independently derived angular normalization $\theta_{\rm sig}/\theta_{\rm SQL,d}$.
Detuning can mix both quadratures, and every rear vacuum field has its own
$C_q$. Thus the coefficients above identify physical roles; they are
not substituted for the multi-input model. In particular, the dimensionless
$\mathcal K_{\rm sag}$ is not the torque-per-angle feedback
$\mathcal K_\theta$ below.

\subsection{Optical circulation and rotational feedback}
For $r=b,d$, the two equations for the returning field have coefficients
\begin{equation}
 \begin{aligned}
 u_b&=1-sr_{\rm PR}F_{\rm PR}^2\cos\psi_b,\\
 v_b&=sr_{\rm PR}F_{\rm PR}^2\sin\psi_b,\\
 u_d&=1-sr_{\rm SR}F_{\rm SR}^2\cos\psi_d,\\
 v_d&=sr_{\rm SR}F_{\rm SR}^2\sin\psi_d,\\
 \Delta_b&=1-2sr_{\rm PR}F_{\rm PR}^2\cos\psi_b+s^2r_{\rm PR}^2F_{\rm PR}^4,\\
 \Delta_d&=1-2sr_{\rm SR}F_{\rm SR}^2\cos\psi_d+s^2r_{\rm SR}^2F_{\rm SR}^4,
 \end{aligned}
 \label{eq:scalar-optical-factors}
\end{equation}
where $s=-e^{-i\Omega\tau_{\rm link}}r_{\rm cav}^2$,
$\psi_b=\pi/2-\Phi_{\rm PR}$, and $\psi_d=\pi/2+\Phi_{\rm SR}$.
Thus $u_r,v_r$ contain only mirror reflectivities, recycling phases, and
propagation. In particular, $\Delta_r=u_r^2+v_r^2$, not
$|u_r|^2+|v_r|^2$.

The scalar torque fed back per unit bar angle is
\begin{equation}
 \begin{aligned}
 \mathcal K_\theta={}&\kappa_0+
 \sum_{r=b,d}\frac{u_r(f_{r1}g_{r1}+f_{r2}g_{r2})}{\Delta_r}\\
 &+\sum_{r=b,d}\frac{v_r(f_{r2}g_{r1}-f_{r1}g_{r2})}{\Delta_r},\\
 d_\theta&=1-\chi\mathcal K_\theta,\qquad
 M=\Delta_b\Delta_d d_\theta.
 \end{aligned}
 \label{eq:scalar-rotational-denominator}
\end{equation}
Here $g_{r\alpha}$ is the returning-field response to angle,
$f_{r\alpha}$ converts that returning quadrature to torque, and
$\kappa_0$ is the torque response at fixed returning fields.
The two fractions describe angle--light--torque propagation through the
bright and dark ports. The restoring-sign rigidity used in
Appendix~\ref{app:sensing} is $K_{\rm opt}=-\mathcal K_\theta$.

Define $\mathcal T^{(0)}_{q\alpha}$ as the driving torque per
$\sqrt{\hbar\omega_0}\hat n_{q\alpha}$ with the bar held fixed but all
optical cavities closed. Its explicit input-dependent expressions are given
below. The angle is then
\begin{equation}
 \hat\theta_j=
 \frac{s_j\theta_{\rm sig}
 +\chi\sqrt{\hbar\omega_0}\sum_{q,\alpha}
       \mathcal T^{(0)}_{q\alpha}\hat n_{q\alpha}}{d_\theta}.
 \label{eq:scalar-driven-angle}
\end{equation}
This equation separates the prescribed signal angle from the rotation driven
by each vacuum. The same feedback denominator acts on both.

\subsection{The two output quadratures and their Chen coefficients}
Let $\mathcal O_{q,\mu\alpha}$ be the fixed-bar optical transfer from
input quadrature $\alpha$ to outgoing SRM quadrature $\mu$, in vacuum
units. Equations below give these scalars directly for each injection site.
Writing $o=t_{\rm SR}F_{\rm SR}$, substitution of the SRM transmission
$oJ$, the two optical equations, and Eq.~\eqref{eq:scalar-driven-angle}
gives
\begin{equation}
 \begin{aligned}
 \hat b^{\rm ext}_1={}&
 \sum_{q,\alpha}\mathcal O_{q,1\alpha}\hat n_{q\alpha}
 -\frac{o(v_dg_{d1}+u_dg_{d2})}{\Delta_d\sqrt{\hbar\omega_0}}
       \hat\theta_j,\\
 \hat b^{\rm ext}_2={}&
 \sum_{q,\alpha}\mathcal O_{q,2\alpha}\hat n_{q\alpha}
 +\frac{o(u_dg_{d1}-v_dg_{d2})}{\Delta_d\sqrt{\hbar\omega_0}}
       \hat\theta_j.
 \end{aligned}
 \label{eq:scalar-output-quadratures}
\end{equation}
In each line, the first term is light reaching the output with the bar fixed.
The second is phase modulation produced by the actual bar angle. Substituting
Eq.~\eqref{eq:scalar-driven-angle} into this term supplies both the signal
and quantum back action. For the same input $q$, these two optical amplitudes
must be added before taking an absolute square.

Reading off all four noise coefficients gives
\begin{equation}
 \begin{aligned}
 C^{\rm ext}_{q,11}=e^{2i\beta_{\rm sag}}\bigl[&M\mathcal O_{q,11}
 -o\chi\Delta_b(v_dg_{d1}+u_dg_{d2})\mathcal T^{(0)}_{q1}\bigr],\\
 C^{\rm ext}_{q,12}=e^{2i\beta_{\rm sag}}\bigl[&M\mathcal O_{q,12}
 -o\chi\Delta_b(v_dg_{d1}+u_dg_{d2})\mathcal T^{(0)}_{q2}\bigr],\\
 C^{\rm ext}_{q,21}=e^{2i\beta_{\rm sag}}\bigl[&M\mathcal O_{q,21}
 +o\chi\Delta_b(u_dg_{d1}-v_dg_{d2})\mathcal T^{(0)}_{q1}\bigr],\\
 C^{\rm ext}_{q,22}=e^{2i\beta_{\rm sag}}\bigl[&M\mathcal O_{q,22}
 +o\chi\Delta_b(u_dg_{d1}-v_dg_{d2})\mathcal T^{(0)}_{q2}\bigr].
 \end{aligned}
 \label{eq:C-elements}
\end{equation}
The two signal coefficients simplify to
\begin{equation}
 \begin{aligned}
 D^{\rm ext}_1&=-\frac{e^{i\beta_{\rm sag}}\theta_{\rm SQL,d}s_j}
 {\sqrt{\hbar\omega_0}}\,o\Delta_b(v_dg_{d1}+u_dg_{d2}),\\
 D^{\rm ext}_2&=+\frac{e^{i\beta_{\rm sag}}\theta_{\rm SQL,d}s_j}
 {\sqrt{\hbar\omega_0}}\,o\Delta_b(u_dg_{d1}-v_dg_{d2}).
 \end{aligned}
 \label{eq:D-elements}
\end{equation}
These expressions have the form of Eq.~\eqref{eq:chen-multiport-output} at
the external SRM plane. The factors $e^{2i\beta_{\rm sag}}$ and
$e^{i\beta_{\rm sag}}$ cancel the extracted phases in that equation;
they do not add a propagation phase to the physical field.
Setting $\chi=0$ removes the torque-mediated noise term and leaves the
fixed-bar optical coefficients, providing a direct check of this separation.

\subsection{Main-input reflection and rear vacuum-field transmission}
First consider a unit vacuum entering the external SRM port of one axis.
Prompt reflection and circulation through the interferometer give
\begin{equation}
 \begin{aligned}
 \mathcal O_{11}=\mathcal O_{22}
 &=r_{\rm SR}\cos\Phi_{\rm SR}+\frac{s o^2v_d}{\Delta_d},\\
 \mathcal O_{12}=-\mathcal O_{21}
 &=r_{\rm SR}\sin\Phi_{\rm SR}+\frac{s o^2u_d}{\Delta_d},\\
 \mathcal T^{(0)}_1
 &=t_1-so\frac{u_df_{d1}+v_df_{d2}}{\Delta_d},\\
 \mathcal T^{(0)}_2
 &=t_2-so\frac{-v_df_{d1}+u_df_{d2}}{\Delta_d}.
 \end{aligned}
 \label{eq:main-scalar-injection}
\end{equation}
The $r_{\rm SR}$ terms are prompt reflection; the terms proportional to
$s o^2$ have crossed the SRM twice. The torque coefficients $t_{1,2}$
refer to the first injection with the returning fields held fixed, as in
Appendix~\ref{app:chen-derivation}. The remaining terms sum its optical
returns. The main input at OBS is shared between the two such axis responses,
as specified below; it is not two independent SRM vacua.

For the rear vacuum fields, we use $d_{\rm ring}=1-r_Ir_Er_Mz_1z_2z_3$ and define
the single-pass-to-Sagnac amplitude $\mathcal R_q$ by
\begin{equation}
 \begin{aligned}
 \mathcal R_{E,i3}&=-\frac{t_It_E z_bz_1}{\sqrt2\,d_{\rm ring}},&
 \mathcal R_{E,i4}&=-\frac{t_It_Er_M z_bz_2z_3}{\sqrt2\,d_{\rm ring}},\\
 \mathcal R_{M,L3}&=-\frac{t_It_Mr_E z_bz_1z_2}{\sqrt2\,d_{\rm ring}},&
 \mathcal R_{M,R3}&=+\frac{t_It_Mr_E z_bz_1z_2}{\sqrt2\,d_{\rm ring}},\\
 \mathcal R_{M,i4}&=-\frac{t_It_M z_bz_3}{\sqrt2\,d_{\rm ring}}.
 \end{aligned}
 \label{eq:rear-ring-coefficients}
\end{equation}
Here $i=L,R$; rear paths 3 and 4 feed directions B and A, respectively.
The factor $1/\sqrt2$ is the Sagnac beam-splitter amplitude. A rear field
injected into direction A subsequently traverses the other ring, adding
$r_{\rm cav}z_c$; one injected into direction B exits toward the beam
splitter immediately. 

\begin{table}[tb]

\caption{Explicit fixed-bar rear vacuum-field coefficients. Each entry is
multiplied by $o\mathcal R_q/\Delta_d$, with the mirror-specific amplitude
already evaluated in Eq.~\eqref{eq:rear-ring-coefficients}. Columns give
output quadrature first and input quadrature second.}
\label{tab:rear-C-injection}
\begin{ruledtabular}
\begin{tabular}{ccccc}
Site & $\mathcal O_{11}$ & $\mathcal O_{12}$ & $\mathcal O_{21}$ & $\mathcal O_{22}$\\
\hline
L, 3 & $-u_d$ & $v_d$ & $-v_d$ & $-u_d$\\
L, 4 & $r_{\rm cav}z_cv_d$ & $r_{\rm cav}z_cu_d$ & $-r_{\rm cav}z_cu_d$ & $r_{\rm cav}z_cv_d$\\
R, 3 & $-v_d$ & $-u_d$ & $u_d$ & $-v_d$\\
R, 4 & $-r_{\rm cav}z_cu_d$ & $r_{\rm cav}z_cv_d$ & $-r_{\rm cav}z_cv_d$ & $-r_{\rm cav}z_cu_d$\\
\end{tabular}
\end{ruledtabular}
\end{table}
Table~\ref{tab:rear-C-injection} replaces both the ring-injection matrices
and their multiplication by the SR response. There is no prompt-reflection
term for a rear vacuum field. ETM and MTM inputs use the same four path patterns but
different $\mathcal R_q$ and driving torques.

To make the torque substitution equally explicit, the following pairs give
$(\mathcal T^{(0)}_{q1}-t_{q1},\mathcal T^{(0)}_{q2}-t_{q2})$ for each
site. For L, 3 they are
\begin{equation}
 \begin{aligned}
 \mathcal T^{(0)}_{q1}-t_{q1}
 &=\mathcal R_q\left[\frac{u_bf_{b1}+v_bf_{b2}}{\Delta_b}
 +\frac{-v_df_{d1}+u_df_{d2}}{\Delta_d}\right],\\
 \mathcal T^{(0)}_{q2}-t_{q2}
 &=\mathcal R_q\left[\frac{-v_bf_{b1}+u_bf_{b2}}{\Delta_b}
 -\frac{u_df_{d1}+v_df_{d2}}{\Delta_d}\right].
 \end{aligned}
 \label{eq:rear-torque-L3}
\end{equation}
For R, 4, multiply the right-hand sides of
Eq.~\eqref{eq:rear-torque-L3} by $r_{\rm cav}z_c$ and use that input's
$\mathcal R_q$ and $t_{q\alpha}$. For R, 3 the pair is
\begin{equation}
 \begin{aligned}
 \mathcal T^{(0)}_{q1}-t_{q1}
 &=\mathcal R_q\left[\frac{-v_bf_{b1}+u_bf_{b2}}{\Delta_b}
 +\frac{u_df_{d1}+v_df_{d2}}{\Delta_d}\right],\\
 \mathcal T^{(0)}_{q2}-t_{q2}
 &=\mathcal R_q\left[-\frac{u_bf_{b1}+v_bf_{b2}}{\Delta_b}
 +\frac{-v_df_{d1}+u_df_{d2}}{\Delta_d}\right].
 \end{aligned}
 \label{eq:rear-torque-R3}
\end{equation}
For L, 4, multiply the right-hand sides of
Eq.~\eqref{eq:rear-torque-R3} by $-r_{\rm cav}z_c$, again using that input's
$\mathcal R_q$ and $t_{q\alpha}$. These sign and delay rules account for
all eight rear vacuum fields on one axis. The coefficients $f_{r\alpha}$ and
$t_{q\alpha}$ are calculated from the local force in
Eq.~\eqref{eq:force-transmission-scalar}, not fitted: hold the angle fixed,
propagate a unit returning quadrature or a unit injected quadrature through
the ring relations, and take $\ell(F_L-F_R)$. The field-to-force conversion
there contains only carrier components, $r_E,t_E$, and $c$.

For example, inserting the ETM L, 3 amplitude directly into the first optical
term of $\hat b^{\rm ext}_1$ gives
\begin{equation}
 \left.\hat b^{\rm ext}_1\right|_{E,L3,\,\theta=0}
 =\frac{o t_It_E z_bz_1}{\sqrt2\,d_{\rm ring}\Delta_d}
       (u_d\hat n^E_{L3,1}-v_d\hat n^E_{L3,2}).
 \label{eq:rear-worked-scalar}
\end{equation}
The factors identify, in order, transmission into the ring, cavity
circulation, propagation to BS$_j$, and SRC circulation. Its back action
contribution follows by inserting the same input's
Eq.~\eqref{eq:rear-torque-L3} into Eq.~\eqref{eq:C-elements}.

\subsection{Combining the axes and projecting the signal}
Let $z_O$ describe the SRM-to-OBS propagation on either symmetric output
path, and $z_h$ the OBS-to-homodyne propagation. With the output-mirror
phases used here, the fields obey the scalar relations
\begin{equation}
 \begin{aligned}
 \hat b_{{\rm sig},1}
 &=z_hz_O(-t_{\rm OBS}\hat b^{\rm ext}_{X,2}
                         +r_{\rm OBS}\hat b^{\rm ext}_{Y,1}),\\
 \hat b_{{\rm sig},2}
 &=z_hz_O(t_{\rm OBS}\hat b^{\rm ext}_{X,1}
                         +r_{\rm OBS}\hat b^{\rm ext}_{Y,2}).
 \end{aligned}
 \label{eq:scalar-obs-output}
\end{equation}
For the single main input at OBS, the incident SRM quadratures are
$z_Or_{\rm OBS}(\hat a_1,\hat a_2)$ on X and
$z_Ot_{\rm OBS}(-\hat a_2,\hat a_1)$ on Y. Thus the two axis amplitudes
from $a$ interfere in Eq.~\eqref{eq:scalar-obs-output}; their PSDs cannot
be added separately. Each rear vacuum field originates on only one axis.

For an arbitrary scalar detection row $(\lambda_1,\lambda_2)$ applied
at the external SRM plane, the angular signal response of that axis is
\begin{equation}
 \begin{aligned}
 H_{\theta,j}=\frac{s_j o\Delta_b}{M\sqrt{\hbar\omega_0}}
 \bigl[&\lambda_2(u_dg_{d1}-v_dg_{d2})\\
       &-\lambda_1(v_dg_{d1}+u_dg_{d2})\bigr].
 \end{aligned}
 \label{eq:scalar-sensing-axis}
\end{equation}
For the signal path, Eq.~\eqref{eq:scalar-obs-output} gives the effective
rows $z_hz_Ot_{\rm OBS}(\lambda_{{\rm sig},2},-\lambda_{{\rm sig},1})$
on X and $z_hz_Or_{\rm OBS}(\lambda_{{\rm sig},1},\lambda_{{\rm sig},2})$
on Y. Summing the two signal responses yields the signal-path sensing
function. Any finite-LO contribution is projected separately and added
coherently using Appendix~\ref{app:detector}. Appendix~\ref{app:sensing}
then factorizes the same frequency dependence into optical phase factors;
its subsequent pole--zero approximation is not used in this scalar reduction.

\section{Trigonometric form of the sensing function}
\label{app:sensing}

This appendix starts from the explicit sensing response in
Eq.~\eqref{eq:scalar-sensing-axis}. The first two subsections factor its
optical denominator and project its numerator without approximation. The
last two introduce a low-frequency expansion for calibration. Keeping this
distinction separates algebraic identities, which preserve the sensitivity
curves, from approximations whose band of validity must be checked.

\subsection{Factorization of the optical denominator}
\label{sec:sensing-trigonometric}
We rearrange the sensing function to display two optical response factors and a
rotational term. For the symmetric reference geometry, define
$\vartheta=\beta_{\rm sag}+\Omega L_{\rm SR}/c$ and $\psi_d=\pi/2+\Phi_{\rm
SR}$. The SR round-trip amplitude magnitude is $\varrho=|r_{\rm cav}|^2r_{\rm
SR}$, so $sr_{\rm SR}F_{\rm SR}^2=\varrho e^{-2i\vartheta}$.
With $x_\pm=\vartheta\mp\psi_d/2$ and $\Lambda_{\rm
SR}=(1+\varrho)/(1-\varrho)$,
\begin{equation}
 \begin{aligned}
 \mathcal F_\pm&=\cos x_\pm+i\Lambda_{\rm SR}\sin x_\pm,\\
 d_\pm&=1-\varrho e^{-2ix_\pm}
       =(1-\varrho)e^{-ix_\pm}\mathcal F_\pm
 \end{aligned}
 \label{eq:sensing-optical-factors}
\end{equation}
This is an exponential identity, with no approximation to frequencies within
the cavity bandwidth. The projectors $\mathsf E_\pm=(\mathbb I_2\mp iJ)/2$
satisfy $J\mathsf E_\pm=\pm i\mathsf E_\pm$, giving $\mathsf A_d=d_+\mathsf
E_++d_-\mathsf E_-$, $\Delta_d=d_+d_-$, and $\operatorname{Cof}(\mathsf A_d)=d_-\mathsf
E_++d_+\mathsf E_-$.

Eliminating the bright port and defining $\kappa_{\rm
eff}=\kappa_0+h_b/\Delta_b$ transforms Eq.~\eqref{eq:chen-common-denominator}
into
\begin{equation}
 \frac M{\Delta_b}=\Delta_d-\chi(\kappa_{\rm eff}\Delta_d+h_d)
 \label{eq:sensing-reduced-denominator}
\end{equation}
With $\mathcal A_{\rm SR}=(1-\varrho)^2e^{-2i\vartheta}$ and $f=\Omega/(2\pi)$,
\begin{equation}
 \begin{aligned}
 \frac M{\Delta_b\mathcal A_{\rm SR}}
       &=\mathcal F_+\mathcal F_- -\frac{\Xi(f)}{f^2},\\
 \Xi(f)&=-\frac{\kappa_{\rm eff}\Delta_d+h_d}
                {4\pi^2I\mathcal A_{\rm SR}}.
 \end{aligned}
 \label{eq:sensing-mechanical-factor}
\end{equation}
The factor $1/f^2$ originates from free rotational susceptibility. In general,
$\Xi$ is complex and frequency dependent; it is not assumed to be the square of
a positive constant. PRC feedback remains in $h_b/\Delta_b$, while
carrier-mediated PRC effects remain in $\vect g_r,\vect f_r,\kappa_0$.

\subsection{Signal numerator and homodyne projection}
The numerator can be written using the same optical factors. Let $\Gamma_j$
propagate the external SRM field on axis $j$ to the signal readout port. For
the zero-phase output folds, $\Gamma_X=z_{\rm out}t_{\rm OBS}J$,
$\Gamma_Y=z_{\rm out}r_{\rm OBS}\mathbb I_2$, and $z_{\rm out}=e^{-i\Omega
L_{\rm out}/c}$. Substituting the SRM transmission
$t_{\rm SR}F_{\rm SR}J$ and detector row gives

\begin{equation}
 g_\pm=\sum_{j=X,Y}\frac{s_jt_{\rm SR}F_{\rm SR}}{\sqrt{\hbar\omega_0}}
 \lambda_{\rm sig}\Gamma_jJ\mathsf E_\pm\vect g_{d,j}.
 \label{eq:sensing-projected-gains}
\end{equation}

With common
$M,\Delta_b$, the signal response is $H_\theta=(g_+d_-+g_-d_+)/(M/\Delta_b)$.
This yields Eq.~\eqref{eq:sensing-trigonometric}, with gains
\begin{equation}
 G_1=\frac{g_+e^{ix_+}}{1-\varrho},\qquad
 G_2=-\frac{g_-e^{ix_-}}{1-\varrho}
 \label{eq:sensing-gains}
\end{equation}
The units of $G_1,G_2$ are W/rad, and those of $\Xi$ are Hz$^2$. The
trigonometric phases follow from $\beta_{\rm sag}$, the SRC round-trip phase,
and transit times. The two numerator terms are homodyne projections of the two
rotation eigencomponents. All cavity reflectivities, transmissions, carrier
fields, rotational dynamics, and detection angles remain in these coefficients.
We neither replace $\varrho,G_1,G_2,\Xi$ by frequency-independent constants nor
introduce approximations such as $\sin x\simeq x$. At singular parameter values
with $\Delta_b=0$ or $1-\varrho=0$, the limit must be taken using
Eq.~\eqref{eq:sensing-function-chen} before division.

Equation~\eqref{eq:sensing-trigonometric} is an alternative representation of
the sensing function for the same symmetric configuration, not a new physical
model. Over 0.05 Hz--10 kHz, its maximum relative difference from the original
complex TF is below $2.8\times10^{-10}$. Evaluating
Eq.~\eqref{eq:sensing-function-chen} with $M,\vect D,\lambda$ obtained from the
boundary conditions agrees with the common-denominator calculation within
$6.8\times10^{-16}$. Rearranging the homodyne projection likewise preserves
every input coefficient.

\subsection{Small-phase approximation and calibration pole--zero form}
\label{sec:sensing-pole-zero}
A finite-order calibration model follows from a small-phase expansion of
Eq.~\eqref{eq:sensing-optical-factors}. Near SRC resonance, $x_-$ lies near
$\pi$, so the small quantities are $x_+$ and $x_--\pi$. Substituting
$\Phi_{\rm SR}=\pi/2-2\delta_{\rm SR}$ and retaining the linear ring phase gives
\begin{equation}
 \begin{aligned}
 x_+&\simeq\Omega\tau_{\rm eff}+\delta_{\rm SR},\qquad
 x_--\pi\simeq\Omega\tau_{\rm eff}-\delta_{\rm SR},\\
 \tau_{\rm eff}&=2\left.\frac{d\beta_{\rm ring}}{d\Omega}\right|_0
              +\frac{\tau_{\rm link}}2+\frac{L_{\rm SR}}c.
 \end{aligned}
 \label{eq:small-phase-delay}
\end{equation}
Angles are in radians. Set $\Lambda_0=\Lambda_{\rm SR}(0)$ and use
$\cos x\simeq1$, $\sin x\simeq x$ where
$|\Omega\tau_{\rm eff}\pm\delta_{\rm SR}|\ll1$. To distinguish the Laplace
variable from the optical coefficient $s$, write $s_{\rm L}=i\Omega$. Then
\begin{equation}
 \begin{aligned}
 \mathcal F_+&\simeq\Lambda_0\tau_{\rm eff}(s_{\rm L}-p_+),\\
 \mathcal F_-&\simeq-\Lambda_0\tau_{\rm eff}(s_{\rm L}-p_-),\\
 p_\pm&=-\gamma_c\mp i\Delta_c,\qquad
 \gamma_c=\frac1{\Lambda_0\tau_{\rm eff}},\quad
 \Delta_c=\frac{\delta_{\rm SR}}{\tau_{\rm eff}}.
 \end{aligned}
 \label{eq:small-phase-poles}
\end{equation}
The zeros of $\mathcal F_\pm$ are therefore a conjugate pair of poles of
$1/(\mathcal F_+\mathcal F_-)$. The quantities $\gamma_c$ and $\Delta_c$ are
the damping rate and detuning frequency shift. The frequency variation of
$\Lambda_{\rm SR}$ and the nonlinear ring phase must also remain small.

If $G_1,G_2$ can additionally be treated as constant over the band, the
optical response to a prescribed angle ($H_{\rm opt}$ below) becomes
\begin{equation}
 \begin{aligned}
 H_\theta^{\rm opt}(s_{\rm L})
 &\simeq K_c\frac{s_{\rm L}-z_{\rm opt}}
 {(s_{\rm L}-p_+)(s_{\rm L}-p_-)},\\
 K_c&=\frac{G_1+G_2}{\Lambda_0\tau_{\rm eff}},\qquad
 z_{\rm opt}=\frac{G_1p_-+G_2p_+}{G_1+G_2}
 \end{aligned}
 \label{eq:calibration-optical-zero}
\end{equation}
The zero $z_{\rm opt}$ arises from cancellation between the two homodyne-projected
signal components. When $G_1+G_2=0$, use the unfactored numerator. This is a
conditional optical approximation; it does not justify dropping the
frequency dependence of $G_{1,2}$ that produces a speed-meter response.

For the two reference points, $\tau_{\rm eff}=0.67819\,\mathrm{ms}$ and
$\gamma_c/(2\pi)=41.17\,\mathrm{Hz}$. The detuning shift
$\Delta_c/(2\pi)$ is 12.97 Hz for $J_{\min}$ and 2.89 Hz for $J_{\log}$.
Over 0.05--10 Hz, the maximum complex relative errors of
Eq.~\eqref{eq:small-phase-poles} against the exact $\mathcal F_\pm$ are 0.43\%
and 0.15\%, respectively. These bound the optical-factor approximation,
not the full-TF calibration error. All sensitivity curves and FINESSE 3
comparisons in the main text retain the exact expressions.

\subsection{Rotational feedback as a second-order Q filter}
\label{sec:sensing-mechanical-q}
Bar motion modulates the reflected light, and radiation pressure from the
returning field drives the bar again. To separate this feedback from the
optical response to an imposed angle, normalize the optical denominator at
zero frequency, $\mathcal D_0=[\mathcal F_+\mathcal F_-]_{f=0}$. Then
\begin{equation}
 \begin{aligned}
 \mathcal D_{\rm opt}&=\frac{\mathcal F_+\mathcal F_-}{\mathcal D_0},\qquad
 \mathcal X=\frac{\Xi}{\mathcal D_0},\\
 H_{\rm opt}&=\frac{G_1\mathcal F_--G_2\mathcal F_+}
                    {\mathcal F_+\mathcal F_-},\\
 H_\theta&=H_{\rm opt}\mathcal B_\theta,\qquad
 \mathcal B_\theta=\frac{f^2}{f^2-\mathcal X/\mathcal D_{\rm opt}}.
 \end{aligned}
 \label{eq:sensing-feedback-factorization}
\end{equation}
Here $H_{\rm opt}$ is the optical response to a prescribed angle, and
$\mathcal B_\theta$ closes the radiation-pressure feedback on that angle.
This factorization is exact for the symmetric configuration and retains the
frequency dependence of $\mathcal X(f)$.

At the same order as Eq.~\eqref{eq:small-phase-poles}, the optical denominator is
\begin{equation}
 \begin{aligned}
 \mathcal D_{\rm opt}(f)&\simeq
 1+\frac{if}{f_{\rm opt}Q_{\rm opt}}-\frac{f^2}{f_{\rm opt}^2},\\
 f_{\rm opt}&=\frac{\sqrt{\gamma_c^2+\Delta_c^2}}{2\pi},\qquad
 Q_{\rm opt}=\frac{\sqrt{\gamma_c^2+\Delta_c^2}}{2\gamma_c}.
 \end{aligned}
 \label{eq:calibration-optical-q}
\end{equation}
The quantity $Q_{\rm opt}$ summarizes the optical pole pair and is distinct
from the mechanical quality factor. If the numerator admits a gain, delay,
and zero approximation, the optical response takes the familiar form
$g e^{-2\pi if\tau}(1+if/z)/\mathcal D_{\rm opt}$. An origin zero required
by the speed-meter response must instead be retained explicitly.

In the special case that $\mathcal X(f)\simeq\xi^2$ can be treated as a real
constant, a first-order expansion of the inverse optical denominator gives
\begin{equation}
 \begin{aligned}
 \frac1{\mathcal D_{\rm opt}}
 &\simeq1-\frac{if}{f_{\rm opt}Q_{\rm opt}},\\
 \mathcal B_\theta(f)&\simeq
 \frac{f^2}{f^2-\xi^2+if\xi^2/(f_{\rm opt}Q_{\rm opt})}.
 \end{aligned}
 \label{eq:calibration-constant-rigidity}
\end{equation}
This yields a product of optical pole--zero response and a second-order
mechanical filter. Terms of order $f^2/f_{\rm opt}^2$ and
$f^2/(f_{\rm opt}Q_{\rm opt})^2$ have been dropped. Near a resonance, a small
denominator error can be amplified, so the full complex TF must be checked.
The frequency-dependent $\Xi$ of the present model cannot generally be
replaced by a constant $\xi^2$.

For the general case, define optical rigidity with the restoring-torque
sign convention $\tau_{\rm rad}=-K_{\rm opt}\theta$:
\begin{equation}
 \begin{aligned}
 K_{\rm opt}&=-\left(\kappa_0+\frac{h_b}{\Delta_b}
                                  +\frac{h_d}{\Delta_d}\right)
       =\frac{4\pi^2 I\mathcal X}{\mathcal D_{\rm opt}},\\
 \mathcal B_\theta(s_{\rm L})&=
 \frac{Is_{\rm L}^2}{Is_{\rm L}^2+K_{\rm opt}(s_{\rm L})}.
 \end{aligned}
 \label{eq:calibration-optical-rigidity}
\end{equation}
Expand the rigidity, including bright- and dark-port feedback, as
$K_{\rm opt}=K_0+K_1s_{\rm L}+K_2s_{\rm L}^2+\cdots$. The real coefficients
$K_0$, $K_1$, and $K_2$ describe restoring torque, velocity-dependent damping,
and an optical correction to moment of inertia. Thus
\begin{equation}
 \begin{aligned}
 I_{\rm eff}&=I+K_2,\qquad
 \omega_m^2=\frac{K_0}{I_{\rm eff}},\qquad
 \Gamma_m=\frac{K_1}{I_{\rm eff}},\\
 \mathcal B_Q(s_{\rm L})&=
 \frac I{I_{\rm eff}}
 \frac{s_{\rm L}^2}{s_{\rm L}^2+\Gamma_m s_{\rm L}+\omega_m^2},\\
 Q_m&=\frac{\omega_m}{\Gamma_m}\quad(\omega_m>0).
 \end{aligned}
 \label{eq:calibration-mechanical-q}
\end{equation}
This is the second-order Q filter for rotational feedback. A conventional
stable resonance requires $I_{\rm eff}>0$, $K_0>0$, and $\Gamma_m>0$.
With the adopted $e^{+i\Omega t}$ convention, its frequency-domain denominator
is $f^2-f_m^2-if f_m/Q_m$, where $f_m=\omega_m/(2\pi)$.
The $+i$ term in Eq.~\eqref{eq:calibration-constant-rigidity} must therefore
not be identified with positive mechanical damping.

For the present free-bar, symmetric configuration, both reference points
give $|K_0|<10^{-9}\,\mathrm{N\,m/rad}$: the static restoring term vanishes
within numerical precision. Cancelling the common origin factor in this limit gives
\begin{equation}
 \mathcal B_Q(s_{\rm L})\simeq
 \frac I{I_{\rm eff}}\frac{s_{\rm L}}{s_{\rm L}+\Gamma_m}.
 \label{eq:calibration-mechanical-limit}
\end{equation}
A finite resonance frequency and $Q_m$ are consequently not assigned.
For $J_{\min}$ and $J_{\log}$, respectively, $I_{\rm eff}$ is 17.865 and
19.764 $\mathrm{kg\,m^2}$, and $\Gamma_m$ is $-1.437$ and
$-0.06285\,\mathrm{s^{-1}}$. Negative damping indicates antidamping in this
uncontrolled low-frequency model; it does not describe an already
stabilized mechanical resonance.

Keeping the optical response exact and approximating only the feedback by
Eq.~\eqref{eq:calibration-mechanical-limit}, the maximum complex-TF errors
are 0.50\% and 0.024\% over 0.05--1 Hz, and 4.9\% and 0.23\% over
0.05--10 Hz. The exact factorization in
Eq.~\eqref{eq:sensing-feedback-factorization} agrees with the original TF
computed from local fields within a relative difference of $10^{-14}$ at
the verification frequencies. Calibration approximations are checked
against the original model; they do not change the sensitivity curves.

To find poles and zeros of the full TF including mechanics, return to the
common-denominator expression. Express the coefficients using propagation
delays $e^{-s_{\rm L}L/c}$ and susceptibility $1/(Is_{\rm L}^2)$. Writing
$H_\theta=N_H/D_H$ gives the exact rearrangement
\begin{equation}
 \begin{aligned}
 N_H&=Is_{\rm L}^2\Delta_b(g_+d_-+g_-d_+),\\
 D_H&=(Is_{\rm L}^2-\kappa_0)\Delta_b\Delta_d
             -h_b\Delta_d-h_d\Delta_b.
 \end{aligned}
 \label{eq:calibration-exact-numerator-denominator}
\end{equation}
After clearing shared optical denominators and cancelling common factors,
zeros of $N_H$ give signal zeros and zeros of $D_H$ give system poles. In
particular, do not count the origin factor $s_{\rm L}^2$ before cancellation.
A low-order calibration model expands both expressions over the same band,
retaining the leading nonzero speed-meter signal term and rotational
feedback. With a separable pure delay $\tau_d$ and retained zeros and poles
$z_k,p_k$, the resulting rational model has the form
$H_{\rm cal}=K e^{-s_{\rm L}\tau_d}\prod_k(s_{\rm L}-z_k)/\prod_k(s_{\rm L}-p_k)$.
Its calibration accuracy is assessed using the complex ratio
$H_{\rm cal}/H_\theta$.

\section{Detection coefficients with a finite local oscillator}
\label{app:detector}
Balanced homodyne detection mixes the signal and LO fields at a readout beam
splitter and measures the power difference between photodetectors $D_1,D_2$.
We derive the coefficients $\lambda_{\rm sig},\lambda_{\rm LO}$ from this
measured power difference.

Use the signal and LO fields propagated to the mixer,
$\hat{\vect u}_j=\sqrt{\hbar\omega_0}\hat{\vect b}_j$ for
$j=\mathrm{sig},\mathrm{LO}$. Each $\hat{\vect b}_j$ has the
input--output form of Eq.~\eqref{eq:chen-multiport-output}, with the
corresponding coefficients $C_{q,j}$ and $\vect D_j$.
These follow by propagating the output coefficients of
Appendix~\ref{app:chen-components} to the two mixer paths.
The remaining step is to project those fields onto the measured
power difference.

Denote the reflection--transmission matrix from incident path
$j=\mathrm{sig},\mathrm{LO}$ to detector $D_i$ by $(S_{\rm HD})_{D_i,j}$.
This $2\times2$ matrix maps the incident amplitude and phase quadratures to
the two outgoing quadratures at that detector. It is obtained by selecting
the corresponding rows and columns of the beam-splitter matrix in
Appendix~\ref{app:local}. The field fluctuation at each detector and its
interference with the carrier produce
\begin{equation}
 \begin{aligned}
 \hat{\vect v}_{D_i}&=\sum_{j=\mathrm{sig},\mathrm{LO}}
                 (S_{\rm HD})_{D_i,j}\hat{\vect u}_j,\\
 \delta\hat P_{D_i}&=\bar{\vect v}_{D_i}^{\mathsf T}\hat{\vect v}_{D_i},\qquad
 \hat o=\delta\hat P_{D_1}-\delta\hat P_{D_2}.
 \end{aligned}
\end{equation}
Collecting the coefficient of each incident field in this difference gives
\begin{equation}
 \lambda_j=\sqrt{\hbar\omega_0}
 \left(\bar{\vect v}_{D_1}^{\mathsf T}(S_{\rm HD})_{D_1,j}
       -\bar{\vect v}_{D_2}^{\mathsf T}(S_{\rm HD})_{D_2,j}\right).
\end{equation}
Thus $\lambda_j$ is a $1\times2$ coefficient that converts a two-quadrature
field into a measured power difference. The detected output is
$\hat o=\lambda_{\rm sig}\hat{\vect b}_{\rm sig}
+\lambda_{\rm LO}\hat{\vect b}_{\rm LO}$. The readout beam splitter is fixed
and satisfies $R_h+T_h=1$. This derivation includes fluctuations of a finite
LO and requires no additional empirical homodyne gain.

For the balanced readout used here, $r_h=t_h=1/\sqrt2$.
Substituting the reflection phase $\Phi_h$ and transmission rotation $J$
into the preceding expression removes the beam-splitter matrices altogether:

\begin{equation}
 \begin{aligned}
 \lambda_{{\rm sig},1}&=\sqrt{2\hbar\omega_0P_{\rm LO}}
                    \sin(\Phi_h-\varphi_{\rm LO}),\\
 \lambda_{{\rm sig},2}&=\sqrt{2\hbar\omega_0P_{\rm LO}}
                    \cos(\Phi_h-\varphi_{\rm LO}),\\
 \lambda_{{\rm LO},1}&=\sqrt{2\hbar\omega_0P_{\rm sig}}
                    \sin(\Phi_h+\varphi_{\rm sig}),\\
 \lambda_{{\rm LO},2}&=-\sqrt{2\hbar\omega_0P_{\rm sig}}
                    \cos(\Phi_h+\varphi_{\rm sig}).
 \end{aligned}
 \label{eq:balanced-readout-scalars}
\end{equation}
Here $P_{\rm LO},\varphi_{\rm LO}$ and
$P_{\rm sig},\varphi_{\rm sig}$ are the carrier powers and phases
at the two incident mixer paths. In this reference-plane convention,
$\zeta=\Phi_h-\varphi_{\rm LO}$. The signal-path readout is therefore
$\sqrt{2\hbar\omega_0P_{\rm LO}}\,
(\hat b_1\sin\zeta+\hat b_2\cos\zeta)$,
the usual homodyne quadrature multiplied by its power gain.

The first two coefficients are set by the LO carrier and select
the measured signal quadrature. The last two are proportional to $\sqrt{P_{\rm sig}}$ and
weight LO fluctuations; they vanish when the signal carrier is zero.
The measured operator is the sum of the four products
$\lambda_{{\rm sig},\alpha}\hat b_{{\rm sig},\alpha}$ and
$\lambda_{{\rm LO},\alpha}\hat b_{{\rm LO},\alpha}$, with $\alpha=1,2$.
This is the scalar readout of the field coefficients in
Appendix~\ref{app:chen-components}.

\section{Numerical checks and additional figures}
\label{app:validation}
\begin{figure}[tbp]
\centering\includegraphics[width=0.70\textwidth]{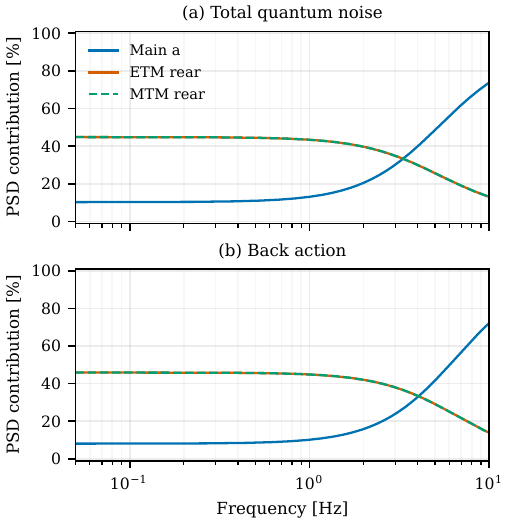}
\caption{Fractions of total quantum PSD and back action PSD supplied by the main input $a$, ETM rear vacuum fields, and MTM rear vacuum fields. Each panel is normalized by the sum over the three retained groups.}
\label{fig:portbudget}
\end{figure}

\begin{figure}[tbp]
\centering\includegraphics[width=0.70\textwidth]{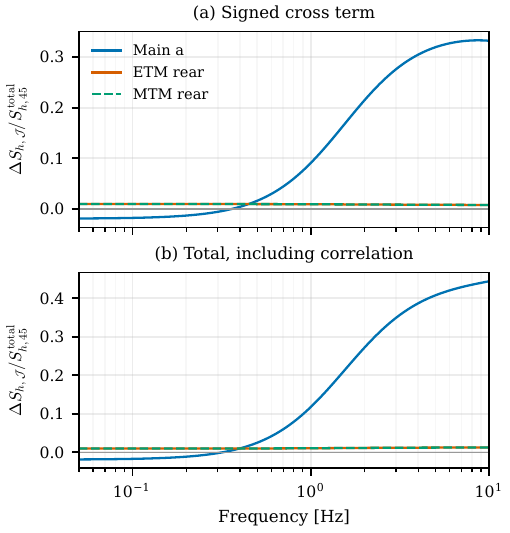}
\caption{Input-resolved PSD changes when $\delta_{\rm PR}$ is changed from $45^\circ$ to $0^\circ$ or $90^\circ$ at matched ETM incident power $P_E$ and fixed SR and LO settings. The carrier phase rotation is not compensated by the LO. The upper panel shows changes in the cross term of Eq.~\eqref{eq:quantum-budget}; the lower panel shows changes in the total contribution including shot noise, back action, and their correlation. All terms use the signal TF of their respective configuration. The denominator is the total quantum PSD at $45^\circ$; the curves show normalized changes, not absolute input fractions.}
\label{fig:prports}
\end{figure}

Comparisons with FINESSE 3 use maximum relative errors in the complex TF and each
ASD; correlation-PSD differences are normalized by the sum of the shot-noise and back action PSDs. Quantum noise at each photodetector and the sum of input covariances were
also checked at representative frequencies. The differential PSD is constructed
from the complex responses to shared inputs as
$S_{11}+S_{22}-2\operatorname{Re}S_{12}$. Figure~\ref{fig:portbudget} shows the input-group fractions.
Figure~\ref{fig:prports} resolves the matched-power comparison of
Sec.~\ref{sec:prdetuning} by input. Matching the summed ETM incident power
$P_E$ while changing PR detuning at fixed SR and LO optical settings leaves
the change in readout phase relative to the carrier. For each input group,
changes in the signed cross PSD and total quantum PSD are normalized to the
total quantum PSD at $45^\circ$. The main input $a$ supplies the largest
change between 1 and 10 Hz. At 0.1 Hz, its decrease partly offsets the
increase from the rear vacuum fields. Thus the noise fractions in
Fig.~\ref{fig:portbudget} and the response to a phase change in
Fig.~\ref{fig:prports} are different quantities.
\begin{figure}[!tp]
\includegraphics[width=\textwidth]{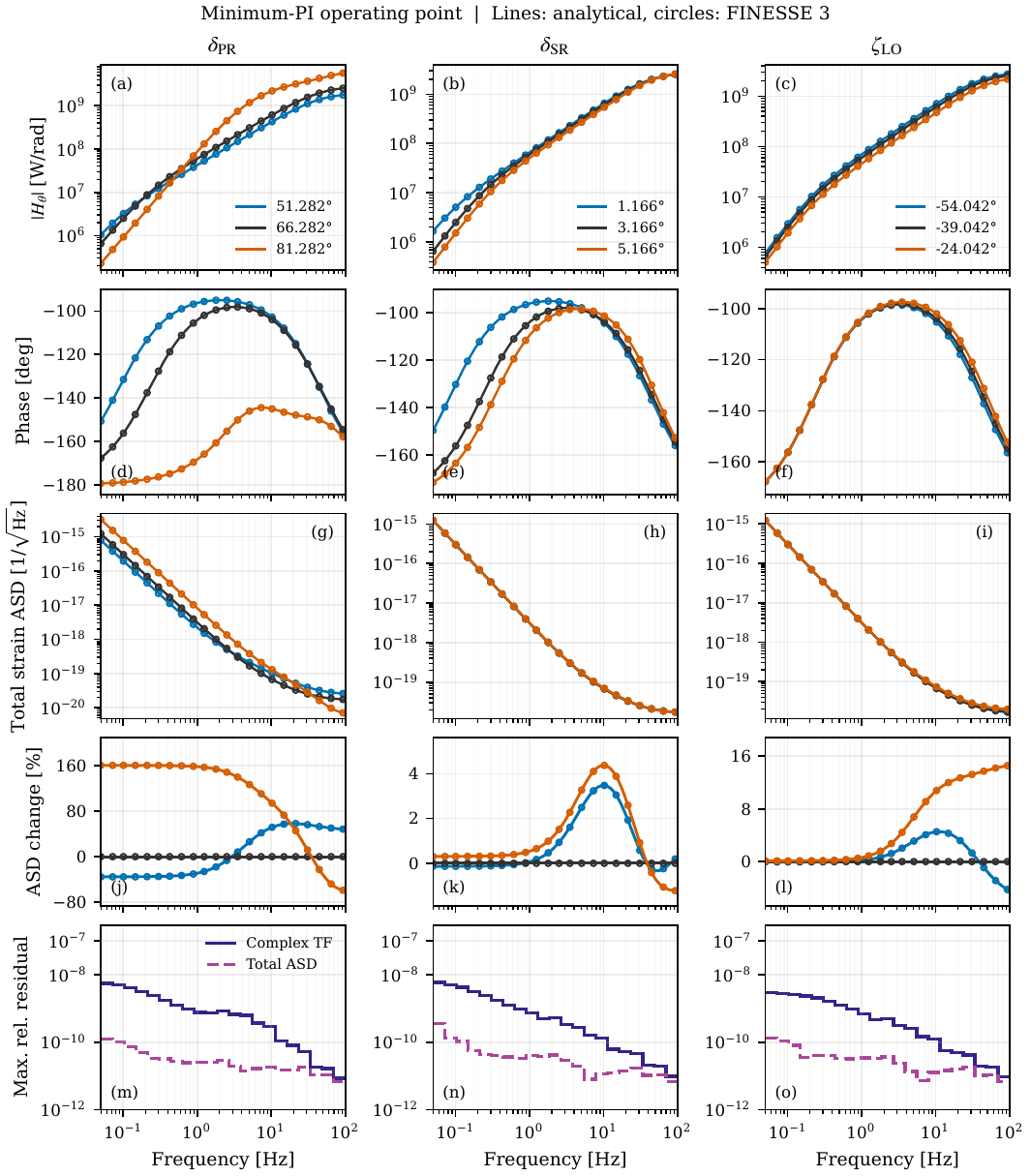}
\caption{Neighborhood of the minimum-PI reference point (0.05--100 Hz). Columns vary PR detuning, SR detuning, and LO angle. Rows show angular TF amplitude, phase, total quantum strain ASD, ASD change from the reference, and interval-maximum relative residuals. In the top four rows, black denotes the reference; blue and orange denote negative and positive shifts (PR $\pm15^\circ$, SR $\pm2^\circ$, LO $\pm15^\circ$). Solid curves are analytical results and circles are selected FINESSE 3 points. The bottom row bounds the complex-TF residual (solid dark purple) and total-quantum-ASD residual (dashed magenta).}
\label{fig:angle-minimum}
\end{figure}

\begin{figure}[!tp]
\includegraphics[width=\textwidth]{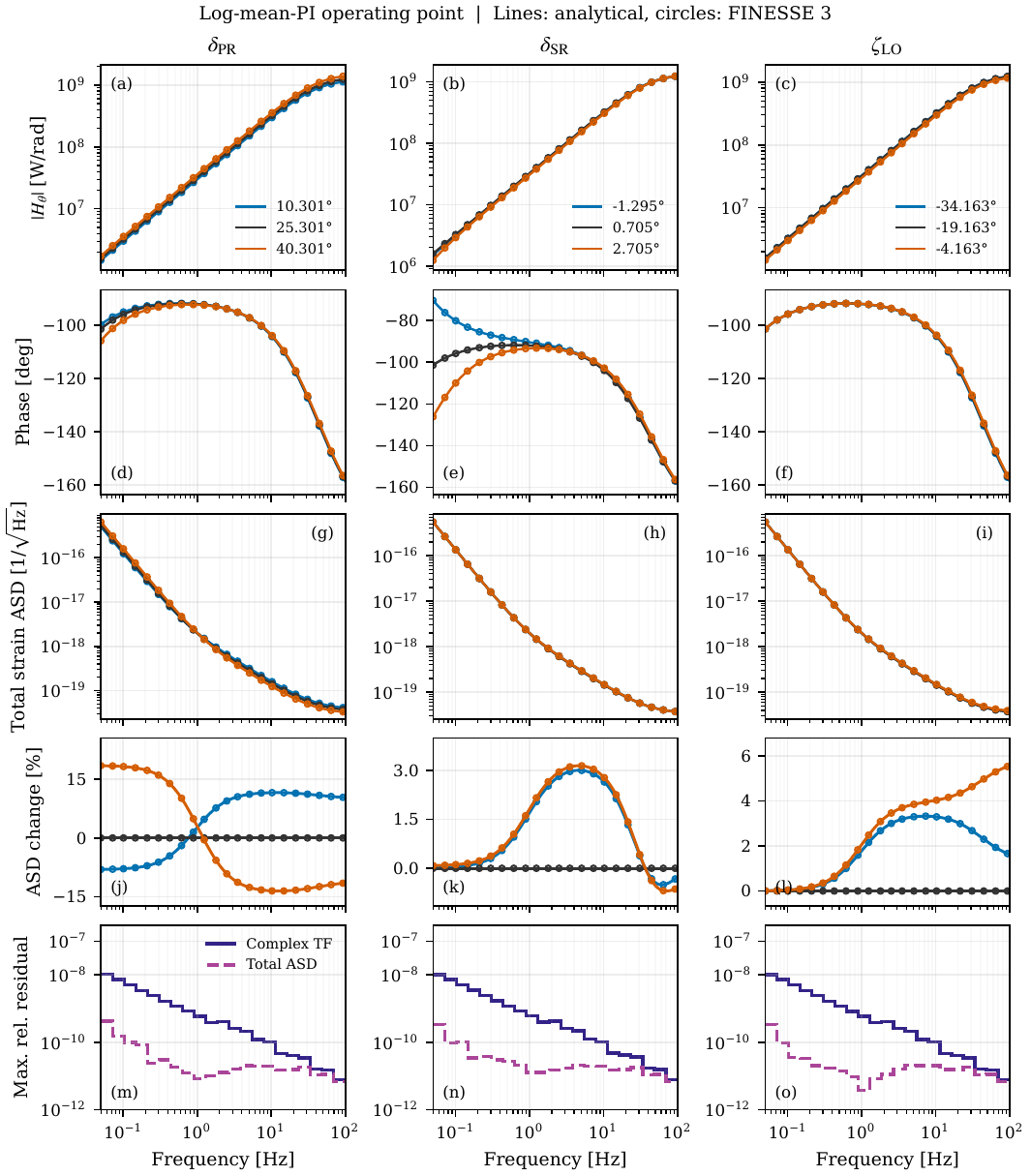}
\caption{Neighborhood of the logarithmic-mean PI reference point. The display and angle shifts match Fig.~\ref{fig:angle-minimum}, with each angle varied independently from this reference point.}
\label{fig:angle-logmean}
\end{figure}

\subsection{Detuning and homodyne-angle variations}
In Figs.~\ref{fig:angle-minimum} and \ref{fig:angle-logmean}, only one
specified angle is changed from each PI reference point. The fourth row shows
$100[\sqrt{S_h}/\sqrt{S_{h,0}}-1]$, where 0 denotes the corresponding
reference point. We define the frequency-dependent relative residuals by
\begin{equation}
 \begin{aligned}
 \epsilon_H(f)&=\frac{|H_\theta^{\rm an}(f)-H_\theta^{\rm Fin}(f)|}
                         {|H_\theta^{\rm Fin}(f)|},\\
 \epsilon_N(f)&=\frac{|\sqrt{S_h^{\rm an}(f)}-\sqrt{S_h^{\rm Fin}(f)}|}
                         {\sqrt{S_h^{\rm Fin}(f)}}.
 \end{aligned}
 \label{eq:validation-residuals}
\end{equation}
Here an and Fin denote the analytical and FINESSE 3 results. The residuals
$\epsilon_H$ and $\epsilon_N$ compare the complex TF and total quantum
ASD, respectively. The bottom row divides 0.05--100 Hz into 21
intervals of equal logarithmic width and shows the maximum residual over all
computed frequencies and all three angle conditions in each interval. Residuals
are not averaged.

Changing the LO angle by $-15^\circ$ and $+15^\circ$ changes the $J_{\min}$
total quantum ASD by $+4.52\%$ and $+10.68\%$ at 10 Hz, but only $-0.039\%$ and
$+0.115\%$ at 0.1 Hz. In this symmetric configuration, the back action output
and signal TF have the same LO projection, so input-referred angular back action
is numerically invariant. Shot and correlation terms change, but have little
effect on the total ASD where low-frequency back action dominates.

\subsection{Omitted inputs and coefficient representations}

Across all 14 conditions over 0.05 Hz--10 kHz, the omitted inputs contribute
less than $2\times10^{-20}$ of the differential detection PSD. The three-group
ASD differs from the complete-model ASD by less than $10^{-12}$, while the TF
is unchanged. This omission accuracy applies to the symmetric differential
readout and not to the driving torque on an individual bar.
Figure~\ref{fig:torquecancellation} uses the same 17 retained inputs.

Direct, common-denominator, component, and trigonometric solutions agree
on the same frequency grid.

\bibliographystyle{apsrev4-2}
\enlargethispage{2\baselineskip}
\bibliography{references}
\end{document}